\documentclass{article}

\usepackage{arxiv}

\usepackage[utf8]{inputenc} 
\usepackage[T1]{fontenc}    
\usepackage[colorlinks=true,urlcolor = black,citecolor=black,linkcolor=blue]{hyperref}
\usepackage{hyperref}
\usepackage{url}            
\usepackage{booktabs}       
\usepackage{amsfonts}       
\usepackage{nicefrac}       
\usepackage{microtype}      

\usepackage{lipsum}         
\usepackage{graphicx}
\usepackage{doi}

\usepackage[utf8]{inputenc}
\usepackage{amsmath}
\usepackage{amssymb}
\usepackage{color}
\usepackage{amssymb,amsfonts,amsmath,amsthm,amsbsy} 
\usepackage{tabularx} 
\usepackage{mathtools,mathrsfs}
\usepackage{mhchem}
\usepackage{enumerate}
\usepackage{booktabs}
\usepackage{bm}
\usepackage{bbm}
\usepackage{multirow}
\usepackage{threeparttable}
\usepackage{float}

\usepackage[table,dvipsnames, svgnames]{xcolor}

\usepackage[stable]{footmisc}
\usepackage[sort&compress]{natbib}

\usepackage{indentfirst}
\usepackage{etoolbox,siunitx}

\usepackage{afterpage}
\robustify\bfseries
\usepackage{adjustbox}
\usepackage{chngpage}
\usepackage{rotating}
\usepackage{hvfloat}
\usepackage{soul}
\usepackage{soul}
\RequirePackage{color}

\usepackage{subcaption}
\usepackage{caption}

\usepackage{cleveref}       

\usepackage{placeins}

\title{Comparative analysis of five $NH_3$/air oxidation mechanisms}

\author{ \href{https://orcid.org/0000-0002-0104-2160}{\includegraphics[scale=0.06]{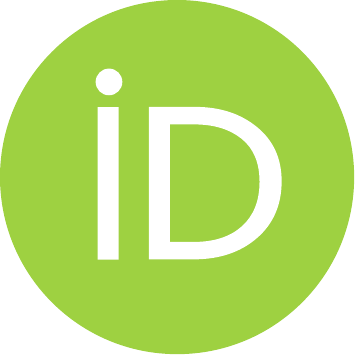}\hspace{1mm}Shahid Rabbani}\\
	Department of Mechanical Engineering\\
	Khalifa University of Science and Technology\\
	Abu Dhabi, UAE 127788 \\
	\texttt{shahid.rabbani@ku.ac.ae} \\
	\And
\href{https://orcid.org/0000-0002-4788-9877}{\includegraphics[scale=0.06]{orcid.pdf}\hspace{1mm}Dimitris M. Manias}\\
	Department of Electrical Engineering and Computer Science\\
	Khalifa University of Science and Technology\\
	Abu Dhabi, UAE 127788 \\
	\texttt{dimitris.manias@ku.ac.ae} \\
	\And
	\href{https://orcid.org/0000-0003-2870-9145}{\includegraphics[scale=0.06]{orcid.pdf}\hspace{1mm}Dimitrios C. Kyritsis}\\
Research and Innovation Center on CO$_2$ and H$_2$\\
  Department of Mechanical Engineering\\
	Khalifa University of Science and Technology\\
 Abu Dhabi 127788, United Arab Emirates\\
 \texttt{dimitrios.kyritsis@ku.ac.ae} \\
	\And
	\href{https://orcid.org/0000-0002-9674-9491}{\includegraphics[scale=0.06]{orcid.pdf}\hspace{1mm}Dimitris A. Goussis\thanks{Corresponding Author: dimitris.goussis@ku.ac.ae}}\\
Research and Innovation Center on CO$_2$ and H$_2$\\
  Department of Mechanical Engineering\\
	Khalifa University of Science and Technology\\
 Abu Dhabi 127788, United Arab Emirates\\
 \texttt{dimitris.goussis@ku.ac.ae} \\
}

\hypersetup{
pdftitle={Comparative analysis of five $NH_3$/air oxidation mechanisms},
pdfsubject={physics.chem-ph},
pdfauthor={Dimitris M. Manias},
pdfkeywords={ammonia, ignition control, CSP diagnostics},
}

\begin{document}
\maketitle

\begin{abstract}
Five recently developed chemical kinetics mechanisms for ammonia oxidation are analysed and compared, in the context of homogeneous adiabatic autoignition.~The analysis focuses on the ignition delay and is based on the explosive mode that is shown to drive the process.~Using algorithmic tools based on the Computational Singular Perturbation algorithm, the reactions responsible for the generation of the explosive mode are identified, along with the variables (species mass fractions and temperature) that associate the most to this mode.~Comparison of these sets of reactions and variables, obtained for each mechanism, allows to correlate the differences in the predictive outcomes from the mechanisms with specific reactions.~The major differences identified, which lead to different ignition delay times, relate to (i)  the relative duration of chemical and thermal runaways (a sizeable chemical runaway develops only in some mechanisms) and (ii) the dominant chemistry during the chemical runaway (chemistry involving species with two nitrogen atoms is active only in some mechanisms).~The major similarities identified refer to the thermal runaway and in particular to (i) the chemical activity, which is supported mainly by $OH$-producing reactions and by reactions producing their reactants and (ii) the thermal activity, which is dominated by strongly exothermic $OH$-consuming reactions.
\end{abstract}

\keywords{Ammonia \and Ignition Control \and CSP Diagnostics}

\section{Introduction}
\label{Introduction}

A renewed interest in using ammonia in the power generation and mobility sectors has recently emerged in terms of not only its utility as a hydrogen carrier, but also as a direct fuel.~Ammonia is a carbon-free energy carrier that can provide high energy density and its production, transportation and storage can be delivered through well-established infrastructure.~The use of ammonia as a sustainable fuel was discussed comprehensively by Kobayashi et al.~\citep{kobayashi2019science,valera2021review}.~An economic comparison of utilization of conventional fuels with the one of ammonia showed that ammonia was the least expensive fuel in terms of \$/GJ \citep{kobayashi2019science}.~Ammonia was also ranked third in terms of GJ/m$^3$, which was in agreement with the results of Zamfirescu et al.~\citep{zamfirescu2009ammonia} and Cesaro et al.~\citep{cesaro2021ammonia}, who estimated the levelized cost of electricity from green ammonia to be between \$167 and \$197 per MWH in 2040.~Cardoso et al.~\citep{cardoso2021ammonia}  proposed subsidizing ammonia as a fuel and  Valera-Medina et al.~\citep{valera2021review} suggested that the use of ammonia as a fuel will expand as a result of the continuous decrease in cost of power from renewable sources, something that is supported by the findings of Fasihi et al.~\citep{fasihi2021global} for ammonia production from hybrid PV-wind power plants.

Ammonia has been suggested as fuel for solid oxide fuel cells \citep{siddiqui2018review,lan2014ammonia}, but what is of interest in this paper is its use as a fuel for combustion \citep{frigo2013analysis,grannell2008fuel, reiter2011combustion}, which comes with serious technical challenges.~These relate to low flame speed (less than 7 cm/s for stoichiometric mixture at standard atmospheric conditions \citep{frigo2013analysis}), low heating value (LHV = 18.8 MJ/kg, as opposed to order of 45 MJ/kg for paraffins \citep{grannell2008fuel}), as well as slow kinetics, as evidenced by long ignition delays and flame stability problems \citep{tian2009experimental,mathieu2015experimental,glarborg2018modeling}.~As a result, currently the main interest in ammonia relates to marine engines, which rotate at low speeds and utilize large quantities of fuel \citep{hansson2020potential,Atchison_2021,Linnet_2021}.

Modelling ammonia oxidation kinetics has developed over the last two decades and, for obvious reasons, has recently advanced in a substantial manner.~Among others, mechanisms have been proposed by  by Tian et al.~\citep{tian2009experimental}, Mathieu et al.~\citep{mathieu2015experimental}, Glarborg et al.~\citep{glarborg2018modeling}, Shrestha et al.~\citep{shrestha2018detailed}, Li et al.~\citep{li2019chemical}, Stagni et al.~\citep{stagni2020experimental}, Zhang et al.~\citep{zhang2021combustion}, Okafor et al.~\citep{okafor2019measurement} and Otomo et al.~\citep{otomo2018chemical}.~The relative lack of experimental data has so far limited the validity of the proposed mechanisms to specific combustion configurations as well as ranges of initial temperature, pressure, and equivalence ratio.~For example, the mechanism by Mathieu et al.~\citep{mathieu2015experimental} has been validated in experiments with $NH_3/O_2$ mixtures at high temperatures and pressures.~However, this mechanism failed to produce satisfactory results at intermediate temperature and pressure conditions \citep{wang2021study}.~Similarly, the mechanism by Tian et al.~\citep{tian2009experimental} predicted laminar flame speeds with acceptable accuracy, but it lacked accuracy in predicting ignition delay times.~The results of Glarborg et al.~\citep{glarborg2018modeling} were in good agreement with ignition delay measurements in a variety of initial conditions; however, they exceedingly overestimated reactivity in the $NH_3/H_2$ mixture \citep{wang2021study}.~Similarly, the mechanism by Shrestha et al.~\citep{shrestha2018detailed} performed well on shock-tube data, such as in \citep{mathieu2013shock, burke2017ephemeral, mathieu2015experimental}, but the model underestimated ignition delay time by factor of 2.5 for ignition of $NH_3/H_2$ mixtures  \citep{dai2020experimental}.

It is exactly because of these discrepancies that some recent works have attempted comparisons between the proposed mechanisms, almost exclusively in the context of agreement with experimental data from ignition experiments in shock tubes and rapid compression machines.~Wang et al.~\citep{wang2021study} tested the mechanisms of Otomo et al.~\citep{otomo2018chemical}, Glarborg et al.~\citep{glarborg2018modeling}, Li et al.~\citep{li2019chemical}, Okafor et al.~\citep{okafor2019measurement} and Shrestha et al.~\citep{shrestha2018detailed} against the RCM data of He et al.~\citep{he2019auto} and the shock-tube data of Shu et al.~\citep{shu2019shock} and chose to use Otomo's mechanism for high-pressure autoignition computations in diesel-combustion conditions (100-200 bar).~However, the comparison presented in the paper, could also justify the choice of other mechanisms.~Otomo's mechanism was also supported by a recent comparison of no less than 15 mechanisms that was performed in the context of $NH_3$ pyrolysis by Alturaifi et al.~\citep{alturaifi2022experimental}.~The authors compared the predictions of each of the 15 mechanisms on the speciation of the pyrolysis of 0.5\%-molar $NH_3$ and of a mixture of 0.42\%-molar $NH_3$ and 2\%-molar $H_2$ in $Ar$ in a shock tube, pointed to Otomo's predictions as demonstrating best agreement with the experimental results, and proceeded to present an improved model for the pyrolysis.~Then, they incorporated the improved pyrolysis model in all oxidation mechanisms they considered.~This modification reduced the initial discrepancy in predicted laminar flame speeds from a factor of three to a factor of 1.5.~The scatter of laminar flame speed results by a factor of two among mechanism predictions and  experimental data was in agreement with the early findings of Xiao et al.~\citep{xiao2017chemical}.~A similar methodology in terms of comparing with experimental results was followed by da Rocha et al.~\citep{da2019chemical} who compared the predictions of ten mechanisms in terms of ignition delay times, flame speeds, and $NOx$ generation.~The flame-speed data were also compared with experimental measurements.~The authors described the state of the art very accurately by stating that ``\emph{currently available chemical kinetic mechanisms predict rather scattered ignition delay times, laminar flame speeds, and $NOx$ concentrations in $NH_3$ flames, indicating that improvements in sub-mechanisms of $NH_3$ and $NH_3/H_2$ oxidation are still needed}".~The conclusion is compounded by a recent quantitative assessment of the relative accuracy with which eight $NH_3$ oxidation mechanisms could capture experimental ignition-delay-time results \citep{kawka2020comparison}, which showed that the accuracy of the computational prediction can exceed the standard deviation of the experimental measurements by more than a factor of ten.~This was probably a conclusion more important than pointing to the mechanisms by Glarborg  et al.~\citep{glarborg2018modeling} and Shrestha et al.~\citep{shrestha2018detailed} as “best” in terms of the particular comparison and, of course, it is nothing new in combustion research.~Rather, it is reminiscent of the decades of gradual progress that it took in order to converge to widely acceptable mechanisms for the (chemically much more complicated) oxidation of methane, and indicative of how much is ``\emph{still needed}" \citep{da2019chemical}.

Much as the comparison of the predictions of the proposed mechanisms with experimental data is valuable, especially at what seems to be an early stage in the construction of high-fidelity models for the oxidation of  $NH_3$, what seems to be missing from the published literature is a comparison of the proposed mechanisms in terms of the chemical dynamics they imply.~In this paper, we aspire to cover this shortage by examining in detail the explosive dynamics of the adiabatic, isochoric autoignition of $NH_3$/air mixtures as this is described by the mechanisms of Glarborg et al.~\citep{glarborg2018modeling}, Shrestha et al.~\citep{shrestha2018detailed}, Li et al.~\citep{li2019chemical}, Stagni et al.~\citep{stagni2020experimental}, and Zhang et al.~\citep{zhang2021combustion}.~The choice of the particular mechanisms was determined mainly by indications of “good performance” in the comparisons cited above and much more by practicalities in terms of availability of the mechanisms to the authors and size of the analysis.~Clearly, the methodology can be applied to other mechanisms that have appeared or will appear in the future.

In order to analyze the chemical dynamics described by each of the mechanisms considered here, we used the algorithmic tools of Computational Singular Perturbation algorithm \citep{Lam1988,Lam1994,hadjinicolaou1998asymptotic}.~CSP allows for the identification of (i) the mode that drives the process and (ii) the reactions and variables that relate the most to this mode \citep{Valorani2003,SIAM2006,kazakov2006computational,lu2008analysis,Prager2011,Diamantis2015b}.~The driving mode can be of either explosive or dissipative character; i.e., the reactions responsible for the emergence of this mode tend to lead the process away or towards equilibrium \citep{Lam1988}.~CSP can handle both types of modes and provide all relevant diagnostics.~This work is following up on our previous CSP analyses of $NH_3$ oxidation that were focused on the determination of additives for ignition control and $NO$ reduction \citep{khalil2019,khalil2021no}, as well as on highlighting differences with methane oxidation \citep{manias2021nh3}.

In what follows, we start with a presentation of the physical problem of the homogeneous, isochoric, adiabatic autoignition and we present ignition-delay computations with the use of the five mechanisms under consideration for a wide range of initial temperatures, pressures and equivalence ratios.~Then, we use CSP diagnostics in order to discuss the predictions of the employed mechanisms in terms of the relative duration of chemical vs. thermal runaway and we point to a conclusion that seems to be unanimous among the employed mechanisms, despite their serious differences in terms of dynamic behavior; i.e. that $NH_3$ autoignition has a relatively short chemical runaway and the longest part of the ignition delay comprises a thermal runaway.~Additionally, we compare the mechanisms in terms of predictions for the initiation of the autoignition, the chemical activity that supports the relatively long thermal runaway, and the generation of heat through exothermic reactions.~In this manner, we aspire to demonstrate the chemical features that each of the mechanisms highlights, which can guide model selection for specific engineering applications.~In order to help the reader focus, we present the detailed CSP diagnostics for each mechanism in appendices and reserve the main body of the paper for the results that relate to the comparison of the ways in which the five mechanisms under consideration capture the underlying chemical fundamentals.

\section{Chemical Kinetics Mechanisms}
\label{Mechanisms}

The chemical kinetics mechanisms considered are those developed by Glarborg et al. \citep{glarborg2018modeling}, Shrestha et al. \citep{shrestha2018detailed}, Li et al. \citep{li2019chemical}, Stagni et al. \citep{stagni2020experimental} and Zhang et al. \citep{zhang2021combustion}:
\begin{enumerate}
\item Glarborg$\_$2018  \citep{glarborg2018modeling}: was compiled by extending previous studies of this group \citep{dagaut2008oxidation,mendiara2009reburn,mendiara2009ammonia,lopez2009experimental} and using experimental data.~The mechanism was validated for initial temperatures and pressures and equivalence rations T$_0$ = 285 - 2500 K,  p$_0$ = 0.045 - 1.38 atm, $\phi$ = 0.002 - 2.0.

\item Shrestha$\_$2018 \citep{shrestha2018detailed}:   was developed on the basis of previous ones \citep{skreiberg2004ammonia,baulch2005evaluated,rasmussen2008experimental,mendiara2009reburn,hong2010shock} and of experimental data.~The validity range of this mechanism is T$_0$ = 298 - 2455 K, p$_0$ = 0.045 - 30 atm, $\phi$ = 0.12 - 2.0.

\item Li$\_$2019  \citep{li2019chemical}:  was developed by assembling three different mechanisms  \citep{aramcomech,shrestha2018detailed,tian2009experimental} and then using Directed Relation Graph with Error Propagation (DRGEP) for its simplification.~This mechanism is valid in the range T$_0$ = 1000 - 2000 K, p$_0$ = 0.98 - 49.34 atm,  $\phi$ = 0.5 - 2.0.

\item Stagni$\_$2020 \citep{stagni2020experimental}:  was generated on the basis of experimental data acquired by the authors and previous theoretical work done by the group \citep{faravelli2003kinetic,ranzi2012hierarchical,metcalfe2013hierarchical,song2019sensitizing,chen2019decomposition}.~The range of validity of the mechanism is T$_0$ = 500 - 2000 K, p$_0$ = 0.8 - 98.7 atm, $\phi$ = 0.5 - 2.0.

\item Zhang$\_$2021 \citep{zhang2021combustion}:   was compiled on the basis of new experiments and previously constructed mechanisms \citep{burke2017ephemeral,hashemi2015hydrogen,mei2019experimental}.~It was validated in the range T$_0$ = 800 - 1280 K, p$_0$ = 1 - 10  atm, $\phi$ = 0.25 - 1.0. 
\end{enumerate}

The number of species and reactions related to nitrogen chemistry in these chemical kinetics mechanisms  are listed in Table \ref{tab:mechanisms}.~In addition, indicative values are listed of the ignition delay time $t_{ign}$ and of the minimum value of the time scale that relates to the explosive mode $\tau_{e,min}$, computed at T$_0$=1100 K, p$_0$=2 atm and $\phi$=1.0.~This set of parameters was selected based on the following considerations.~First, this set is within the range of validity of all mechanisms.~Second, the selected values of T$_0$ and $\phi$ are relevant to the operation of gas turbines and reciprocating internal combustion engines.~Finally, the selected value of p$_0$ is relatively low, because in Ref.~\citep{glarborg2018modeling} the Glarborg$\_$2018 mechanism was validated at low pressures, although it was later suggested that it is valid at higher pressures as well \citep{valera2021review}.~

It is shown in Table \ref{tab:mechanisms} that there are no significant differences in the number of species and reactions of the various mechanisms.~However, significant differences are recorded for the computed $t_{ign}$, although the values of $\tau_{e,min}$ do not vary as much.

\vspace{0cm}
\begin{table}[t]
\caption{Number of species and reactions in the five $NH_3/air$ chemical kinetics mechanisms, along with $t_{ign}$ and $\tau_{e,min}$ at T$_0$=1100 K, p$_0$=2 atm and $\phi$=1.0.}
\begin{center}
\begin{tabular}{@{~}@{~}l@{~}c@{~}c@{~}c@{~~}c}
\hline
Mechanism                                   				& ~~Species~ 	& ~~Reactions~  	& ~~$t_{ign}$ (s)~		& ~~$\tau_{e,min}$ (s)                       \\
\hline
Glarborg$\_$2018      	& 33               	& 211                      	& 0.27             & 1.32$\times10^{-6}$ \\
Shrestha$\_$2018      	& 34              	& 261                    	& 0.14             & 1.85$\times10^{-6}$ \\
Li$\_$2019            		& 34             	& 252                    	& 0.13             & 1.88$\times10^{-6}$ \\
Stagni$\_$2020    		& 31             	& 203                      	& 0.64             & 2.15$\times10^{-6}$ \\
Zhang$\_$2021       		& 34             	& 262                    	& 0.38             & 2.23$\times10^{-6}$ \\
\hline
\end{tabular}
\label{tab:mechanisms}
\end{center}
\end{table}

All five mechanisms include the species $Ar$, $H$, $H_2$, $H_2NN$, $H_2NO$, $H_2O$, $H_2O_2$, $He$, $HNO$, $HNO_2$, $HNOH$, $HO_2$, $HONO$, $HONO_2$, $N$, $N_2$, $N_2H_2$, $N_2H_3$, $N_2H_4$, $N_2O$, $NH$, $NH_2$, $NH_2OH$, $NH_3$, $NNH$, $NO$, $NO_2$, $NO_3$, $O$, $O_2$ and $OH$.~In addition,  Glarborg$\_$2018 includes $HON$ and $O_3$ and  Shrestha$\_$2018, Li$\_$2019 and Zhang$\_$2021 include $HON$, $OH^{*}$ and $HNO_3$.~Regarding the reactions in the five chemical kinetics mechanisms, there are various degrees of overlap.~In terms of nitrogen chemistry, Shrestha$\_$2018 shares the subsets of $H_2NO$ and $NO_3$ of reactions with Glarborg$\_$2018.~Li$\_$2019 shares a very large portion of nitrogen chemistry with Shrestha$\_$2018.~Stagni$\_$2020 shares the the $NH_2$ and part of the $NNH$ subsets of rections with Glarborg$\_$2018 and the $NH_2OH$, $HNOH$ and $NH_2OH$ subsets of reactions with Shrestha$\_$2018.~Finally, Zhang$\_$2021 shares a large part of nitrogen chemistry with Shrestha$\_$2018.~In terms of hydrogen reactions, there is no shared chemistry between Shrestha$\_$2018 and Glarborg$\_$2018.~Similarly, there is no shared chemistry between Li$\_$2019 and both Shrestha$\_$2018 and Glarborg$\_$2018, except the $OH^{*}$ chemistry, which is shared with Shrestha$\_$2018.~Stagni$\_$2020 shares all hydrogen-chemistry reactions with Li$\_$2019, except the $OH^{*}$ chemistry.~Zhang$\_$2021 shares all hydrogen reactions with Glarborg$\_$2018, with the exception of a few reactions.

These five mechanisms have been included in a number of investigations that compared the agreement of computational predictions of various mechanisms with experimental data in several combustion configurations; i.e., shock-tubes, laminar flames, stirred reactors \citep{wang2019comparative,da2019chemical,kawka2020comparison,wang2021study,alturaifi2022experimental,dagaut2022oxidation}.~Apart from comparing profiles (e.g., various variables, reaction rates, ignition delays, flame speeds), the agreement among the various mechanisms was assessed on the basis of sensitivity analysis (SA), rate of production analysis (RoPA) and element flux analysis (EFA).~Such methodologies do identify some reactions as important.~These are reactions that either participate in the equilibria generated due to the fast dynamics or drive the process and research should concentrate on them, in order to improve the validity of the mechanism.~SA of global quantities (ignition delay or flame speed) does not provide information on the progress of the process.~Additionally, SA of both global and  local (species mass fraction) quantities, as well as RoPA and EFA, cannot distinguish between reactions that participate in established equilibria or drive the process.~Moreover, EFA results were provided at one instance only, so the conclusions reached were of limited value.~Although these methodologies do not provide the time-resolved details that CSP does, certain conclusions can be drawn.~A widely accepted conclusion refers to the importance of the sequence $NH_3\rightarrow NH_2 \rightarrow H_2NO\rightarrow HNO$ \citep{wang2019comparative,da2019chemical,kawka2020comparison,wang2021study,dagaut2022oxidation}.~In general, there is an agreement on the reactions that realise this path.~Several works also mention the importance of reactions involving species with two atoms of nitrogen in their molecules (to be hereafter referred to as $N2$ chemistry) \citep{wang2019comparative,da2019chemical,wang2021study,dagaut2022oxidation}.~In this case, there is no set of reactions that is accepted to realise this path.~It is also noted that the importance of the influence of $OH$ is explicitly pointed out in Refs.~\citep{wang2021study,dagaut2022oxidation} and is implicitly supported by the reactions identified in Refs.~\citep{wang2019comparative,da2019chemical}.



\begin{figure*}[t]
\centering
 \includegraphics[scale=0.215]{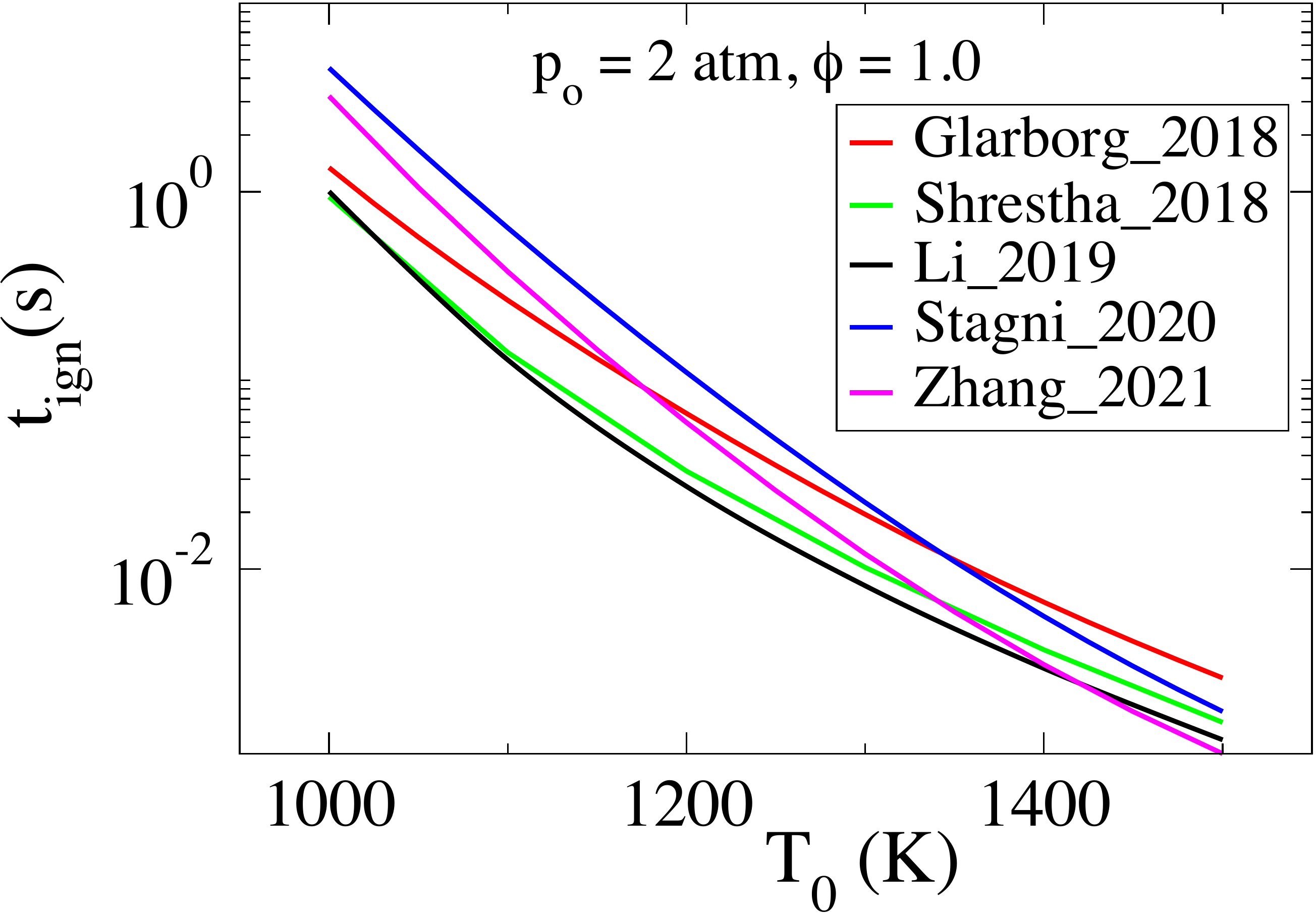}  \hfill  \includegraphics[scale=0.215]{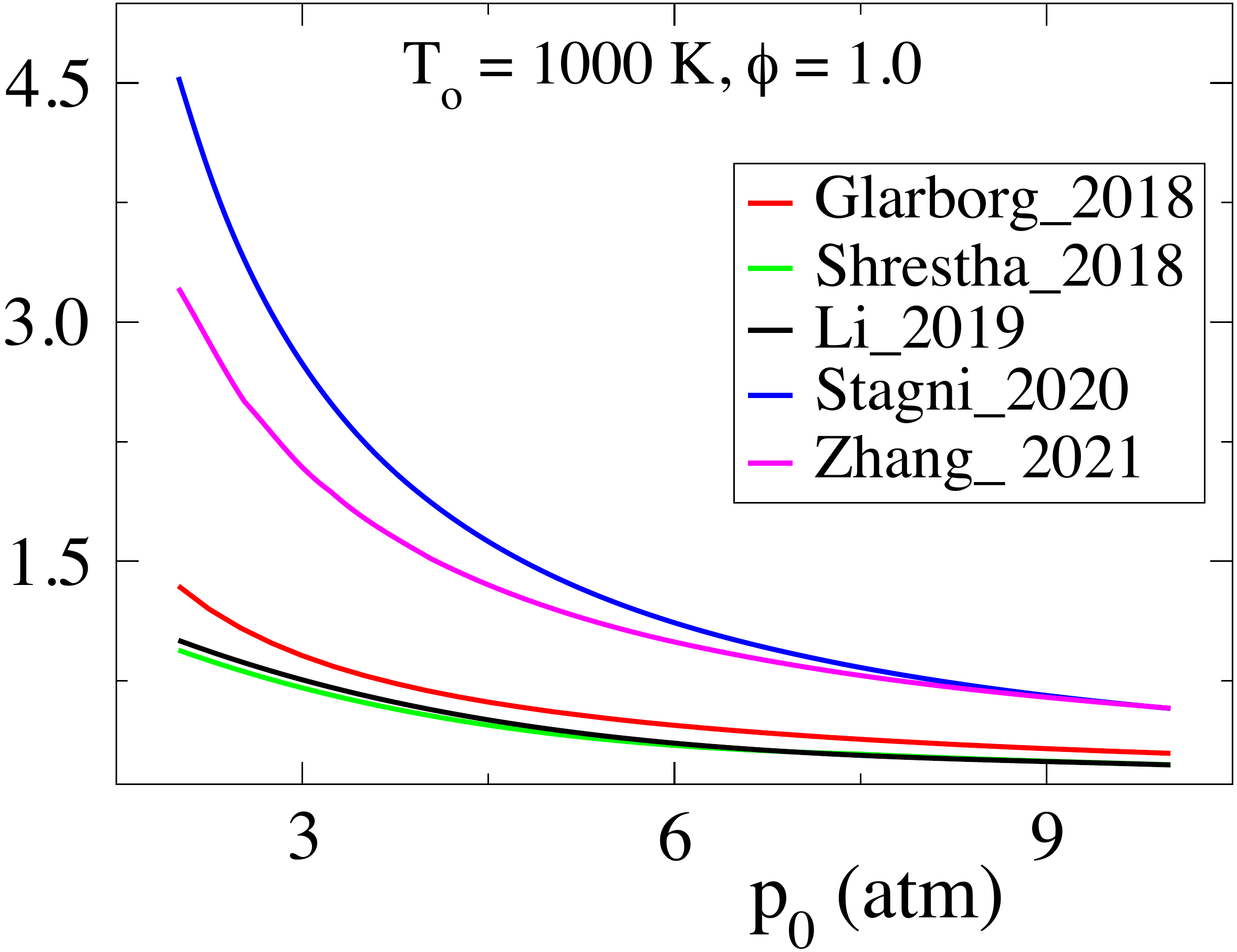}  \hfill  \includegraphics[scale=0.215]{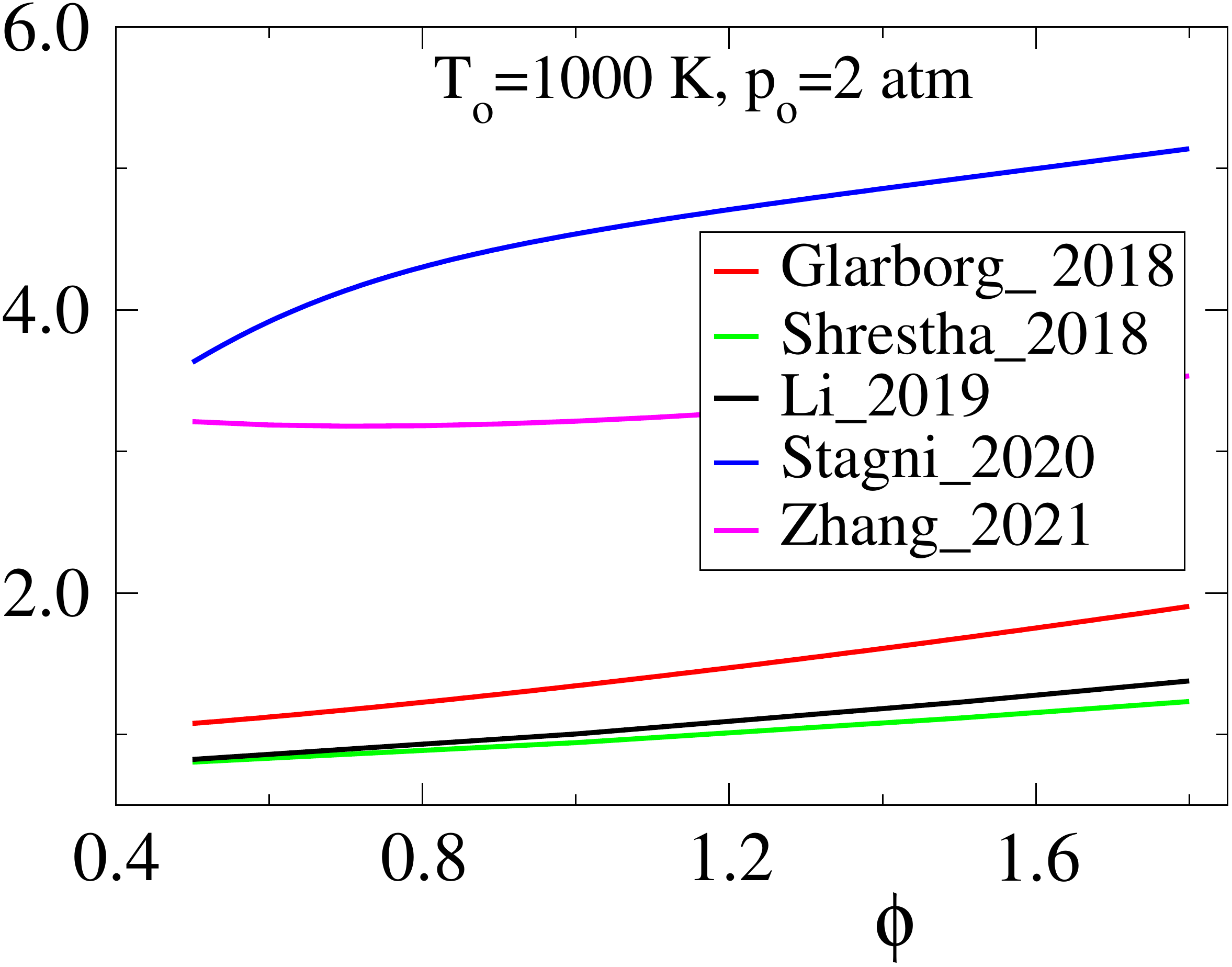}  \\ 
\vspace{10 pt}
 \includegraphics[scale=0.215]{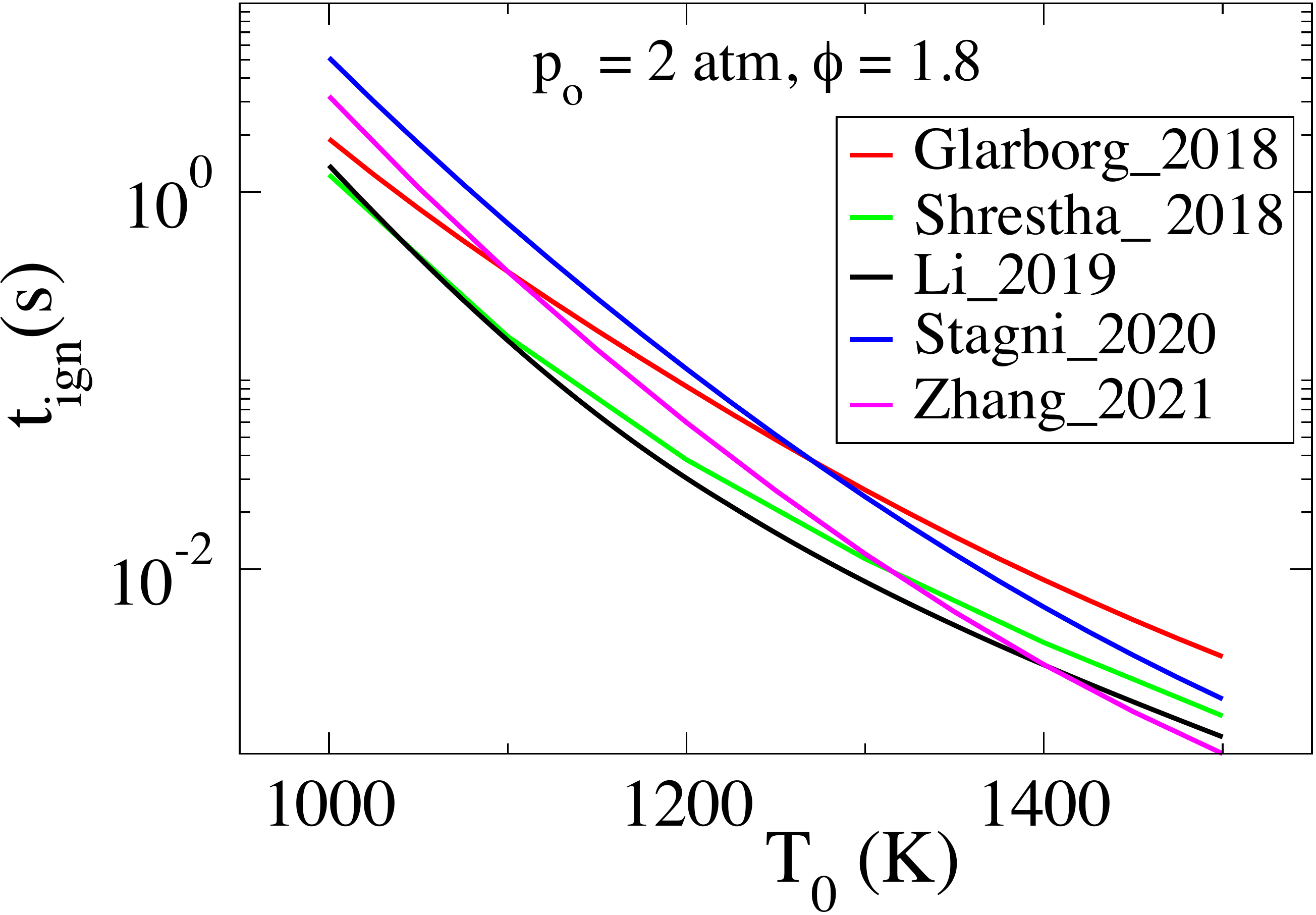} \hfill  \includegraphics[scale=0.215]{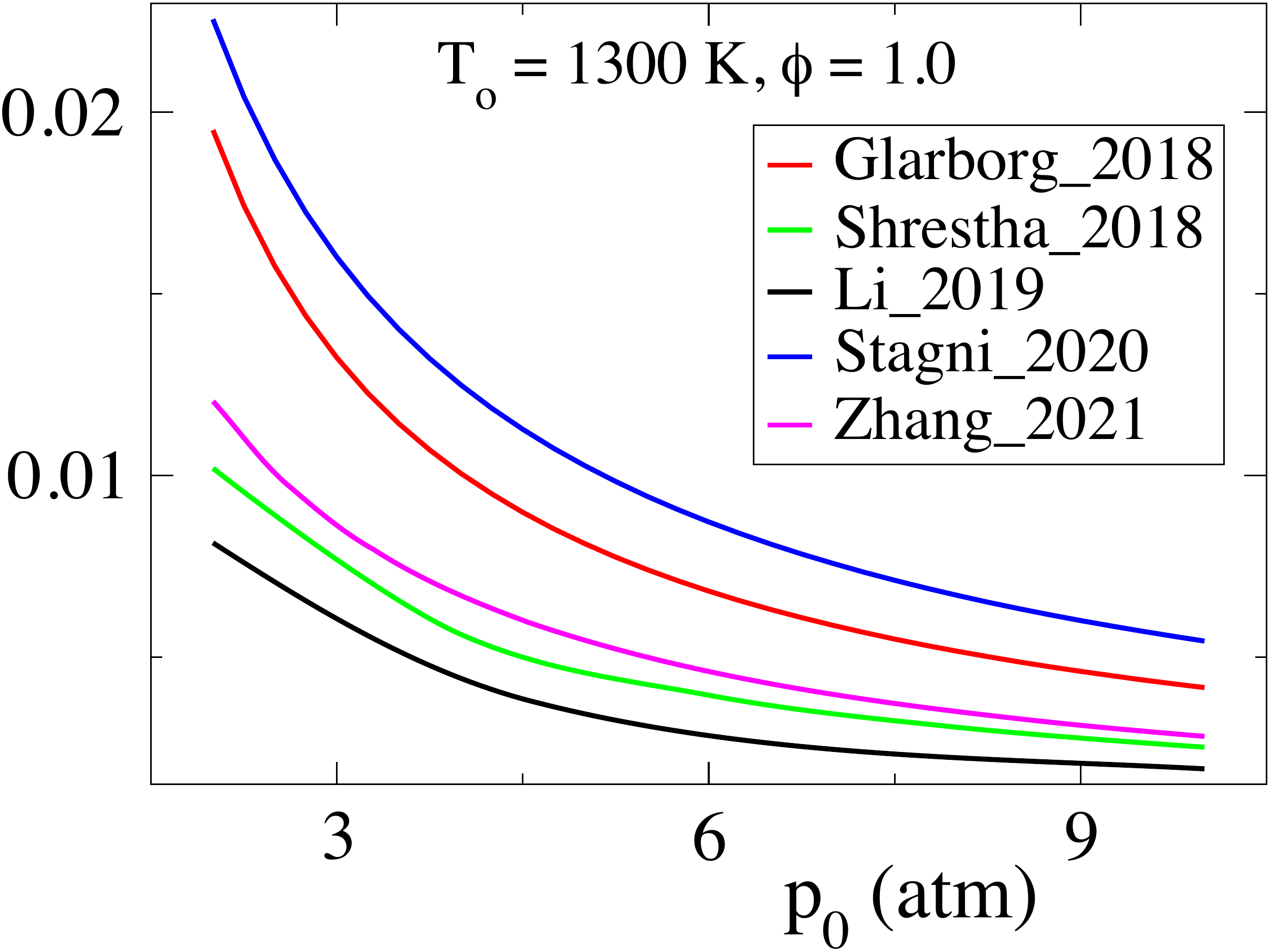}   \hfill \includegraphics[scale=0.215]{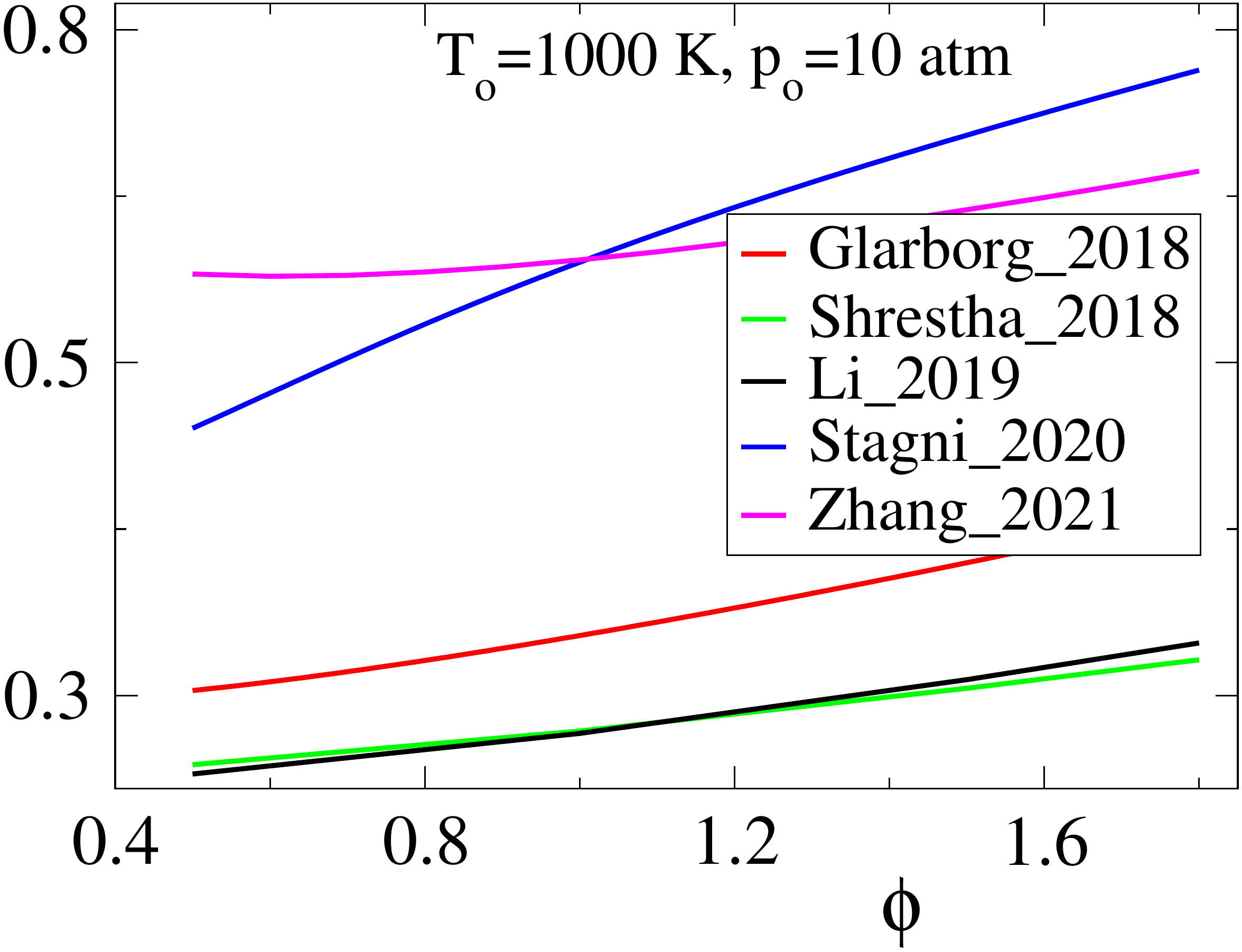}    \\ 
\caption{The Ignition Delay Time $t_{ign}$ as a function of T$_0$, p$_0$ and  $\phi$.}
\label{fig:Ign_Delay}
\end{figure*}

Despite the significant similarities in the mechanisms considered here, Table \ref{tab:mechanisms} shows that there are significant differences in the computed values of IDT, $t_{ign}$, while there is a better agreement on the minimum value of the explosive time scale $\tau_{e,min}$ that occurs at about $t_{ign}$.~A broader picture of the differences in $t_{ign}$ is provided in Fig.~\ref{fig:Ign_Delay}, where the variation of $t_{ign}$ in a wide range of values of T$_0$, p$_0$ and  $\phi$ is displayed.~It is shown that $t_{ign}$ computed from the five mechanisms follows the same trends.~Moreover, it is shown that the values of $t_{ign}$ produced by  Shrestha$\_$2018 and Li$\_$2018  are in general very close.~The profiles of $t_{ign}$ produced by  Glarborg$\_$2018 follow those of Shrestha$\_$2018 and Li$\_$2018, however they over-predict IDT.~Finally, the profiles of $t_{ign}$ produced by  Stagni$\_$2020 and Zhang$\_$2021 are close to the profiles generated by the other mechanisms only for certain range of  values of T$_0$, p$_0$ and  $\phi$.

Figure~\ref{fig:Ign_Delay} shows that the largest drop of $t_{ign}$ with increasing values of T$_o$ is provided by Zhang$\_$2021 and the smallest by Glarborg$\_$2018.~Also, the largest drop of $t_{ign}$ with increasing values of p$_o$ is provided by Stagni$\_$2020 and Zhang$\_$2021 for low values of T$_o$ and by Stagni$\_$2020 and Glarborg$\_$2018 for large values of T$_o$, while the smallest drop is provided by Shrestha$\_$2018 and Li$\_$2018.~Finally, it is shown that the largest sensitivity to $\phi$ is exhibited by Stagni$\_$2020 and the smallest by Zhang$\_$2021.~The increase of $t_{ign}$ with $\phi$, indicated by all mechanisms is in contrast with what happens with hydrocarbons and alcohols \citep{TINGAS201828,rabbani2022chemical,rabbani2022dominant}. 

Given that all five mechanisms considered involve the same species, either exactly or with very few differences, the large differences in the prediction of $t_{ign}$, displayed in Fig.~\ref{fig:Ign_Delay}, have to be attributed to differences in the influence of reactions in each mechanism.~In the following, the species and reactions that influence the most $t_{ign}$ will be identified and compared among the five mechanisms.

\section{CSP methodology}
\label{CSP}

The process of homogeneous autoignition is a multi-scale problem.~The fast time scales become exhausted quickly and the process is then characterised by the slower time scales \citep{Lam1988,goussis1992study,valorani01,Valorani2003}.~The time scale that characterises the process during the ignition delay time (IDT) for each of the five cases considered will be identified and analysed with the tools developed in the context of the Computational Singular Perturbation (CSP) algorithm.

The adiabatic, isochoric and homogenous autoignition is governed by the equations:
\begin{align}
&&\dfrac{d\mathbf{y}}{dt} =& \dfrac{1}{\rho}\mathbf{W}\cdot  \sum_{k=1}^{2K} \mathbf{S}_kR^k & \label{eq:gov1} \\
&&\dfrac{dT}{dt} =& \dfrac{1}{\rho c_v} \left( -\mathbf{h}_c \cdot \mathbf{W} + \mathrm{R}T  \mathbf{U} \right)  \cdot \sum_{k=1}^{2K} \mathbf{S}_kR^k \label{eq:gov2}
\end{align}
\noindent where the vector $\mathbf{y}$ includes the $N$ species mass fractions, $T$ is the temperature, $\rho$ the mixture density and $c_v$ the heat capacity at constant volume \citep{law2010combustion}.~Assuming $K$ reactions in the chemical kinetics mechanism, their forward and backward rates are considered separately in Eqs.~\eqref{eq:gov1} and \eqref{eq:gov2}, where $R^k$ denotes the rate of the $k$-th unidirectional reaction and $\mathbf{S}_k$ represents the related N-dim.~stoichiometric vector.~The N-dim.~vector $\mathbf{h}_c$ denotes the species absolute enthalpies, $\mathbf{W}$ is a $N \times N$ diagonal matrix of the species molecular weights, $\mathrm{R}$  is the universal gas constant and $\mathbf{U} = [1, 1, \cdots, 1]$.

CSP considers the governing Eqs.~\eqref{eq:gov1} and \eqref{eq:gov2} in the vector form:
\begin{equation}
\label{eq:gov2a}\\
\dfrac{d\mathbf{z}}{dt} = \sum_{n=1}^{N+1}\mathbf{a}_nf^n       \hfill \qquad \qquad \qquad \hfill      f^n=\mathbf{b}^n .~\mathbf{g}(\mathbf{z})
\end{equation}
where $\mathbf{z}=[\mathbf{y},T]^T$ is a ($N$+1)-dim.~vector, $\mathbf{a}_n$ is the ($N$+1)-dim.~column CSP basis vector, $\mathbf{b}^n$ is the related ($N$+1)-dim.~dual row vector ($\mathbf{b}^i \cdot \mathbf{a}_j = \delta_j^i$) \citep{Lam1988,Lam1994} and $\mathbf{g}(\mathbf{z})$ represents the summation of the $2K$ terms in Eqs.~\eqref{eq:gov1} and \eqref{eq:gov2}:
\begin{equation}
\label{eq:gov1a}
\mathbf{g}(\mathbf{z}) =\mathbf{\hat{S}}_1R^1+\mathbf{\hat{S}}_2R^2+ \dots + \mathbf{\hat{S}}_{2K}R^{2K}
\end{equation}
where $\mathbf{\hat{S}}_k$ is the generalized stoichiometric vector, the relation of which to $\mathbf{S}_k$ is presented in  \citep{Diamantis2015b}.~In Eq.~\eqref{eq:gov2a} the amplitude $f^n$  is set positive, by appropriately changing the sign of $\mathbf{b}^n$ and $\mathbf{a}_n$.~Each CSP mode $\mathbf{a}_nf^n$ in Eq.~\eqref{eq:gov2a}  is characterized by a single time scale, say $\tau_n$; $n=1$ refers to the fastest time scale ($\tau_1$) and $n=N$ to the slowest one ($\tau_N$).~Assuming the M fastest modes being exhausted, Eq.~\eqref{eq:gov2a} simplifies to the reduced model:
\begin{equation}
\label{eq:gov2aa}\\
\dfrac{d\mathbf{z}}{dt} \approx \sum_{n=M+1}^{N+1}\mathbf{a}_nf^n      \hfill  \qquad \qquad   \qquad   f^m\approx 0 
\end{equation}
where $m=1,\dots,M$ \citep{Lam1988,Lam1994}.~The M algebraic equations in Eq.~\eqref{eq:gov2aa} approximate the equilibria that are generated by the exhausted modes, within which the solution evolves according the system of differential equations.~The availability of the reduced model in Eq.~\eqref{eq:gov2aa} allows for the identification of the reactions and variables that control the evolution of the process.~In summary, the following tools are used here:
\begin{itemize}

    \item \emph{Time scale Participation Index} (TPI): measures the relative contribution of each reaction to the development of the time scale $\tau_n$, that is associated with the $n$-th CSP mode $\mathbf{a}_nf^n$ \citep{diamantis2015reactions,SIAM2006}.~A significant positive/negative value of the TPI suggests that the reaction promotes the explosive/dissipative character of $\tau_n$.~An explosive/dissipative time scale is generated by reactions that tend to drive the system away from/towards equilibrium \citep{tingas2016comparative, tingas2018ch4,song2018computational}.

    \item \emph{Amplitude Participation Index} (API): measures the relative contribution of each reaction to magnitude of the amplitude $f^n$ of the $n$-th CSP mode \citep{massias1999algorithm,Diamantis2015b}.~In the case of an exhausted modes, large APIs are exhibited by reactions that participate in the equilibria generated by the fast dynamics, while in the case of active modes (either explosive or dissipative), a positive/negative value of API is exhibited by a reaction acting towards strengthening/weakening the impact of the $n$-th mode \citep{fotache1997ignition,kazakov2006computational,manias2019topological}.

    \item \emph{Pointer} (Po): identifies the variables (species mass fraction or temperature) that relate the most to the $n$-th mode \citep{goussis1992study,Valorani2003,manias2016mechanism}.
    \end{itemize}
   
Naturally, the modes that drive the autoignition process during IDT are among those represented in Eq.~\eqref{eq:gov2aa}, the dynamics of which involves the slow time scales.~Since all modes in this equation are active, the time scale of a driving mode must be among the fastest of these time scales and its amplitude must be among the largest, so that this mode has a substantial impact.~In order to identify these modes, the time scales and amplitudes of all modes in Eq.~\eqref{eq:gov2a} will be examined next.

\subsection{Time scale of explosive mode}
\label{tign}

\begin{figure}[t!]
\centering
\hfill \includegraphics[scale=0.26]{ 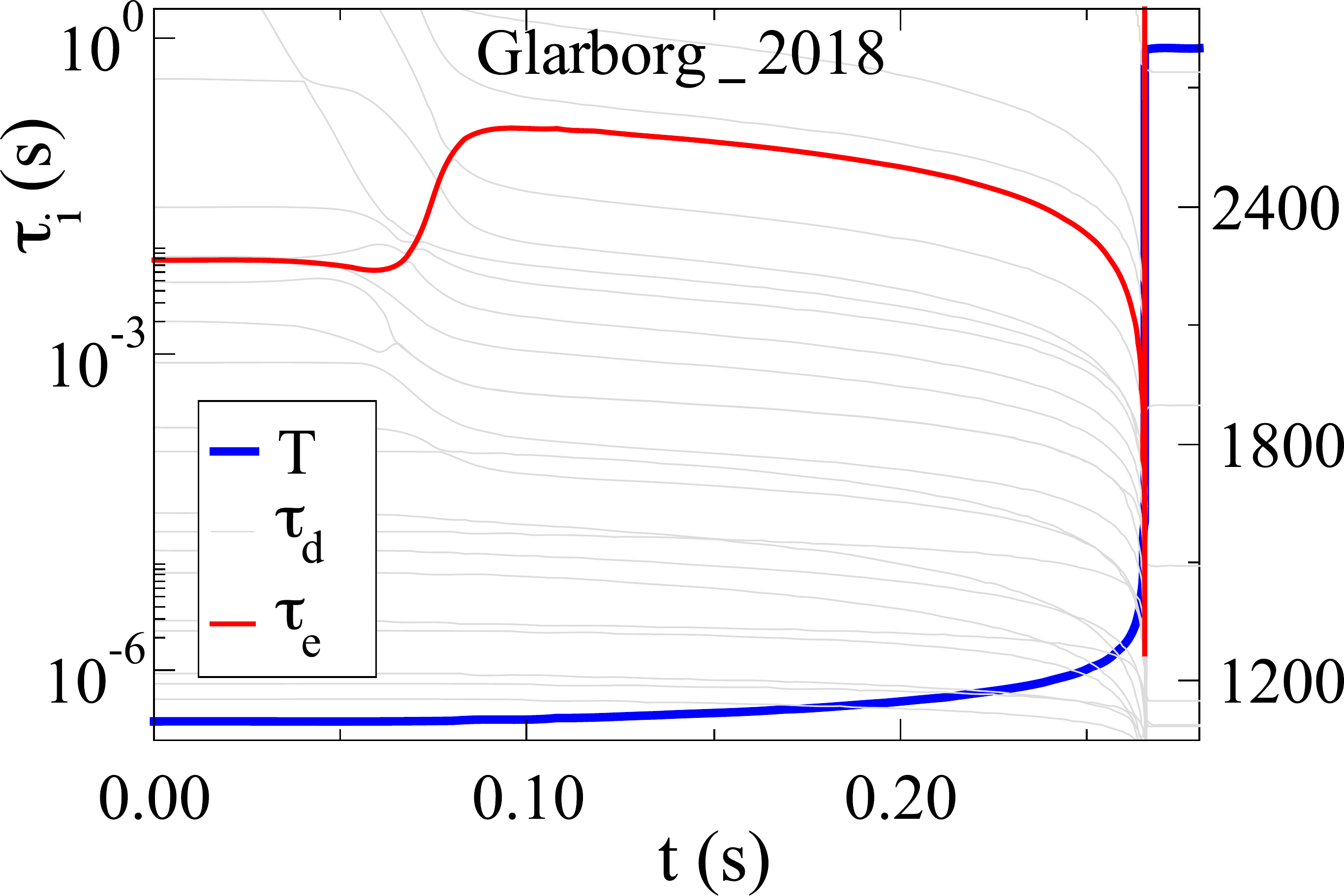}  \hfill \includegraphics[scale=0.26]{ 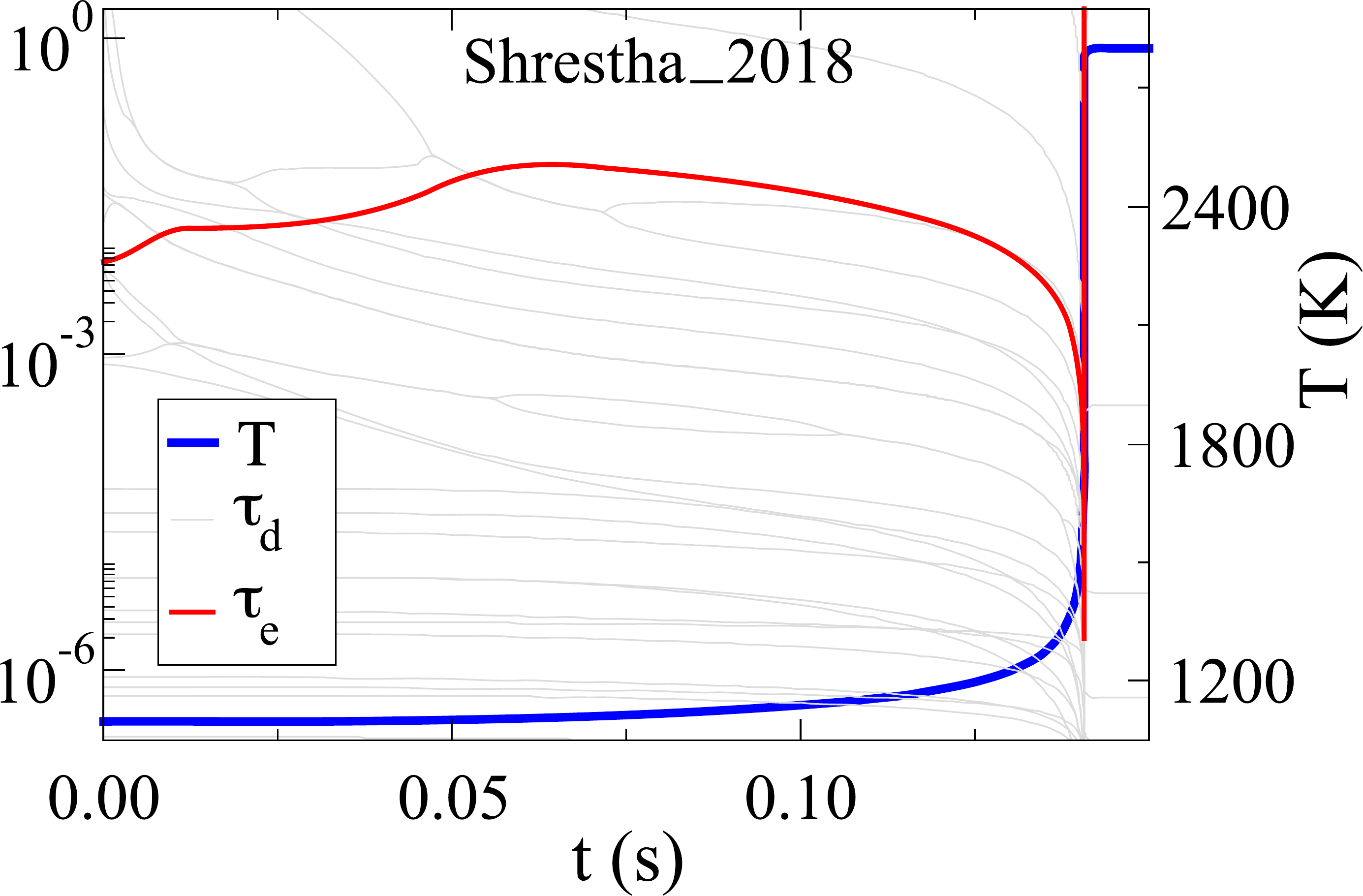} \hfill\hfill\hfill\\
 \vspace{10 pt}
\hfill\includegraphics[scale=0.26]{ 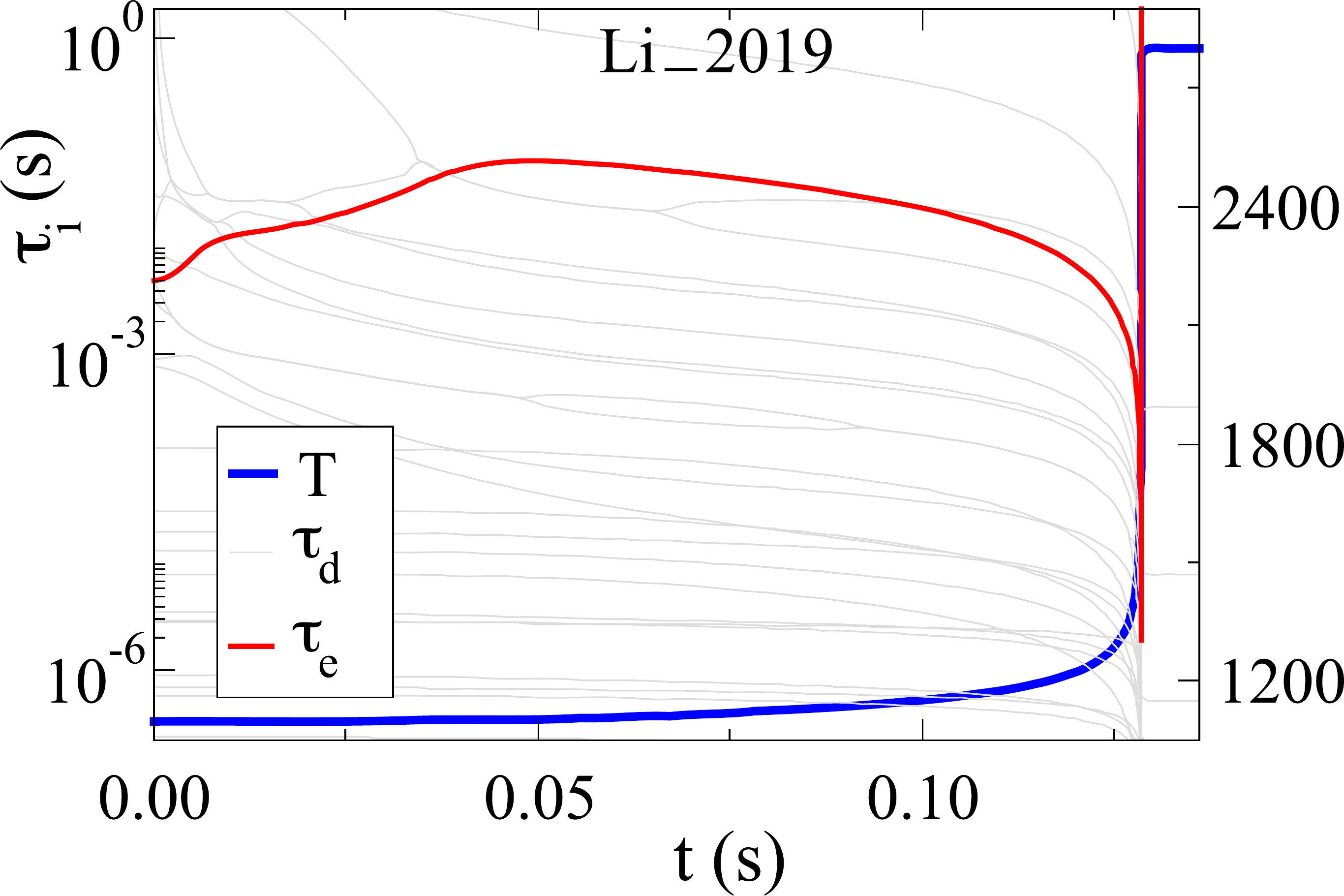} \hfill  \includegraphics[scale=0.26]{ 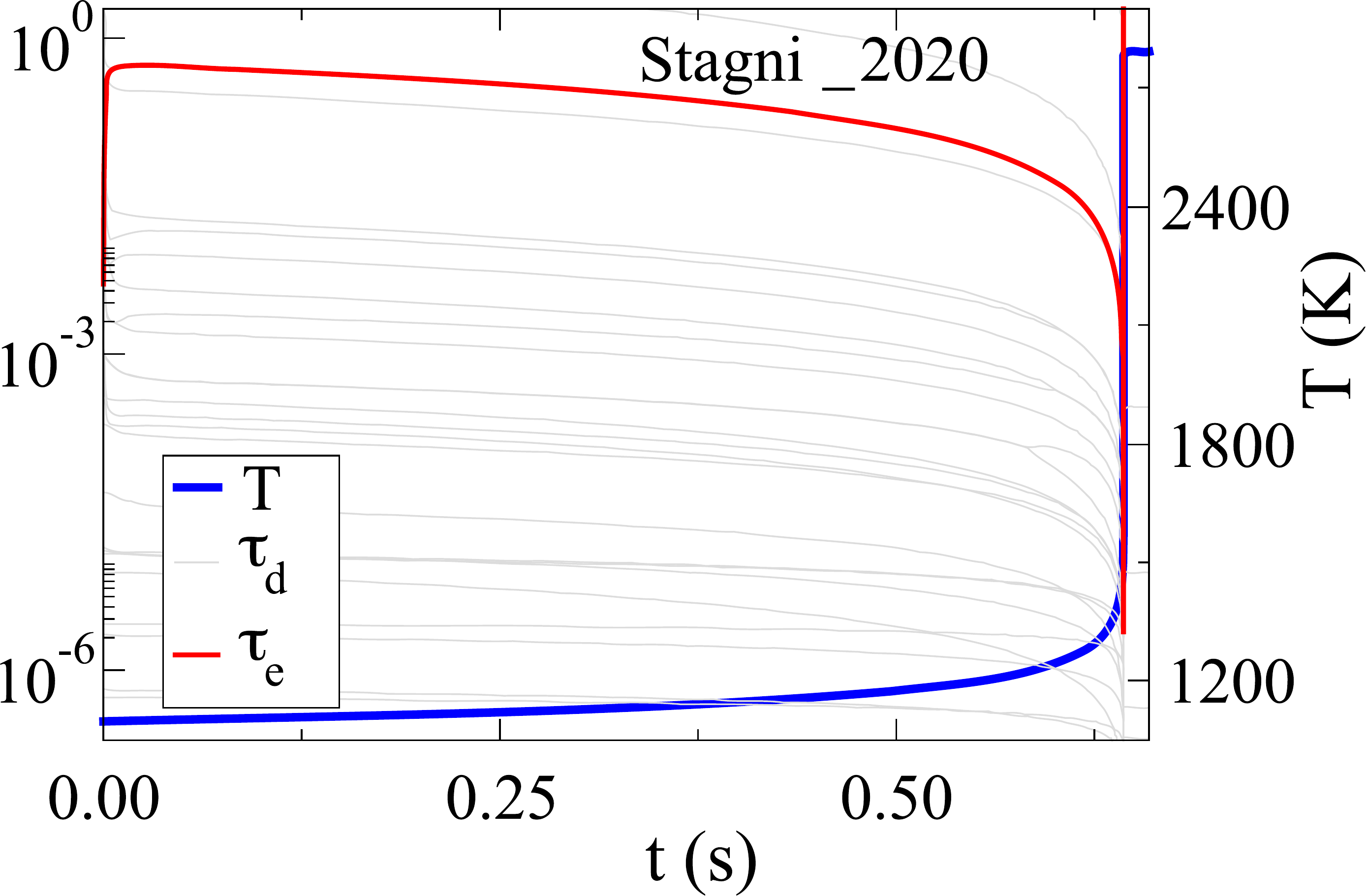} \hfill\hfill\hfill\\
\vspace{10 pt}
\includegraphics[scale=0.26]{ 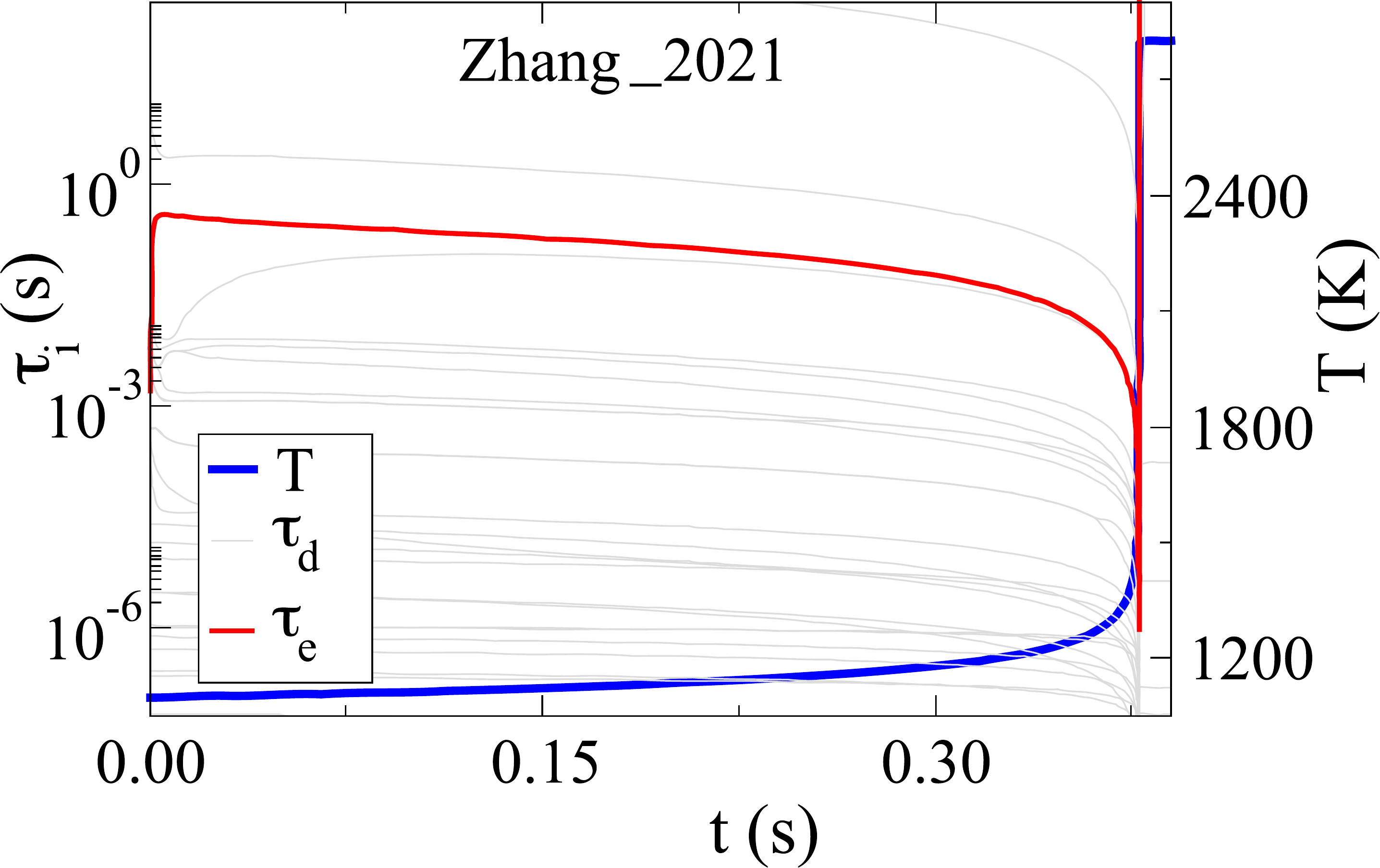}
\vspace{-0 pt}
\caption{The developing dissipative $\tau_{d}$ (grey) and explosive $\tau_{e}$ (red) time scales during ignition delay, overlaid with the temperature profile (blue); T$_0$=1100 K, p$_0$=2 atm and $\phi$=1.0.}
\label{fig:timescales}
\end{figure}

Figure \ref{fig:timescales} displays the time scales $\tau_n$ of all modes for the cases considered here.~The dynamics of all mechanisms exhibit a wide range of time scales.~Among them, there is an explosive time scale $\tau_e$, which extends throughout IDT and disappears at the point where the temperature undergoes a steep increase at ignition.~The minimum value of $\tau_e$ is recorded close to the inflection point of the temperature profile.

In the case of Glarborg$\_$2018, the explosive time scale $\tau_e$ initially remains approximately constant, then decelerates and subsequently starts accelerating at an increasing rate.~In the case of Shrestha$\_$2018 and Li$\_$2019, there is a gradual deceleration from the start, at the end of which $\tau_e$ accelerates similarly to Glarborg$\_$2018.~Finally, in the case of Stagni$\_$2020 and Zhang$\_$2021, $\tau_e$ decelerates immediately after the start of the process and then follows the accelerating pattern of the previous cases.~In all cases considered, Fig.~\ref{fig:timescales} shows that $\tau_e$ is among the slowest time scales.~In the  Stagni$\_$2020 and Zhang$\_$2021 cases, $\tau_e$ is positioned above a time scale gap throughout IDT, which is a clear indication that the explosive mode is among those driving the process, given that the faster modes are all dissipative.~In the cases of Glarborg$\_$2018, Shrestha$\_$2018 and Li$\_$2019 $\tau_e$  is positioned above a time scale bandwidth that is sparse.

\begin{figure}[t!]
\centering
\hfill\includegraphics[scale=0.18]{ 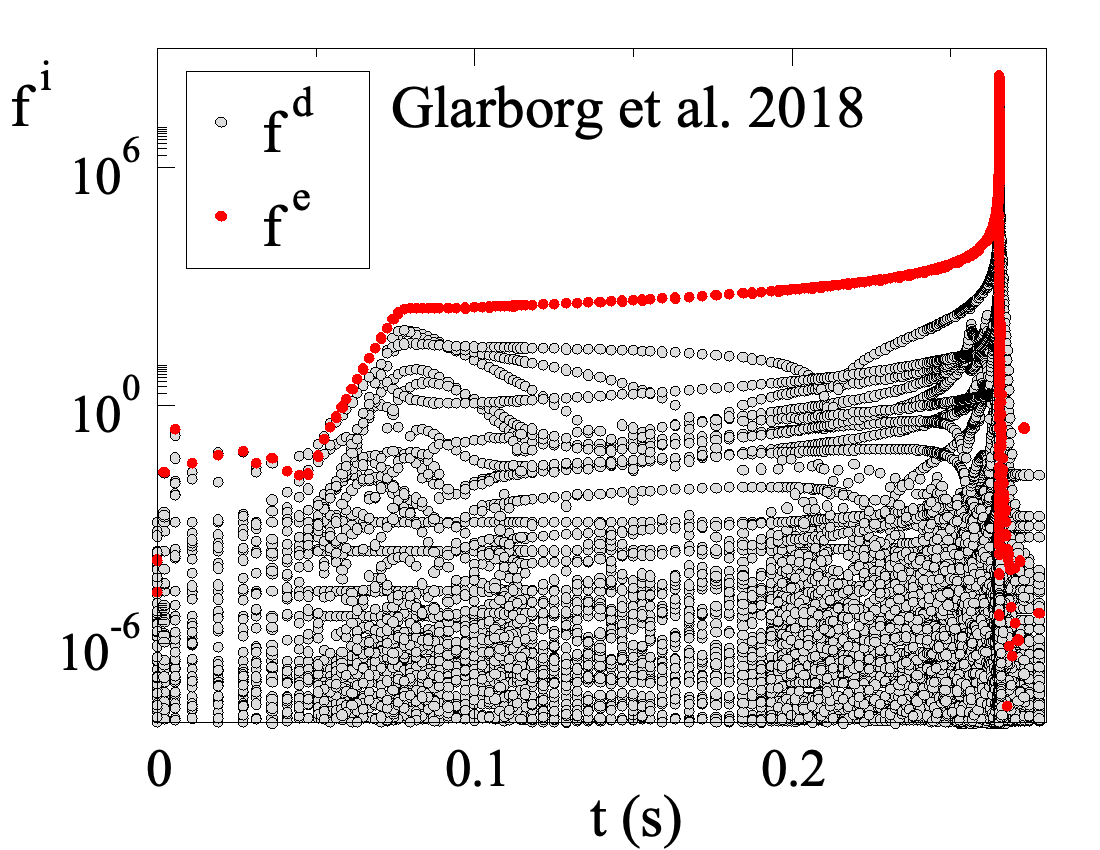}  \hfill  \includegraphics[scale=0.18]{ 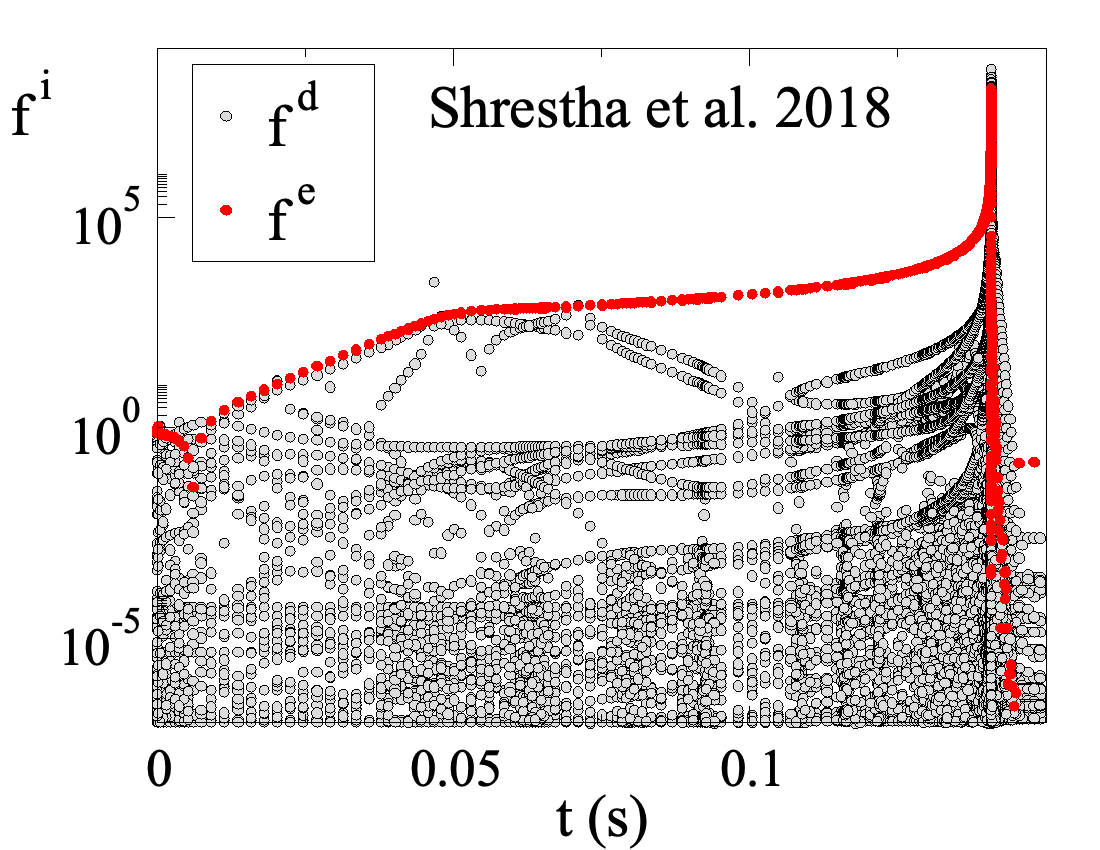} \hfill\hfill\hfill\\
\vspace{5 pt}
\hfill \includegraphics[scale=0.18]{ 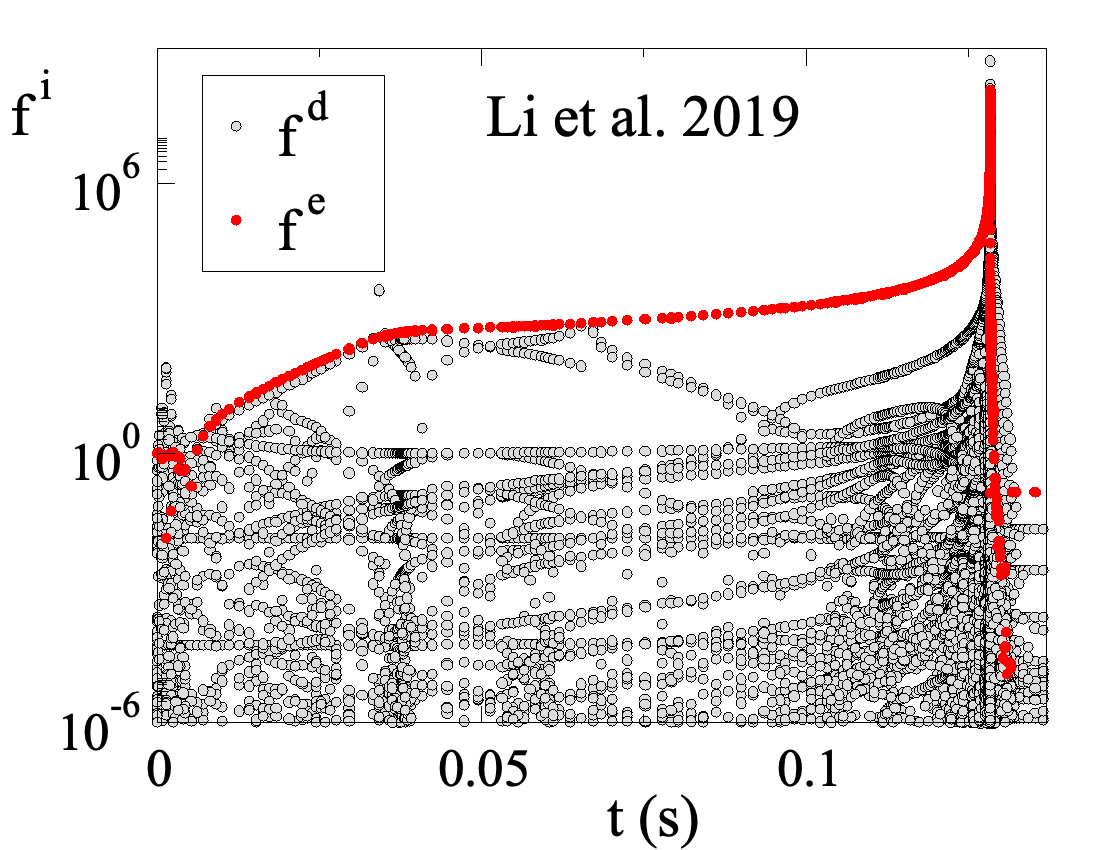}   \hfill    \includegraphics[scale=0.18]{ 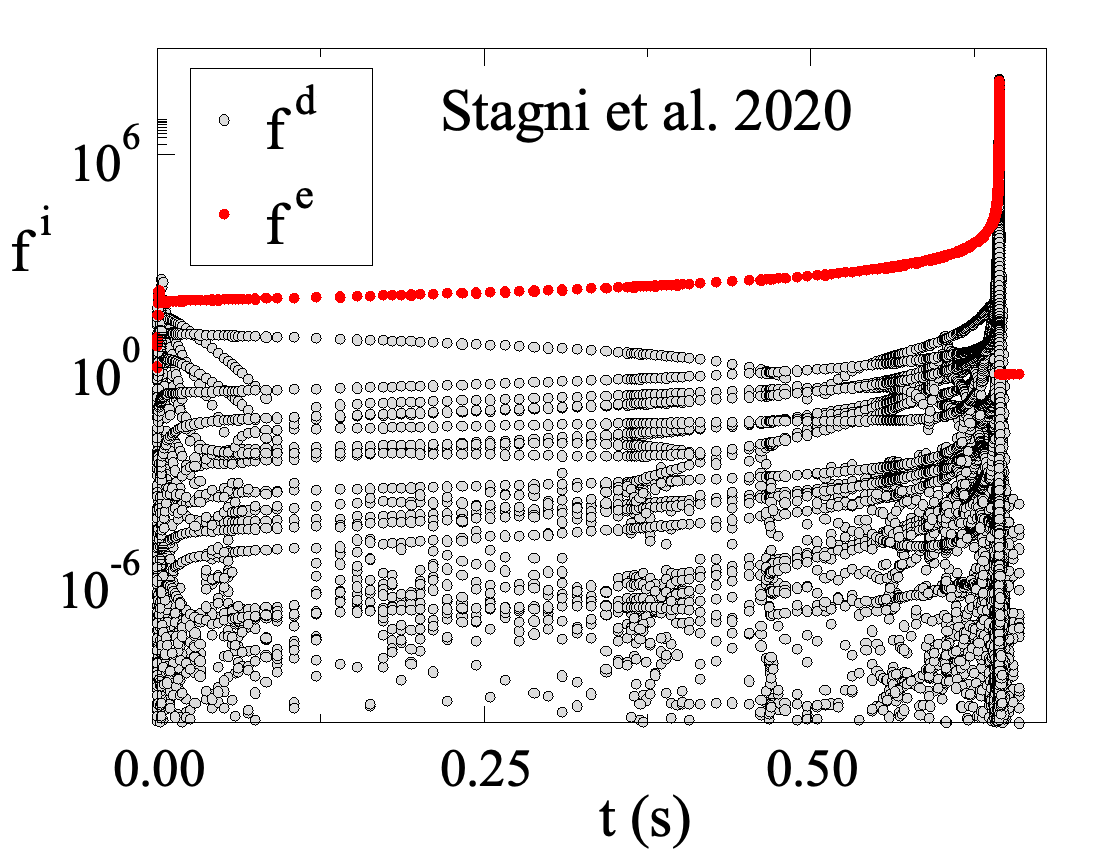} \hfill\hfill\hfill\\
 \vspace{5 pt}
 \includegraphics[scale=0.18]{ 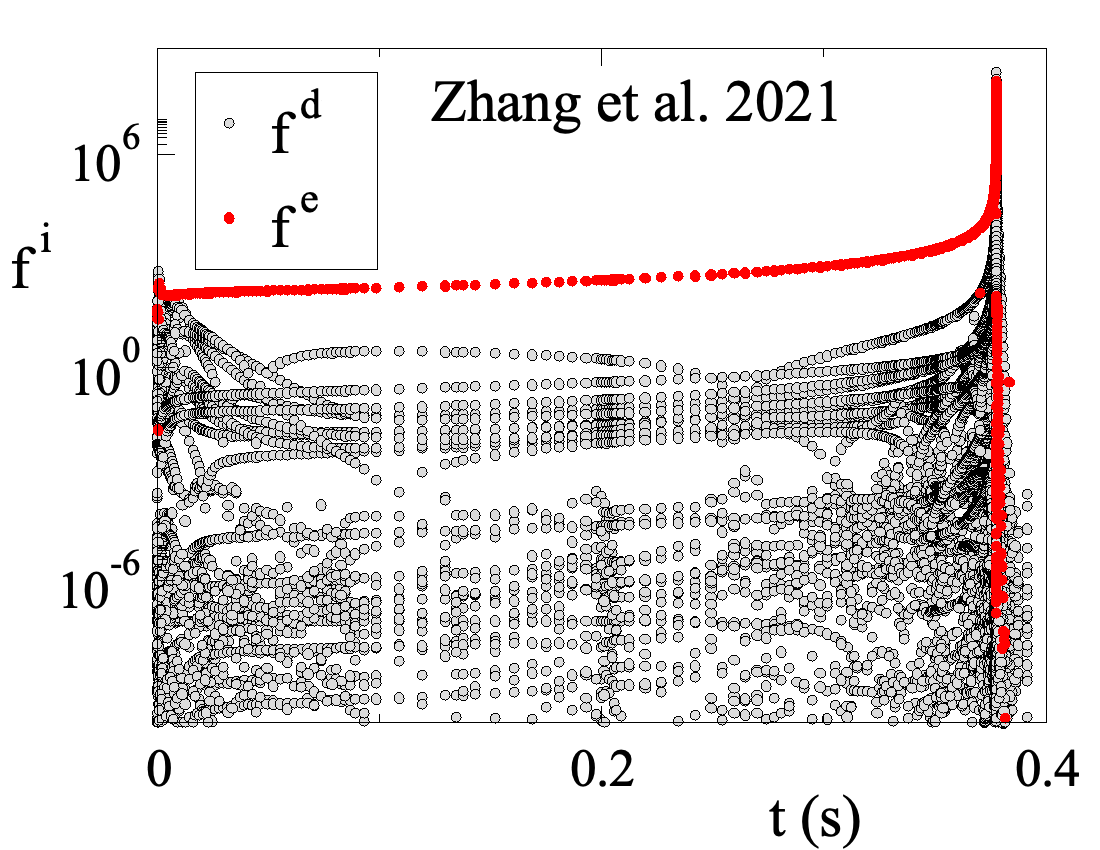} 
\caption{The amplitudes $f^n$ of all modes during IDT for  T$_0$=1100 K, p$_0$=2 atm and $\phi$=1.0; the amplitude of the explosive mode is shown in red.}
\label{fig:Amplitudes}
 \vspace{-15 pt}
\end{figure}

\subsection{Amplitude of explosive mode}
\label{Amplitude}

An additional assessment on whether the explosive time scale is the driving one during IDT can be obtained by comparing its amplitude $f^e$ with that of all other modes.~This comparison is performed through the results shown in Fig.~\ref{fig:Amplitudes}, where the amplitudes $f^n$ of all modes are displayed for all cases considered.~It is shown that the amplitude of the explosive mode is the largest or among the largest  throughout the IDT.~

Both time scale and amplitude considerations suggest that the explosive mode is among the ones driving the process during IDT.~Therefore, the analysis that follows will concentrate on the origin of the explosive dynamics.~The explosive mode was first introduced in the field of chemical kinetics by the first CSP paper \citep{Lam1988} and was subsequently analyzed with CSP tools \citep{fotache1997ignition,lee2005chain,kazakov2006computational,SIAM2006,mittal2008dimethyl,chen2007high,lu2008analysis}.~Later on, after the introduction of \emph{Chemical Explosive Mode Analysis} (CEMA) \citep{lu2010three}, which consists of concepts and tools developed previously by CSP \citep{goussis2021origin}, the explosive mode is analyzed with both CSP and  CEMA.~CSP considers the explosive mode as one of the slow ones, so its dominance is subject to verification \citep{SIAM2006,tingas2016comparative,tingas2018chemical,rabbani2022chemical}.~In contrast, CEMA considers this mode independently of all other slow modes, thus promoting the misconception  that it is always dominant \citep{shan2014bifurcation,steinberg2021structure,huang2022large,zhu2022pulsating,guo2022evolutions,desai2022effects}.






\section{CSP diagnostics} 
\label{sec:Results}

All CSP diagnostics for the explosive mode are displayed in \ref{glarborg} (Glarborg$\_$2018), \ref{shrestha} (Shrestha$\_$2018), \ref{li} (Li$\_$2019), \ref{stagni} (Stagni$\_$2020) and \ref{zhang} (Zhang$\_$2021).~Parts of the discussion that follows are based on the data and findings included in these Appendices.

\subsection{Chemical vs Thermal runaway}
\label{sec:chemtherm}
\begin{figure}[t]
\centering
\hfill\includegraphics[scale=0.26]{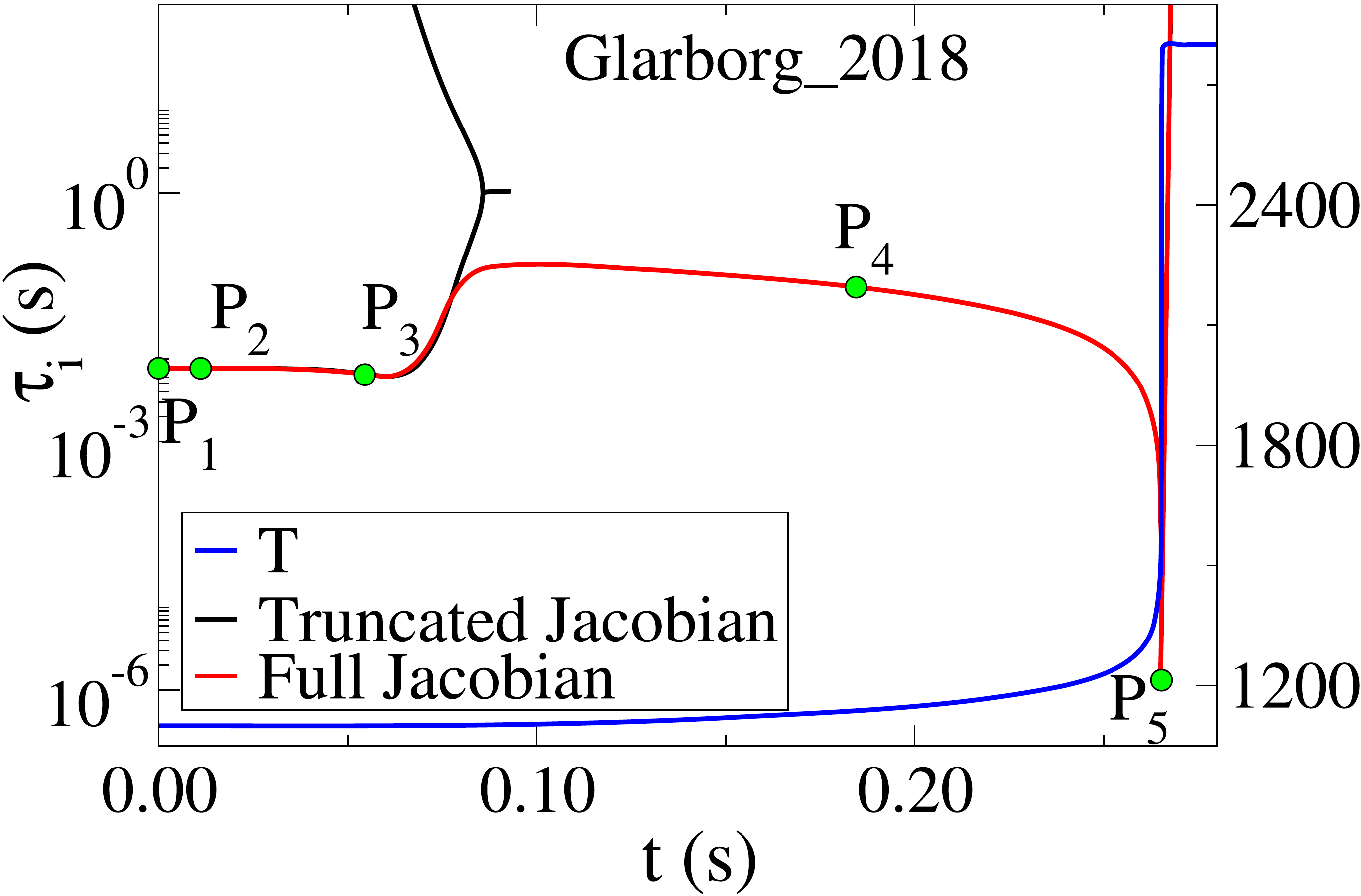}  \hfill  \includegraphics[scale=0.26]{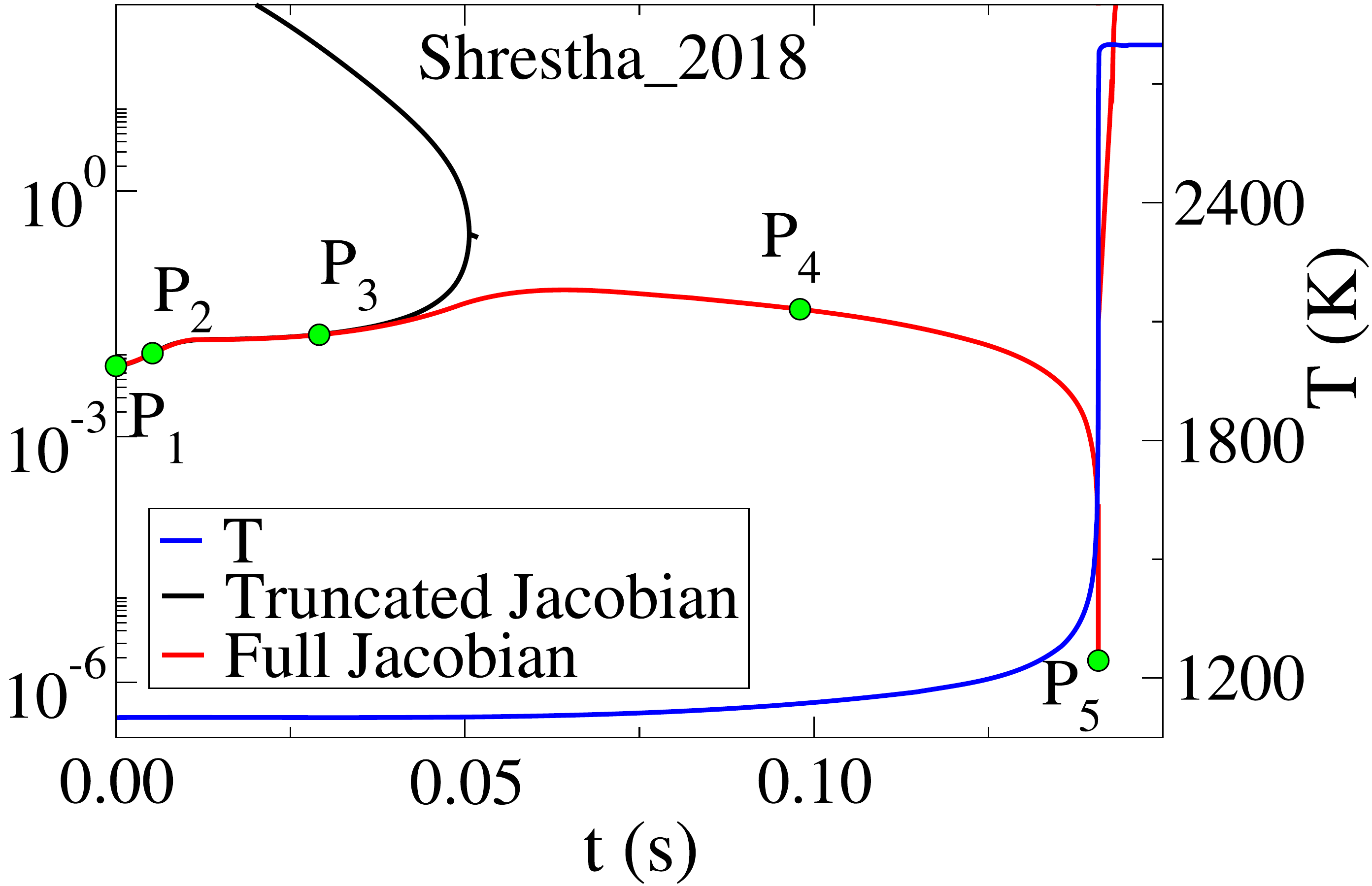} \hfill \hfill \hfill\\
\vspace{7 pt}
\hfill \includegraphics[scale=0.26]{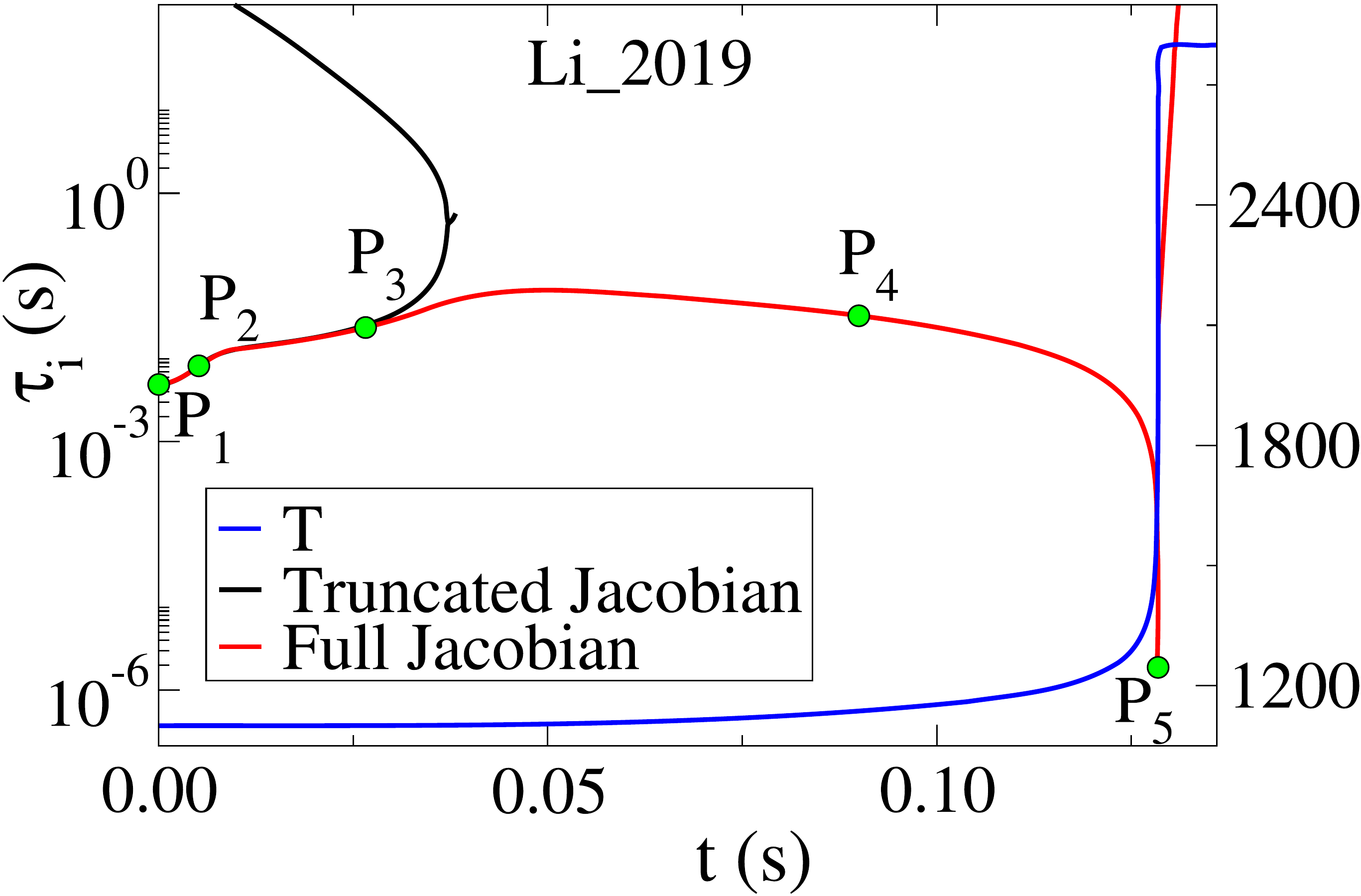}   \hfill  \includegraphics[scale=0.26]{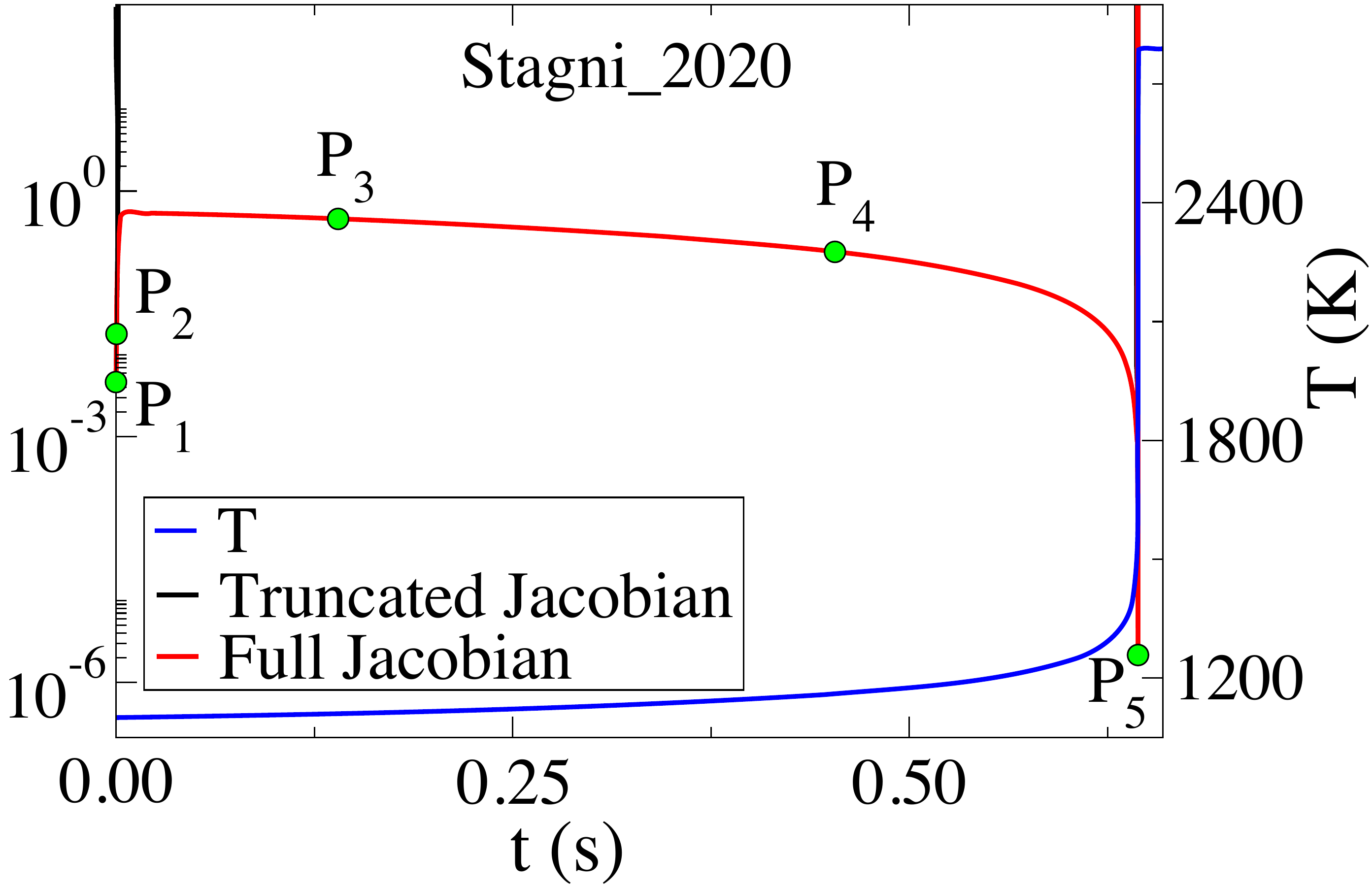} \hfill \hfill \hfill\\
 \vspace{7 pt}
 \includegraphics[scale=0.26]{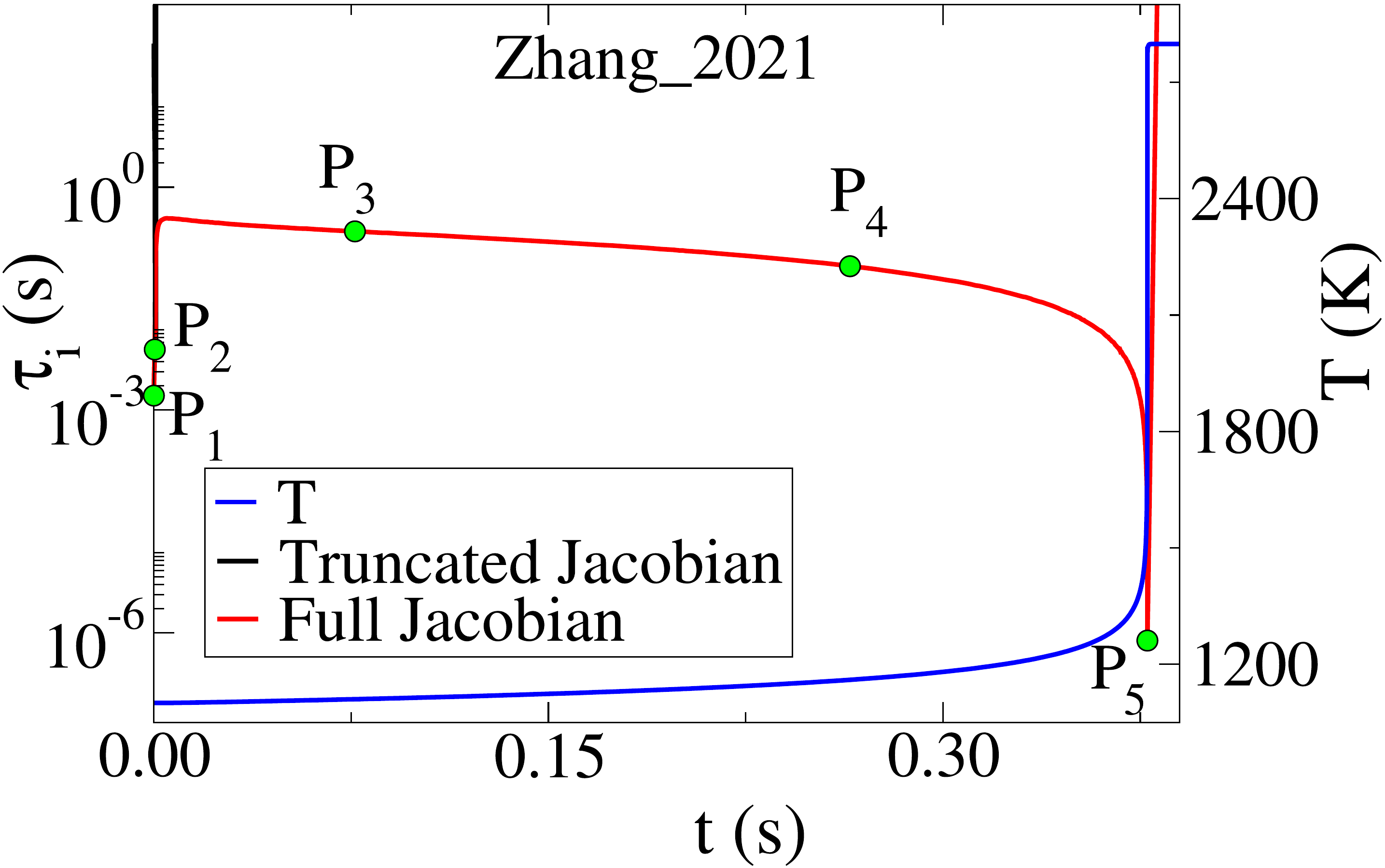} 
\caption{The developing explosive time scale, full (red) and temperature-truncated (black) Jacobian along the ignition delay time,   overlaid with the temperature profile (blue); T$_0$=1100 K, p$_0$=2 atm and $\phi$=1.0.~P$_1$ to P$_5$ denote instances at which CSP diagnostics are computed.}
\label{fig:Trunc_timescales}
\end{figure}

An important characteristic of the oxidation of widely used carbon-containing fuels, such as hydrocarbons and alcohols, is that they demonstrate a chemical runaway, during which an extensive radical pool is generated.~This extends over a substantial part of IDT, because it involves the formation of stable intermediate species, such as aldehydes, $H_2O_2$ and $CO$.~This part of IDT is followed by the thermal runaway, which is mainly driven by strong exothermic reactions.~It is therefore interesting to explore whether these features manifest themselves in ammonia autoignition.

Given that the explosive mode characterises the dynamics during IDT, a simple method was introduced in Ref.~\citep{Diamantis2015b} in order to estimate the parts occupied by the chemical and thermal runaways.~This method consists in comparing the explosive time scale in the dynamics of the species and temperature Eqs.~\ref{eq:gov1} and \ref{eq:gov2} with that in the dynamics of only the species  Eq.~\ref{eq:gov1}.~In the chemical runaway the two explosive time scales are close to each other, since temperature has minor influence there, while in the thermal runaway the two time scales deviate significantly.~This methodology was employed successfully in autoignition studies of of various fuels; e.g., \citep{tingas2018use,tingas2018ch4,TINGAS201828,rabbani2022dominant}.

\begin{figure}[t]
\begin{center}
\centering
\hfill\includegraphics[scale=0.26]{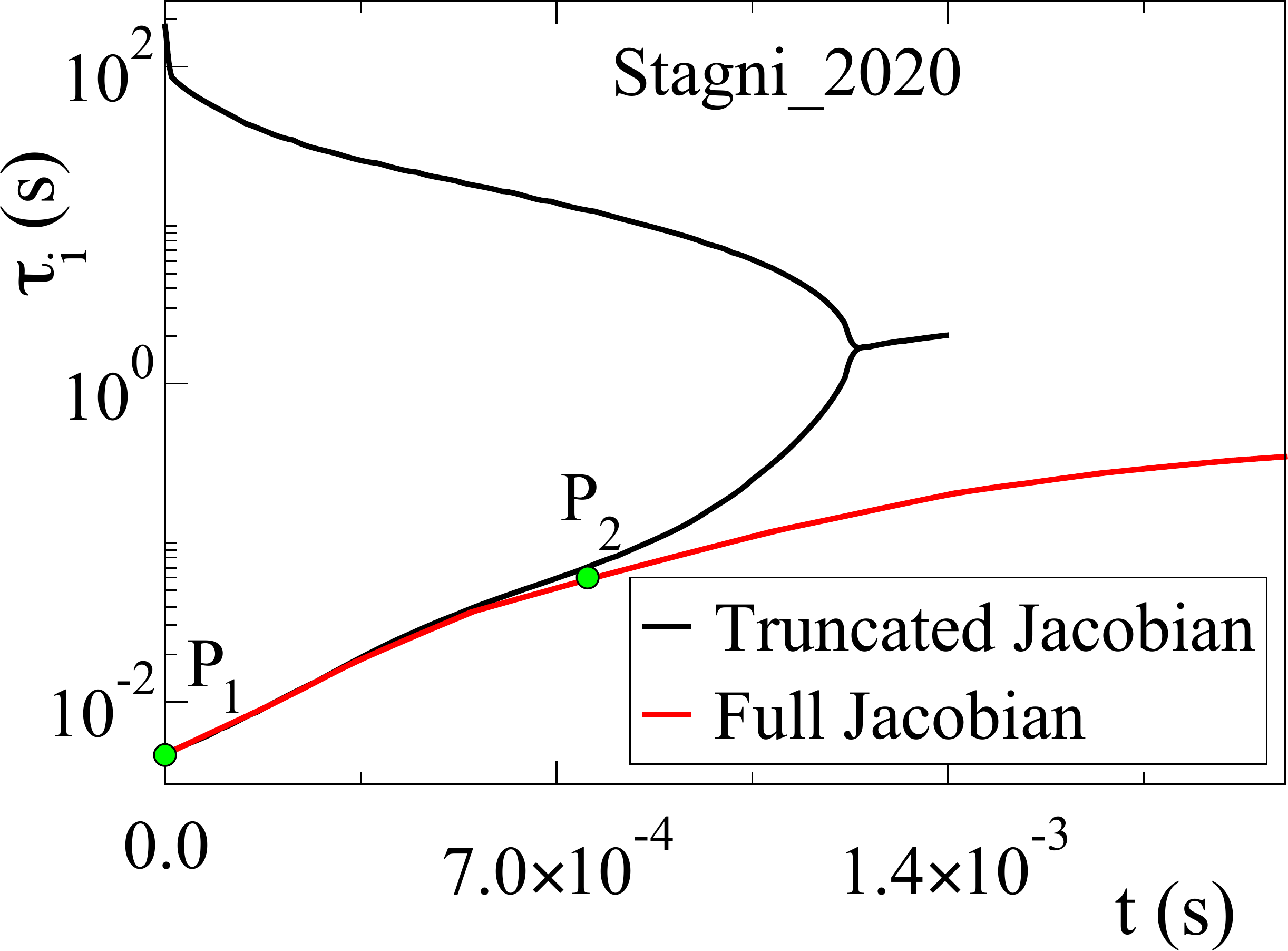}  \hfill  \includegraphics[scale=0.26]{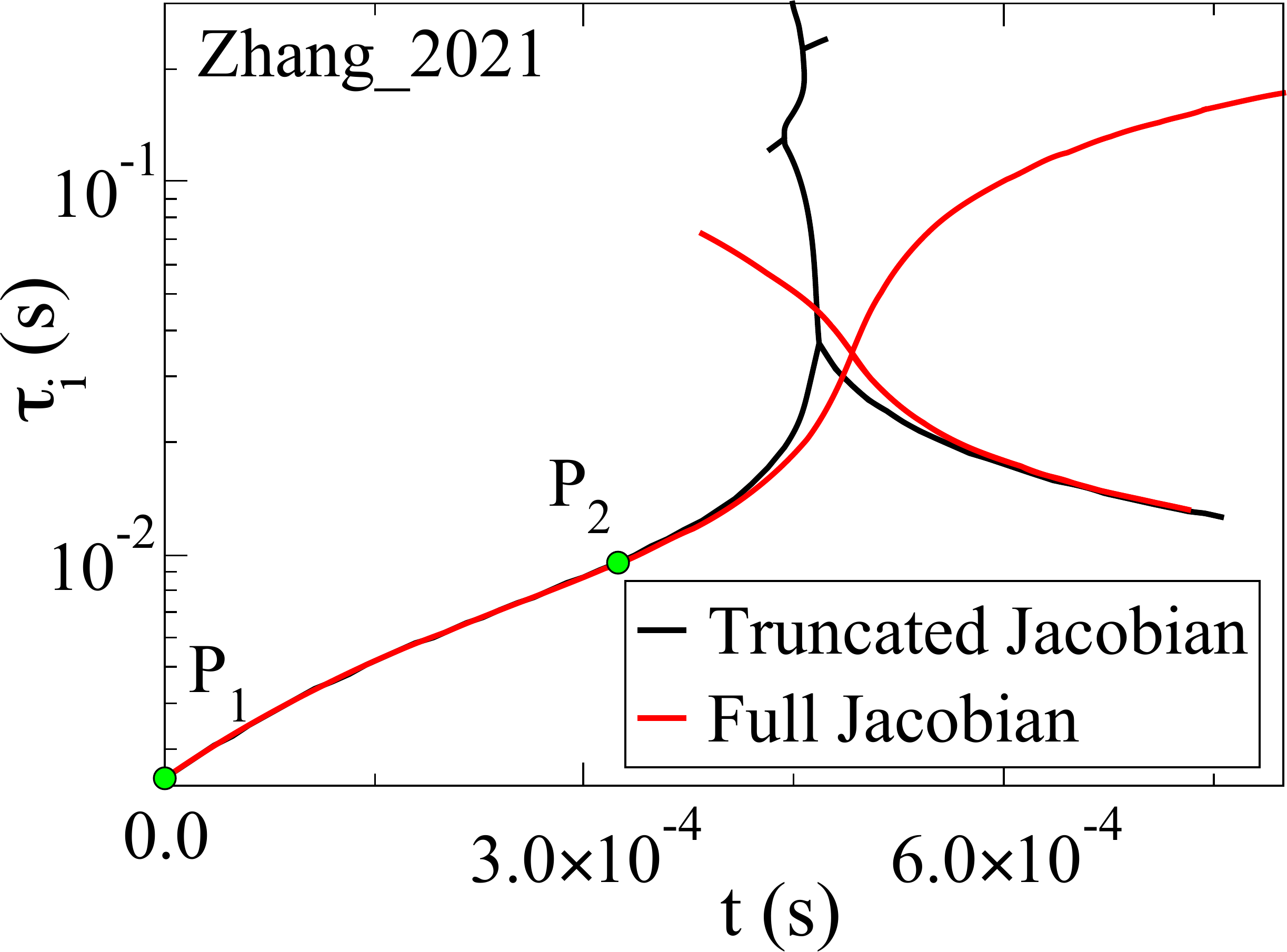} \hfill \hfill \hfill\\
\end{center}
 \vspace{-10 pt}
\caption{The developing explosive time scale, full (red) and temperature-truncated (black) Jacobian along the initiation period; T$_0$=1100 K, p$_0$=2 atm and $\phi$=1.0.}
\label{fig:Stagni_Timescales_t0}
\end{figure}

For each of the five cases considered here, Fig.~\ref{fig:Trunc_timescales} displays the explosive time scale computed on the basis of the full Jacobian of the system in Eqs.~\ref{eq:gov1} and \ref{eq:gov2} and of the truncated Jacobian of the system in Eq.~\ref{eq:gov1}.~It is shown that a relatively large chemical runaway develops in Glarborg$\_$2018, Shrestha$\_$2018 and Li$\_$2019, while in Stagni$\_$2020 and Zhang$\_$2021 it is practically non-existent.~This is more clearly demonstrated in Fig.~\ref{fig:Stagni_Timescales_t0}, where it is shown that a chemical runaway is indeed developing in Stagni$\_$2020 and Zhang$\_$2021, but its length is negligible.~In all cases, the chemical runaway extends in the part of the IDT where $\tau_e$ decelerates and ends before $\tau_e$ starts to accelerate.~It is noted that the two mechanisms that have negligible chemical runaway yield the two longest IDTs; see Table \ref{tab:mechanisms} and Fig.~\ref{fig:Trunc_timescales}.

The differences in the relative extent of chemical and thermal runaway will be examined in the context of the reactions that relate to the explosive mode, which was shown to dominate along the two runaways.~In the following, the reactions that contribute to the impact of this mode will by identified and discussed, along with those that determine the time frame of its action.~The impact of the mode is characterised by its amplitude and the time frame of its action by the explosive time scale.~The API and the TPI indices will identify the reactions that contribute to the amplitude (impact) and time scale (time frame), respectively, of the explosive mode.

\subsection{Initiation of the oxidation process according to the five mechanisms}
\label{sec:t=0}

The reactions producing the largest API indices at P$_1$ ($t=0$) are $NH_2+HO_2\leftarrow NH_3+O_2$  and  $NH_2+H(+M)\leftarrow NH_3(+M)$  for Glarborg$\_$2018 and only  $NH_2+HO_2\leftarrow NH_3+O_2$  for Shrestha$\_$2018, Li$\_$2019, Stagni$\_$2020 and Zhang$\_$2021.~This difference is due to the fact that the rate of reaction $NH_2+HO_2\leftarrow NH_3+O_2$ is much smaller in Glarborg$\_$2018, while that of $NH_2+H(+M)\leftarrow NH_3(+M)$ does not vary much, as shown in Table~\ref{tab:mechanisms1}.~This finding is in agreement with the results in Ref.~\citep{valera2021review}.~Therefore, the resulting smaller relative contribution of the first reaction to the amplitude of the explosive mode $f^e$ in Glarborg$\_$2018, allows the contribution of the second reaction to increase.


\begin{table}[h]
\caption{The Arrhenius constants A [$kmol/m^3\cdot s \cdot K^b$], b and E [$kJ/kmol$] and the reaction rate $R^i$ [$kmol/m^3\cdot s$]; the rates computed at $T(0)=1100~K$.~Numbers in parentheses denote powers of ten.}
\vspace{+5 pt}
\begin{center}
\begin{tabular}{lcccc|cccc}
\hline
                    &\multicolumn{4}{c}{$NH_2+HO_2\leftarrow NH_3+O_2$} &\multicolumn{4}{c}{$NH_2+H(+M)\leftarrow NH_3(+M)$}                     \\
\cline{2-9}
        & \textbf{$A$} & \textbf{$b$}          & \textbf{$E$} & \textbf{$R^i$}    & \textbf{$A$} & \textbf{$b$}          & \textbf{$E$} & \textbf{$R^i$}                       \\
\hline
Glarborg$\_$2018	& 7.72(5)	&	1.50	&	60633	&	3.67(7)	&	5.17(17)	&	-0.55	&	108662	&	7.48(10)    \\
Shrestha$\_$2018	& 2.24(8)	&	1.67	&	58983	&	4.15(10)	&	9.00(16)	&	-0.39	&	110300	&	3.39(10)    \\
Li$\_$2019		& 1.09(8)	&	1.76	&	57676	&	4.54(10)	&	9.00(16)	&	-0.39	&	110300	&	3.39(10)    \\
Stagni$\_$2020		& 1.42(10)	&	1.28	&	55224	&	2.64(11)	&	3.50(30)	&	-5.22	&	111160	&	2.45(9)    \\
Zhang$\_$2021		& 1.08(9)	&	1.71	&	54671	&	4.47(11)	&	2.20(16)	&	0.00	&	~93470	&	8.01(11)    \\

\hline
\end{tabular}
\label{tab:mechanisms1}
\end{center}
\end{table}

\begin{figure}[t!]
\centering
\includegraphics[scale=0.29]{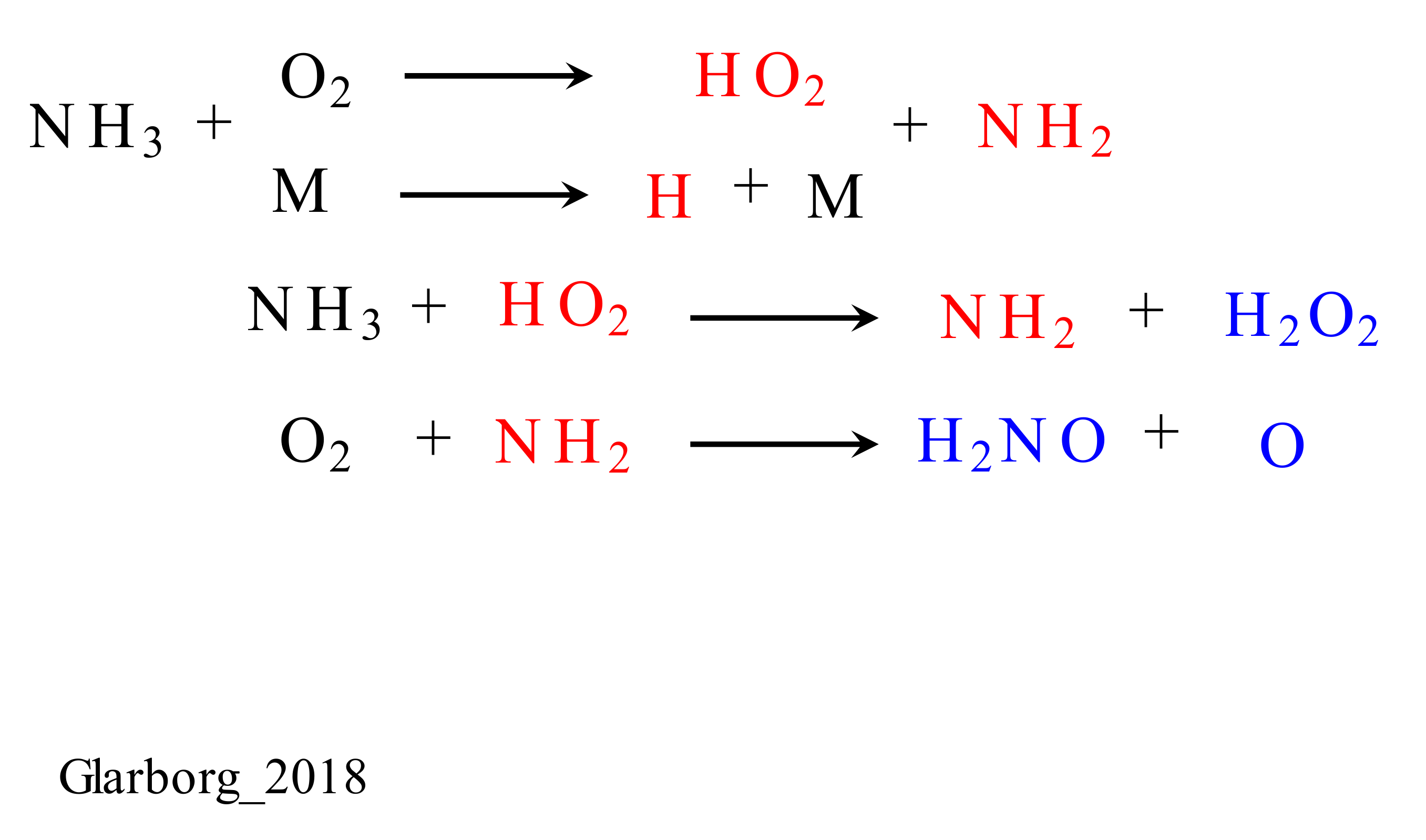}  \hfill  \includegraphics[scale=0.29]{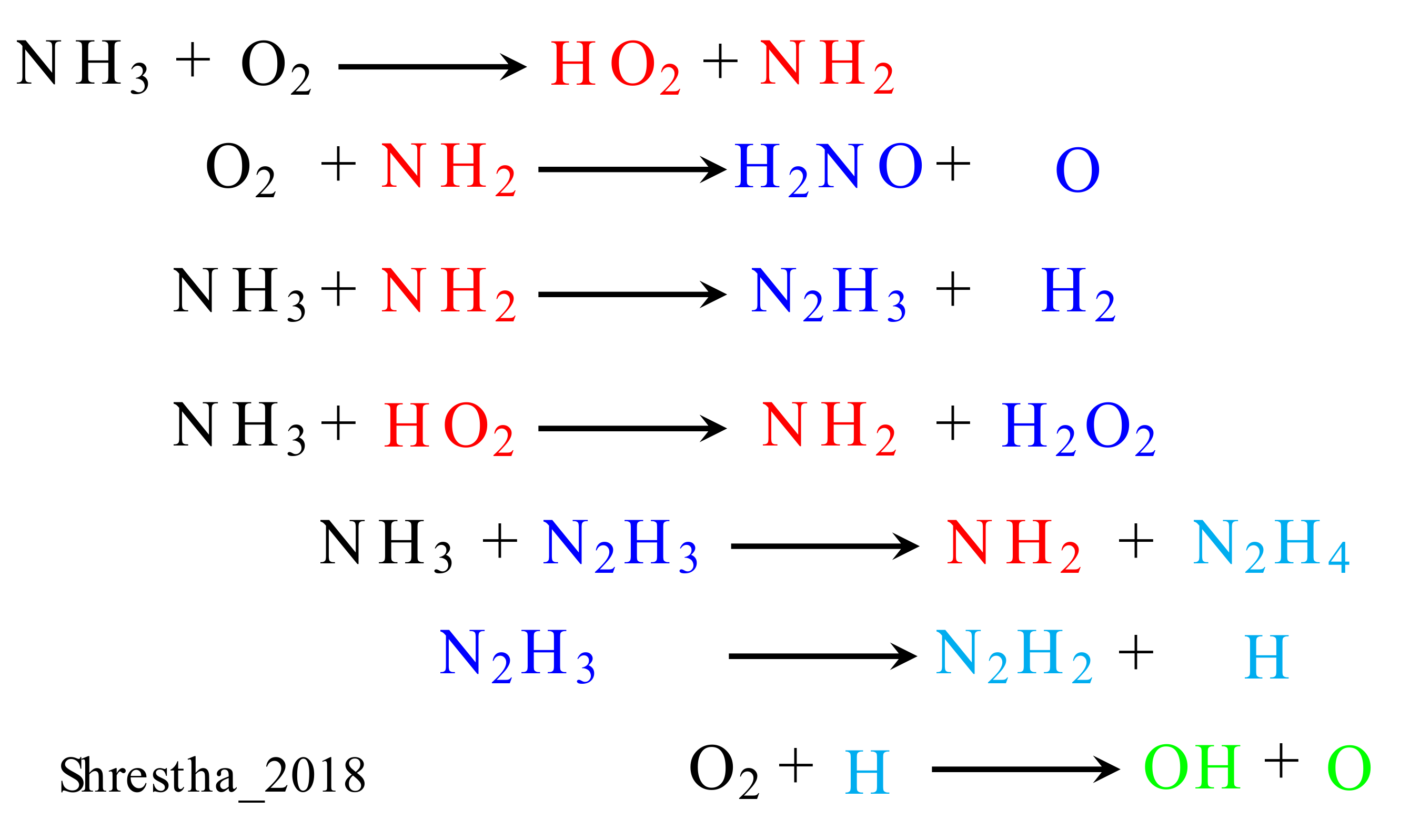} \\  
\vspace{10 pt}
\includegraphics[scale=0.29]{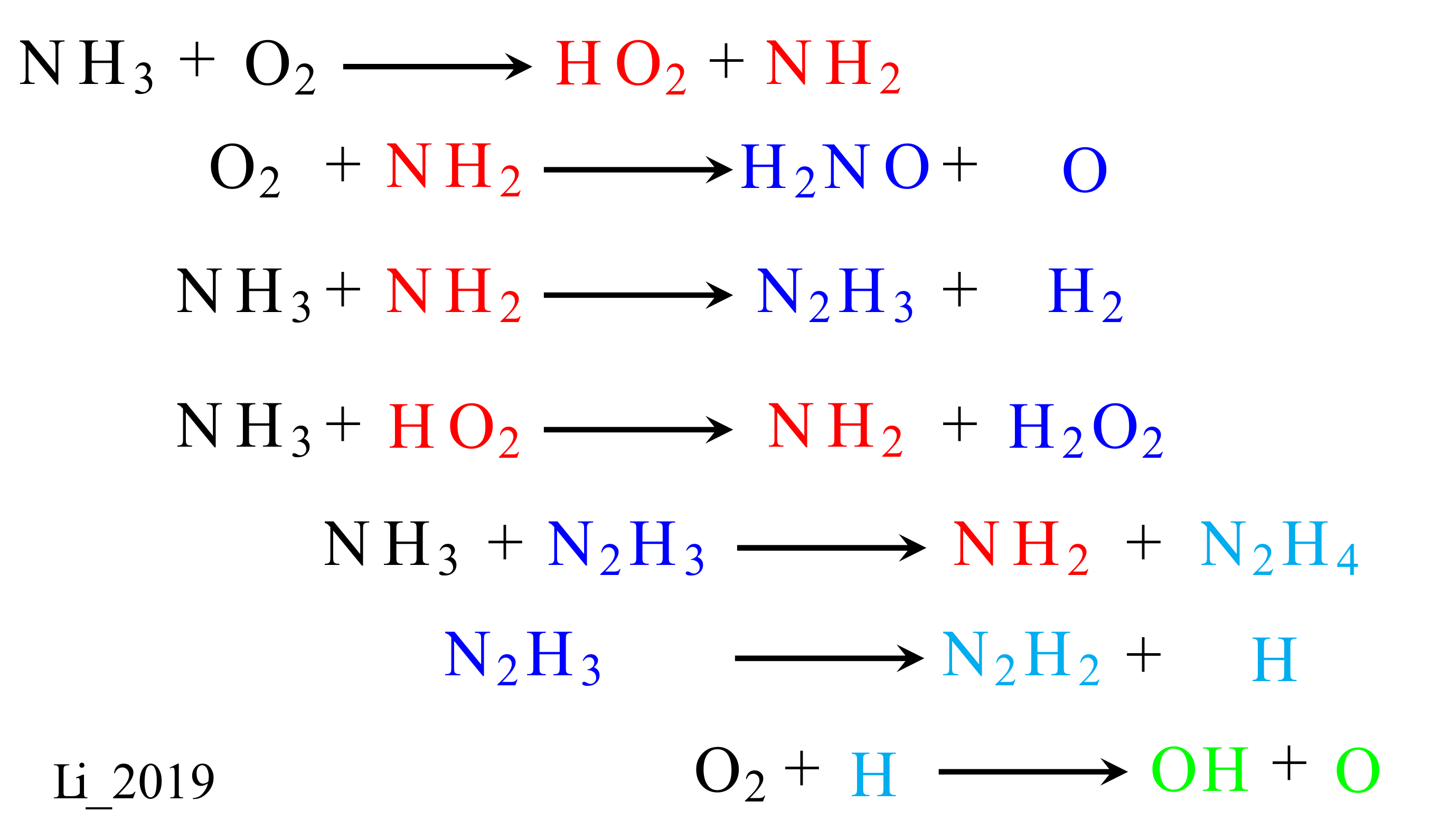}  \hfill  \includegraphics[scale=0.29]{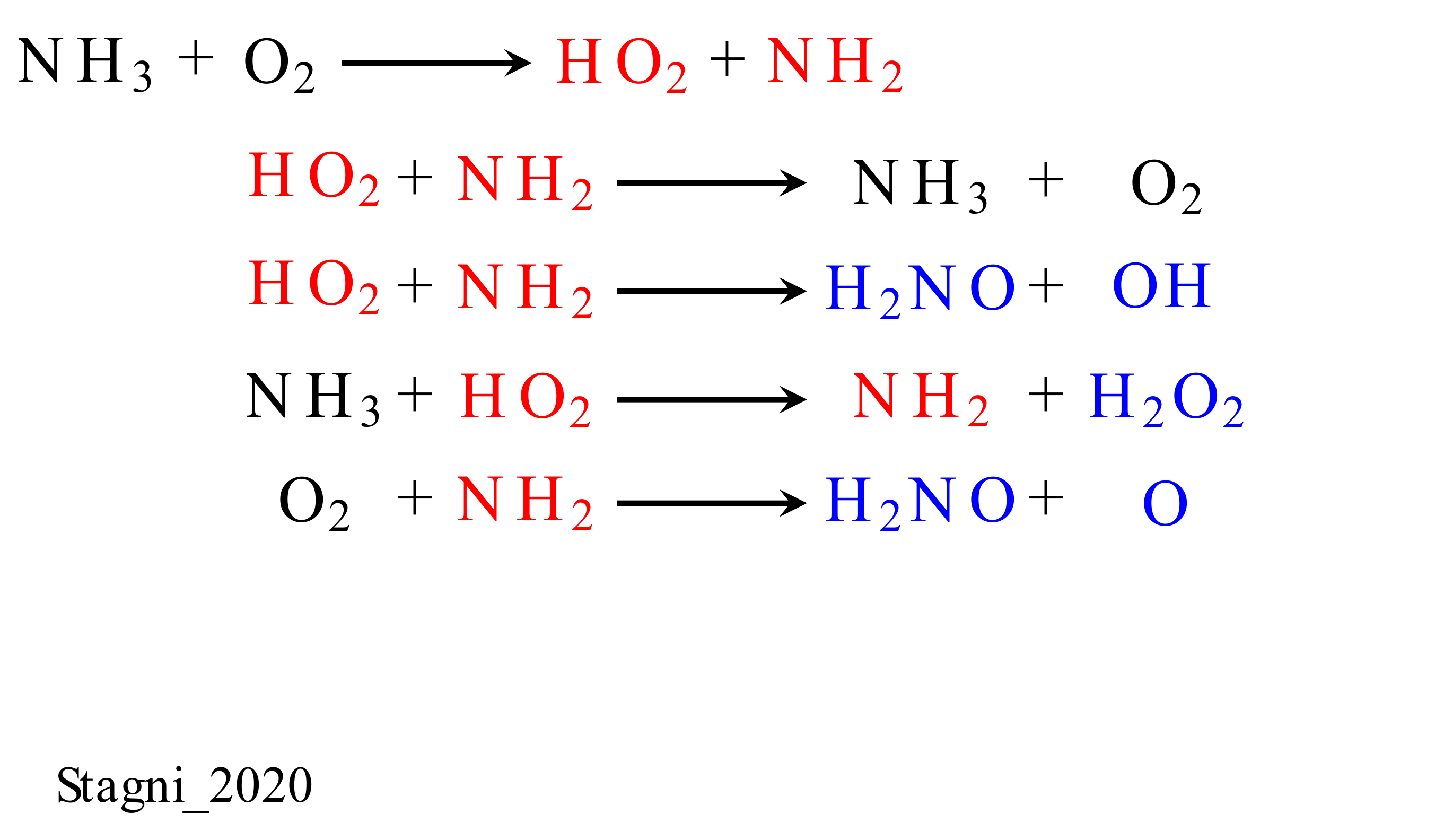} \\
\vspace{10 pt}
 \includegraphics[scale=0.29]{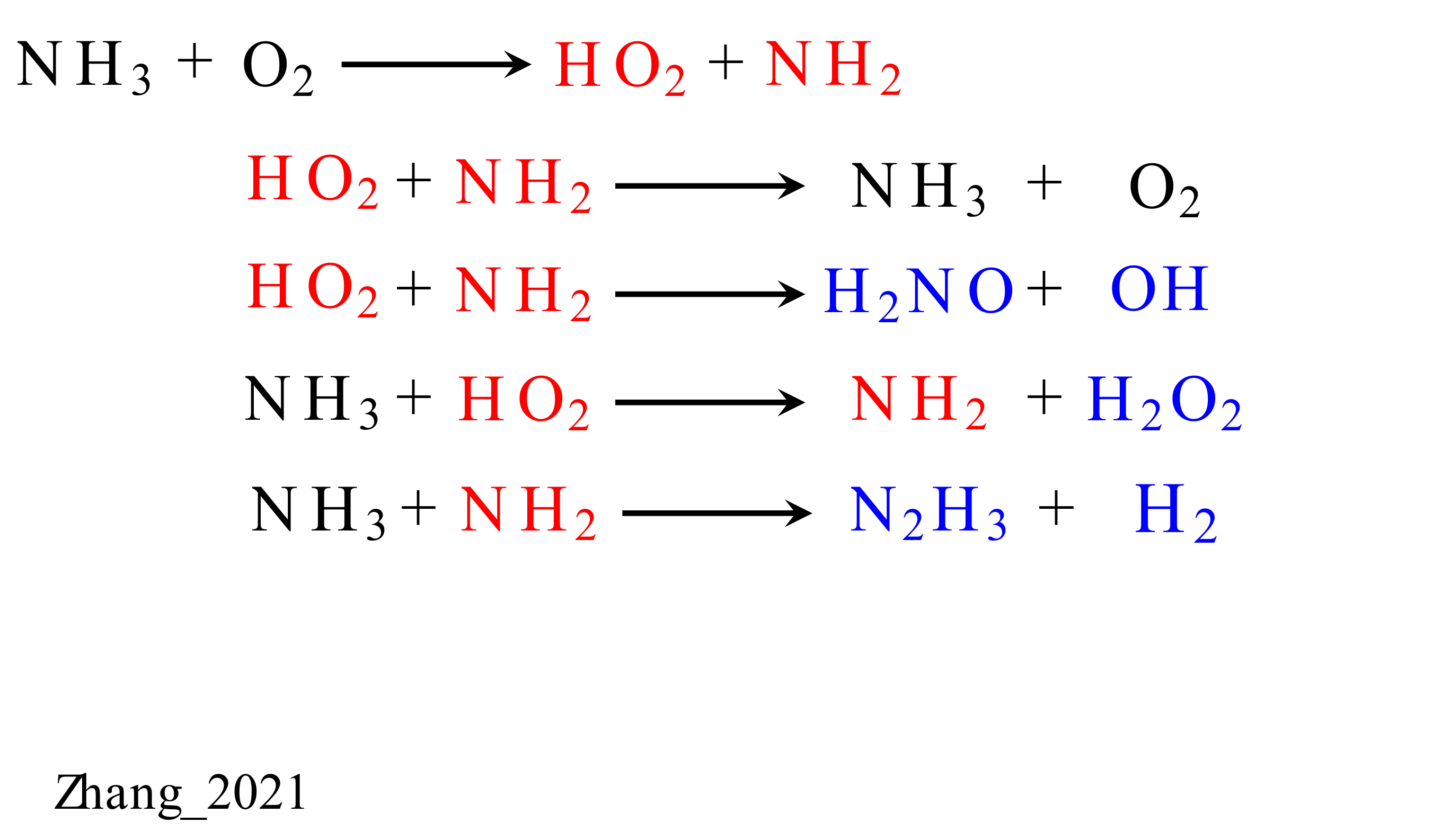} 
\caption{The sequence of reactions that exhibit significant contributions to the amplitude of the explosive mode  (large APIs) at P$_1$ and P$_2$; T$_0$=1100 K, p$_0$=2 atm and $\phi$=1.0.~All reactions tend to increase the impact of the explosive mode (large positive APIs), except for reaction $HO_2+NH_2\rightarrow NH_3+O_2$ in  Stagni$\_$2020 and Zhang$\_$2021, which tends to decrease its impact (large negative API).~Note that $OH$-producing reactions contribute to the explosive dynamics sooner in  Stagni$\_$2020 and Zhang$\_$2021 and later in Glarborg$\_$2018.~The colors black, red, blue, cyan and green are meant to highlight successive phases of the process, as they are described in the text.}
\label{fig:flow}
\end{figure}

Figure \ref{fig:flow} lists the sequence of reactions driving the process at P$_1$ and P$_2$.~It is shown that effectively, in all cases the process starts with the production of $NH_2$ and $HO_2$.~A radical pool starts forming, consisting initially of these two species with subsequent formation of $H_2NO$ and $H_2O_2$ in all mechanisms and $OH$ in all mechanisms except in Glarborg$\_$2018.~In parallel, $N2$ chemistry (involving species with two atoms of nitrogen in their molecule) starts developing in Shrestha$\_$2018, Li$\_$2019 and Zhang$\_$2021.~This chemistry starts with the reaction $NH_3+NH_2\rightarrow N_2H_3+H_2$, leading to the formation of $N_2H_2$ and $N_2H_4$, as shown in Fig.~\ref{fig:flow}.~In Glarborg$\_$2018 and Stagni$\_$2020 there is no $N2$ chemistry manifesting itself at the start of the process.

\begin{figure}[t]
\begin{center}
\centering
\hfill\includegraphics[scale=0.26]{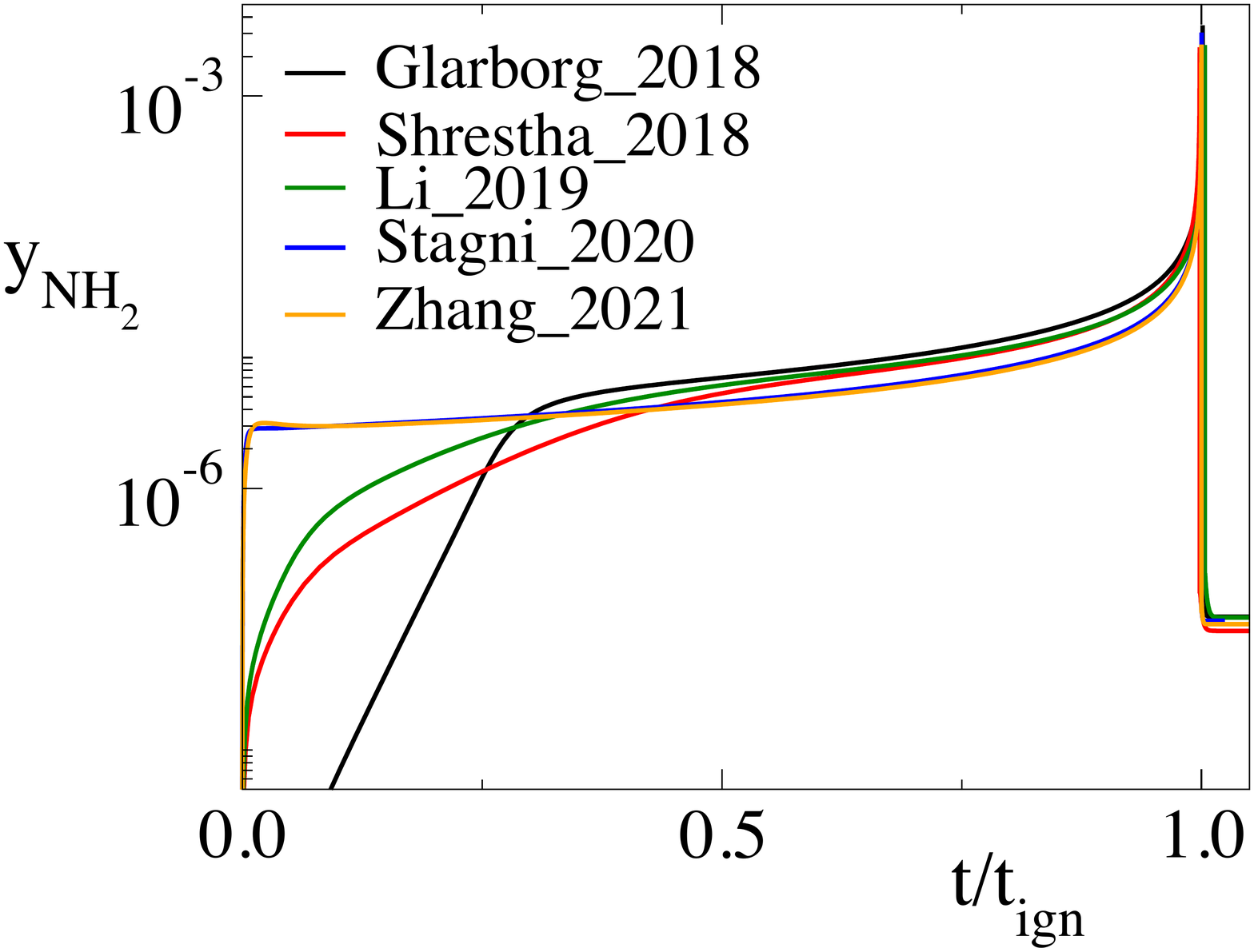}  \hfill  \includegraphics[scale=0.26]{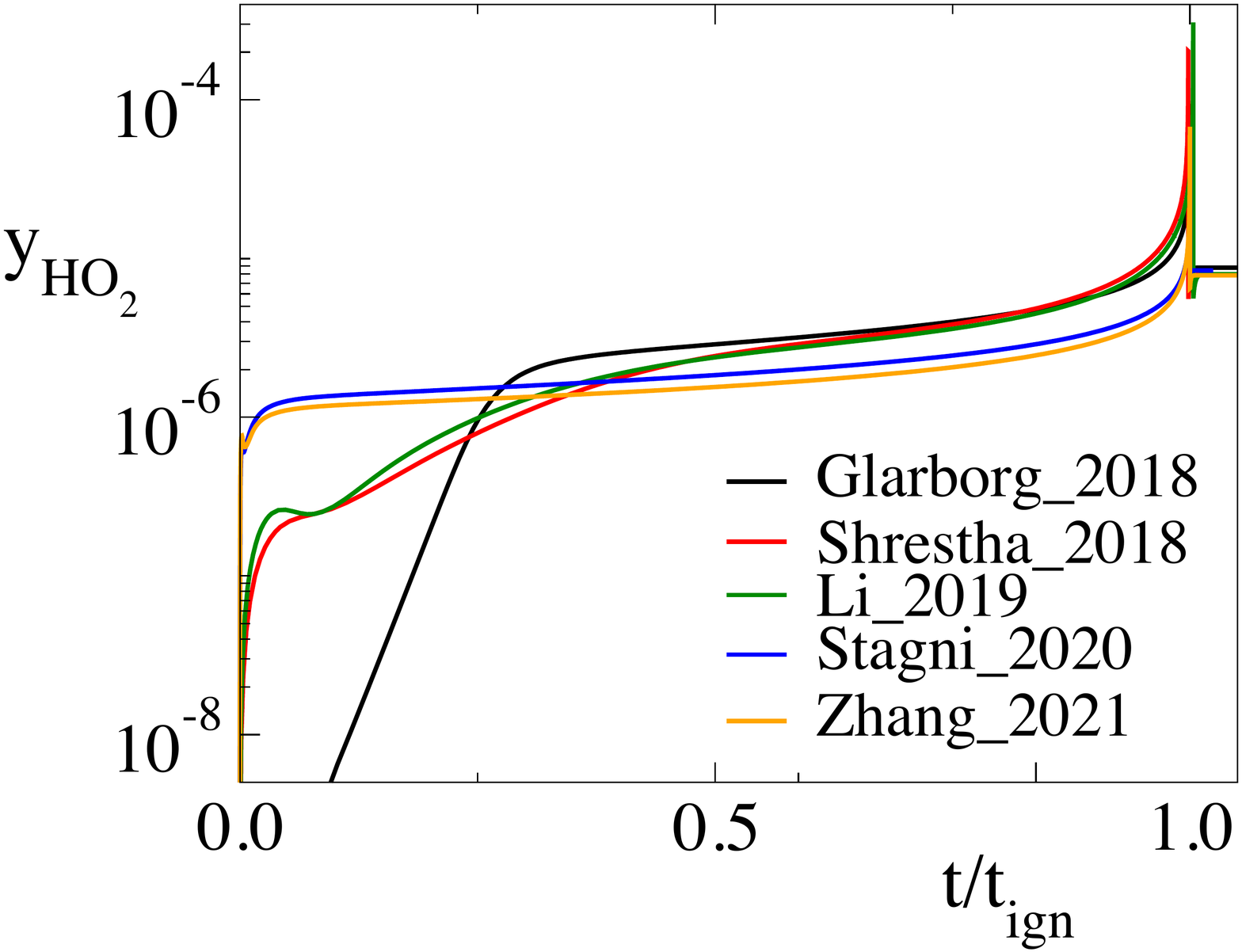} \hfill\hfill\hfill\\
\end{center}
\caption{Mass fraction profiles of $NH_2$ and $HO_2$ vs time scaled with $t_{ign}$; T$_0$=1100 K, p$_0$=2 atm and $\phi$=1.0.~It is shown that in Stagni$\_$2020 and Zhang$\_$2021 $NH_2$ and $HO_2$ attain quasi-steady-state status immediately after the start of the process.}
\label{fig:profiles1}
\end{figure}

The fate of this radical pool differs drastically between two groups of mechanisms.~In Glarborg$\_$2018, Shrestha$\_$2018 and Li$\_$2019 there is no significant opposition to the impact of the explosive mode (no reactions with negative APIs, large in magnitude).~In contrast, for Stagni$\_$2020 and Zhang$\_$2021 there is a substantial opposition from reaction $HO_2+NH_2\rightarrow NH_3+O_2$ leading its reactants to an early steady state, as shown in Fig.~\ref{fig:profiles1}.~The action of this reaction in diminishing the impact of the explosive mode is pronounced only in Stagni$\_$2020 and Zhang$\_$2021 and is due to the relatively large $HO_2$ and $NH_2$ concentrations in these two cases, right after the start of the process, as shown in Fig.~\ref{fig:profiles1}.~These large concentrations are realised because the  rate  of $NH_3 + O_2 \rightarrow NH_2 + HO_2$ is much larger in Stagni$\_$2020 and Zhang$\_$2021, as shown in Table \ref{tab:mechanisms1}.

The reactions contributing the most to the amplitude of the explosive mode (large APIs), listed in Fig.~\ref{fig:flow}, are not necessarily the same with those that contribute to its time scale $\tau_e$ (large TPIs).~For example, although the $NH_2$ and $HO_2$-producing reactions $NH_2$ + $HO_2$  $\leftarrow$ $NH_3$ + $O_2$ and $NH_2$ + $H (+M)$  $\leftarrow$ $NH_3(+M)$ contribute the most to the amplitude $f^e$ at P$_1$ in Glarborg$\_$2018, it is the $NH_2$ and $HO_2$-consuming reactions  $NH_2$ + $ O_2$  $\rightarrow$  $H_2NO$ + $ O $ and $NH_3$ + $ HO_2$  $\rightarrow$ $NH_2$ + $ H_2O_2$ that contribute the most to the explosive time scale $\tau_e$ there; see \ref{glarborg}.~However, as the process progresses, the sets of reactions that contribute the most to $f^e$ and $\tau_e$ become similar.~For example at P$_3$ in Glarborg$\_$2018, the reactions promoting the most $f^e$ and $\tau_e$ are the $NH_2$, $H_2NO$ and $HO_2$-consuming reactions  $NH_2+O_2 \rightarrow H_2NO+O$, $H_2NO+O_2$ $\rightarrow HNO+HO_2$ and $NH_2+HO_2 \rightarrow H_2NO+OH$, while the one opposing the most is $NH_2+NH_2(+M) \rightarrow N_2H_4(+M)$.

In general, the reactions that contribute to most to $\tau_e$ within the chemical runaway are those whose reactants are identified by the CSP Pointer.~Such reactants are $HO_2$,  $NH_2$ and $H_2NO$ for all five mechanisms and $N_2H_3$ and $N_2H_4$ only for Shrestha$\_$2018, Li$\_$2019 and Zhang$\_$2021, that involve $N2$ chemistry.
\begin{itemize}
    \item  The reaction that involves $HO_2$ as a reactant and contributes significantly to promoting $\tau_e$ is $NH_3+HO_2 \rightarrow NH_2+H_2O_2$.~Its influence is mainly manifested at the start of the process; i.e., in the vicinity of P$_1$.

    \item For $NH_2$, the related reaction is $NH_2+O_2 \rightarrow H_2NO+O$ and its influence is exercised up until the middle of the chemical runaway regime in all five mechanisms considered.~In addition, the $NH_2$-consuming reaction $NH_3+NH_2\rightarrow N_2H_3+H_2$  is shown to promote $\tau_e$ during the first half of the chemical runaway in the three mechanisms exhibiting $N2$ chemistry.

    \item As for $H_2NO$, the reaction that contributes the most to $\tau_e$, is $H_2NO+O_2 \rightarrow HNO+HO_2$ and its influence lasts throughout the chemical runaway in Glarborg$\_$2018 (P$_1$-P$_3$) and Stagni$\_$2020 (P$_1$-P$_2$) and in the second half of this period in Shrestha$\_$2018 (P$_2$-P$_3$), Li$\_$2019 (P$_2$-P$_3$) and Zhang$\_$2021 (P$_2$).

    \item In the case of $N_2H_3$, there are two reactions that promote $\tau_e$: $N_2H_3\rightarrow N_2H_2+H$ and $N_2H_4+NH_2 \leftarrow N_2H_3+NH_3$.~Both reactions promote $\tau_e$ in Shrestha$\_$2018 and Li$\_$2019, while only the former acts in this manner in Zhang$\_$2021.~In all three cases, the influence of these reactions appears at the start of the chemical runaway regime; P$_1$ and P$_2$ in Shrestha$\_$201 and P$_1$ in Li$\_$2019 and Zhang$\_$2021.

    \item Finally, the $N_2H_4$-consuming reaction  $NH_2+NH_2(+M)\leftarrow N_2H_4(+M)$ promotes $\tau_e$ and is active only in Shrestha$\_$201 and  Li$\_$2019, at the start of the chemical runaway regime.
\end{itemize}

\begin{figure}[t]
\begin{center}
\centering
\hfill\includegraphics[scale=0.3]{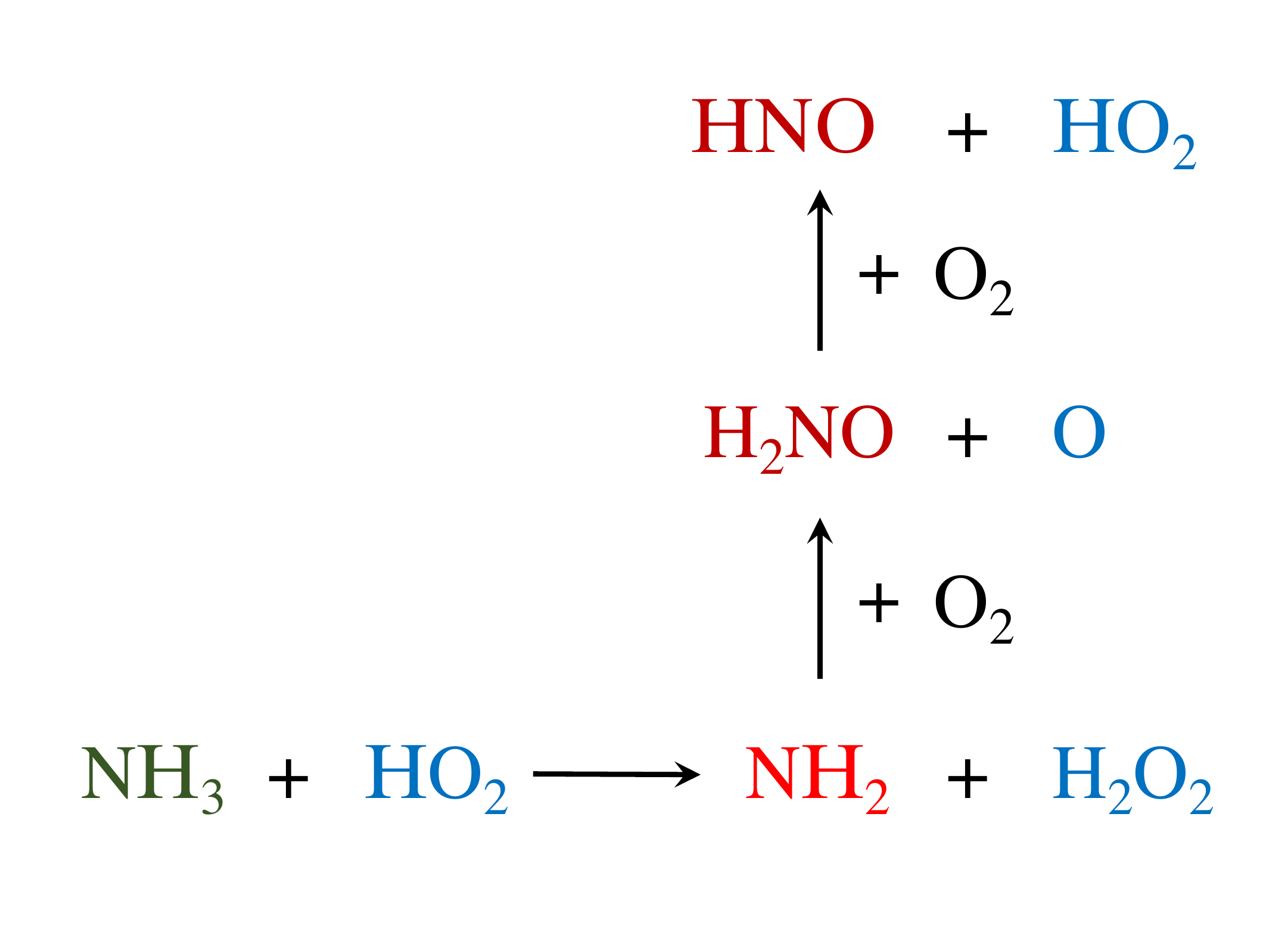}  \hfill  \hfill \includegraphics[scale=0.3]{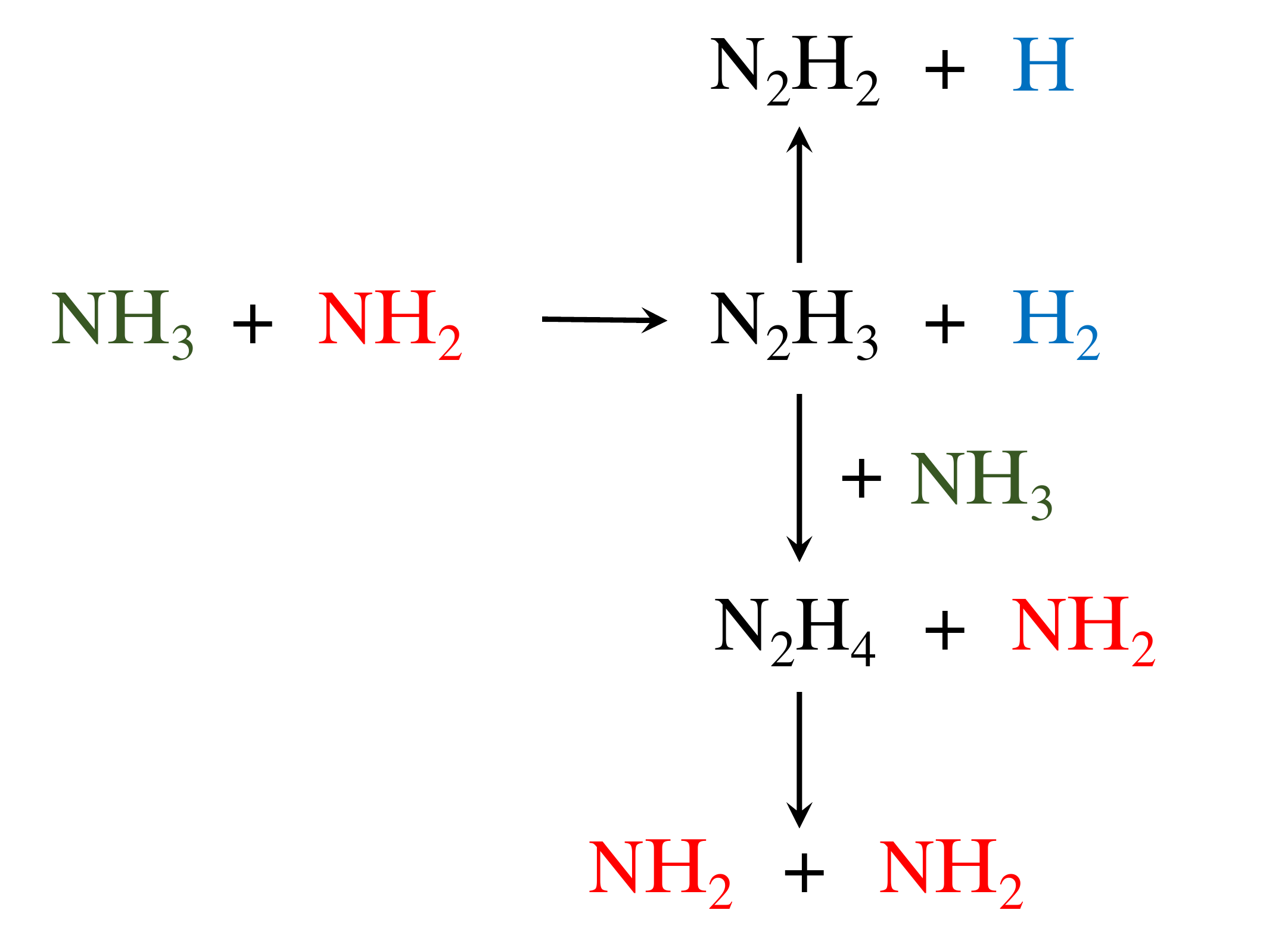} \hfill\hfill\hfill\\
\end{center}
\caption{Schematic representation of the two paths that determine the explosive time scale, which characterises the oxidation process within the chemical runaway.~The path in the left panel is functioning in all five mechanisms, while that in the right panel is functioning only in Shrestha$\_$2018, Li$\_$2019 and Zhang$\_$2021.}
\label{fig:twopaths}
\end{figure}

These findings suggest the existence of two paths that determine the explosive time scale, which characterises the oxidation process within the chemical runaway.~First is the path $NH_3 \rightarrow NH_2 \rightarrow H_2NO \rightarrow HNO$.~Then there is the path that can manifest itself as either $(NH_3, \ NH_2) \rightarrow N_2H_3 \rightarrow N_2H_4 \rightarrow NH_2$ or $(NH_3, \ NH_2) \rightarrow N_2H_3 \rightarrow N_2H_2 ~ \text{and} ~ H$.~Both paths are depicted in Fig.~\ref{fig:twopaths}.~The first path is functional in all five mechanisms, while the second path relates to $N2$ chemistry that is functional only in Shrestha$\_$2018, Li$\_$2019 and Zhang$\_$2021.~It is noted that rection $H+O_2 \rightarrow O+OH$ is shown to promote $\tau_e$ early in the chemical runaway in all three cases involving $N2$ chemistry.~Apparently, this is due to reaction $N_2H_3\rightarrow N_2H_2+H$, which is shown to promote $\tau_e$.

Significant opposition to $\tau_e$ is manifested in Shrestha$\_$2018, Li$\_$2019 and Zhang$\_$2021 via the $H$-consuming reaction $NH_3+H\rightarrow NH_2+H_2$ and, more significantly so in Stagni$\_$2020 and Zhang$\_$2021, via reaction $NH_2+HO_2 \rightarrow NH_3+O_2$.~As discussed earlier, the latter reaction was also shown to oppose the impact of the explosive mode, by  acting towards decreasing its amplitude $f^e$.~Finally, in the case of Shrestha$\_$2018 and Li$\_$2019, in which $N2$ chemistry is present, the $N_2H_3$-consuming reactions $N_2H_3+NH_2\rightarrow H_2NN+NH_3$ and $NH_3+NH_2 \leftarrow N_2H_3+H_2$ provide a minor opposition.

\subsection{Dynamics of the thermal runaway}
\label{sec:IDT}

The reactions contributing towards the generation of $\tau_e$ do not demonstrate as large variety between mechanisms in the thermal runaway as in the chemical runaway.~These reactions can be classified in the following three main sets:
\begin{itemize}
       \item $OH$-producing reactions that support $\tau_e$:  $NH_2+NO\rightarrow NNH+OH$ and $NH_2+HO_2\rightarrow H_2NO+OH$ that are relevant to all five mechanisms,
       \item non-$OH$-producing reactions that support $\tau_e$: $H_2NO+O_2\rightarrow HNO+HO_2$ that is relevant to all five mechanisms, $NH_2+NO_2 \rightarrow H_2NO+NO$ that is relevant to Shrestha$\_$2018 and Li$\_$2019 and $NH_3+O_2 \rightarrow HNO+HO_2$ that is relevant to Stagni$\_$2020 and Zhang$\_$2021.
              \item $NH_2$-depleting reactions that oppose $\tau_e$: $NH_2+NO\rightarrow N_2+H_2O$ that produces the equilibrium products and applies in all five cases, $NH_2+NO_2\rightarrow N_2O+H_2O$ that applies in Shrestha$\_$2018 and Li$\_$2019 and $NH_2+HO_2\rightarrow NH_3+O_2$ that produces the initial reactants and applies in Stagni$\_$2020 and Zhang$\_$2021.
\end{itemize}
These findings suggest that an ubiquitous action supporting  $\tau_e$ originates from the $NH_2$-consuming reactions in the first set that produce $OH$.~The strongest support originates from $NH_2+NO\rightarrow NNH+OH$ in the cases where a sizeable chemical runaway develops (Glarborg$\_$2018, Shrestha$\_$2018 and Li$\_$2019) and from $NH_2+HO_2\rightarrow H_2NO+OH$ in the cases where a chemical runaway practically does not exist (Stagni$\_$2020 and Zhang$\_$2021).~The reactions in the second set recycle species that are reactants in the reactions of the first set; i.e., $H_2NO$, $NO$, $HO_2$.~The action of the reactions in the third set, that oppose $\tau_e$, is due to (i) the competition for $NH_2$ with the promoting ones in the first and second sets and and (ii) the production of stable species, such as $N_2$, $H_2O$ and $N_2O$.

The conclusions stated previously apply throughout the thermal runaway, except the final part where $\tau_e$ decreases rapidly towards its minimum value at $P_5$.~In this final part all five mechanisms provide similar diagnostics.~In particular, in all five cases considered, the following reactions contribute in promoting $\tau_{e}$:
\begin{itemize}
\item The largest contribution originates from the $OH$-producing reaction $H+O_2\rightarrow O+OH$.
\item The two directions of $OH+H_2\leftrightarrow H+H_2O$ provide the second largest contributions to $\tau_{e}$, which cancel each other; the $H$-producing forward direction promoting $\tau_{e}$.
\item The third largest contribution originates from the $H$ and $OH$-producing reaction $O+H_2\rightarrow H+OH$.
\end{itemize} 
The next largest contributions originate from reactions that oppose the explosive character of $\tau_{e}$; i.e., $NH_3+H\rightarrow NH_2+H_2$  in Glarborg$\_$2018, $NH+OH\rightarrow NO+H_2$, $HO_2(+M)\leftarrow OH+O$, $NH+OH\rightarrow HNO+H$  in Shrestha$\_$2018, $NH+OH\rightarrow NO+H_2$, $NH_3+O\rightarrow NH_2+OH$  in Li$\_$2019, $NH+H\rightarrow N+H_2$, $N_2O+H\rightarrow N_2+OH$, $HNO+H\rightarrow NO+H_2$  in Stagni$\_$2020 and $NH_2+H\rightarrow NH+H_2$, $NH_3+H\rightarrow NH_2+H_2$, $NH+OH\rightarrow HNO+H$ in Zhang$\_$2021.

In summary, the explosive activity at the end of the thermal runaway is supported by the $OH$-producing reactions  $H+O_2\rightarrow O+OH$ and $O+H_2\rightarrow H+OH$.~The major opposition to the explosive mode is due to reactions that consume their reactants $H$ and $O$ and that deplete $OH$.~Clearly, the key feature of the explosive mode at this point is the generation of $OH$.



\subsection{Exothermic reactions}
\label{sec:exothermic}

\begin{figure}[t!]
\centering
\hfill \includegraphics[scale=0.35]{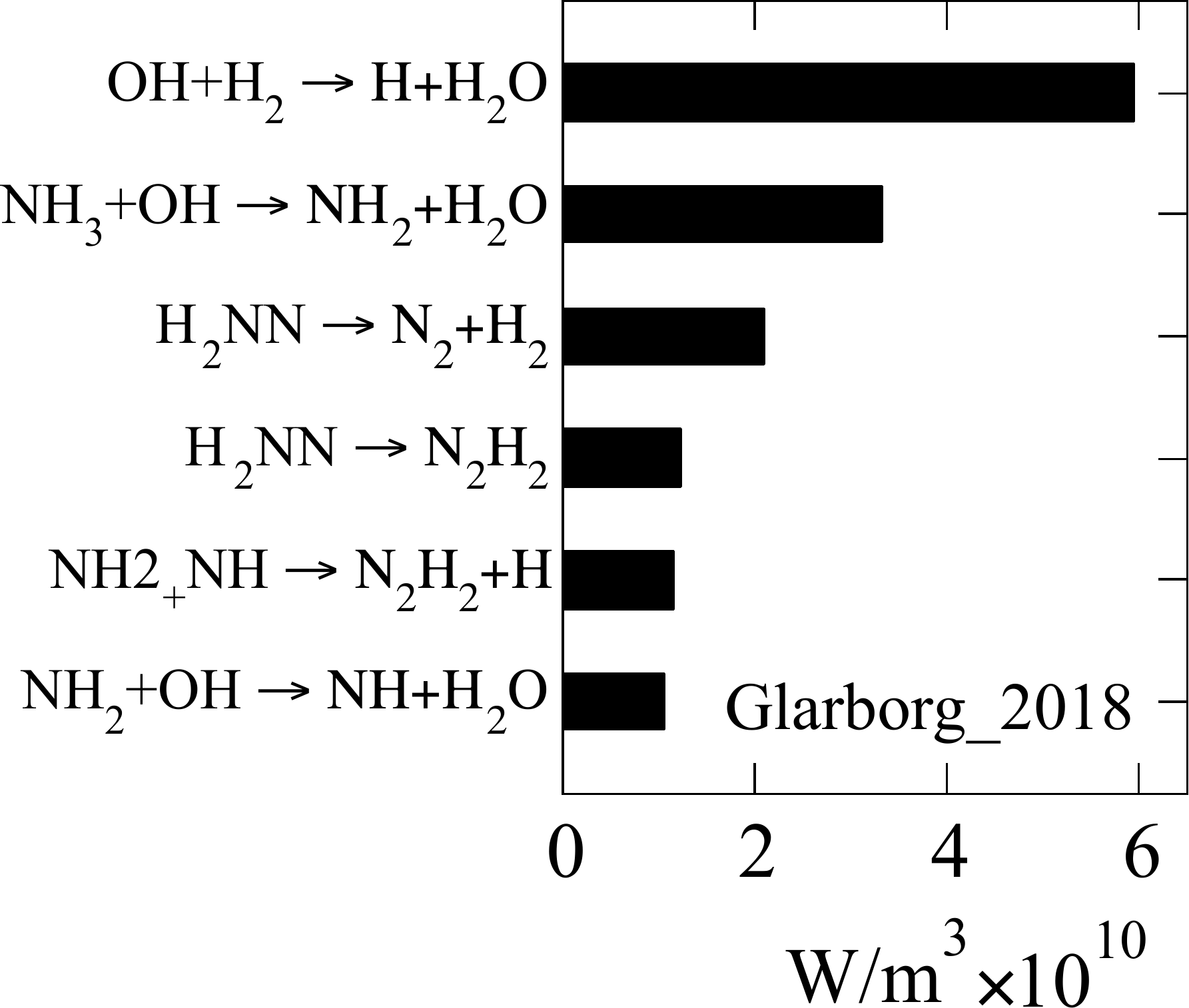}  \hfill  \includegraphics[scale=0.35]{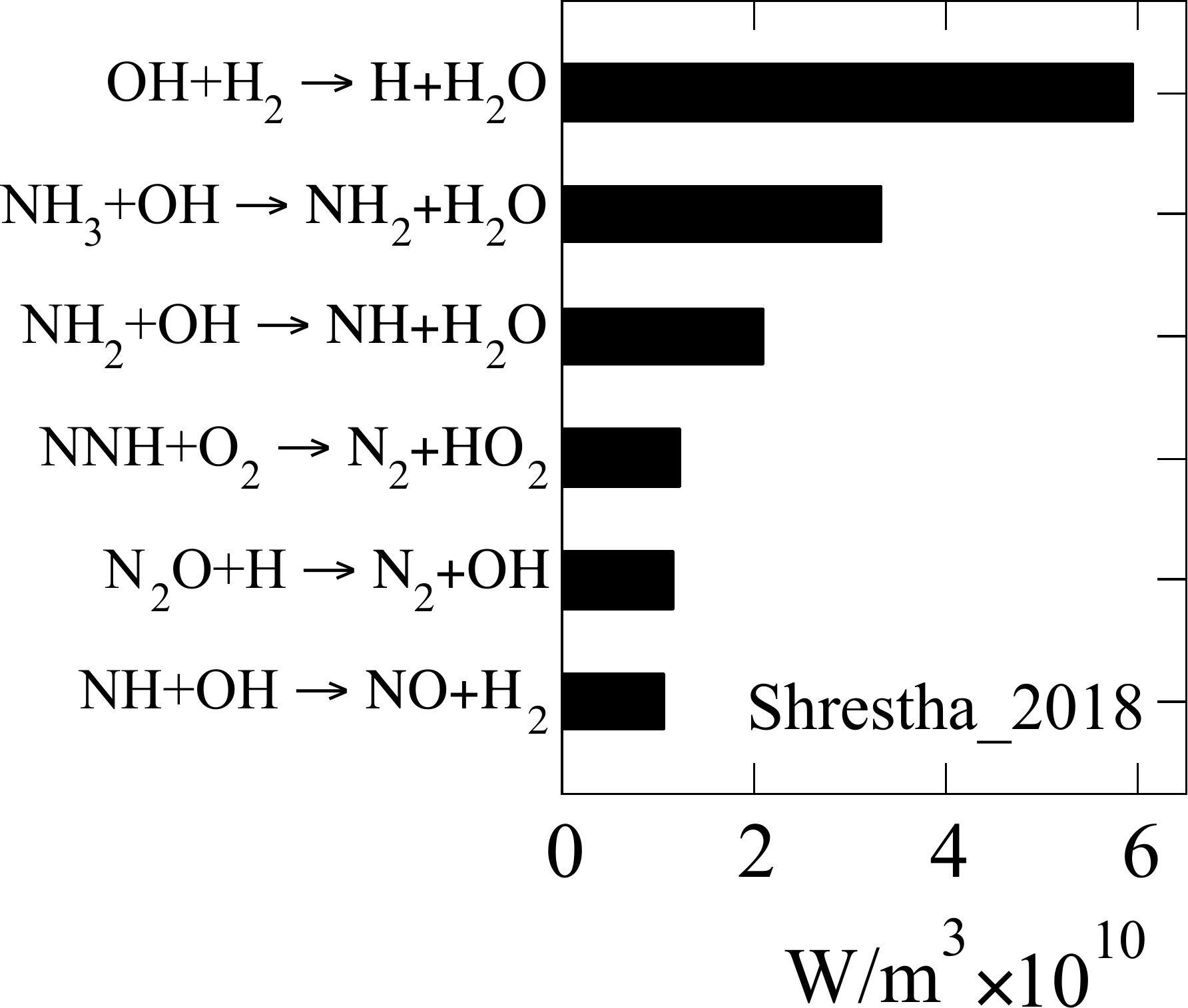}    \hfill \hfill \hfill \\
\vspace{14 pt}
\hfill  \includegraphics[scale=0.35]{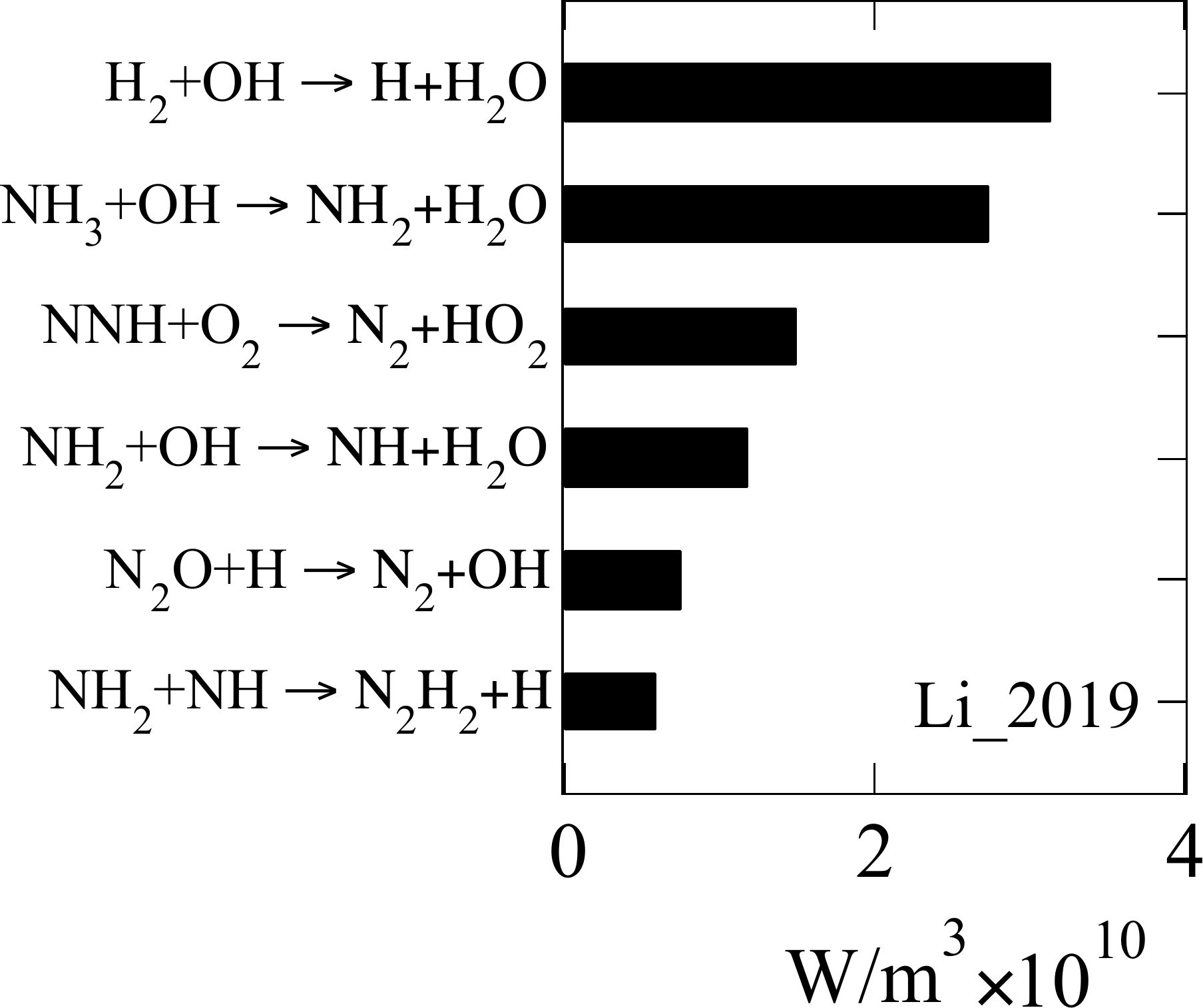}   \hfill  \includegraphics[scale=0.35]{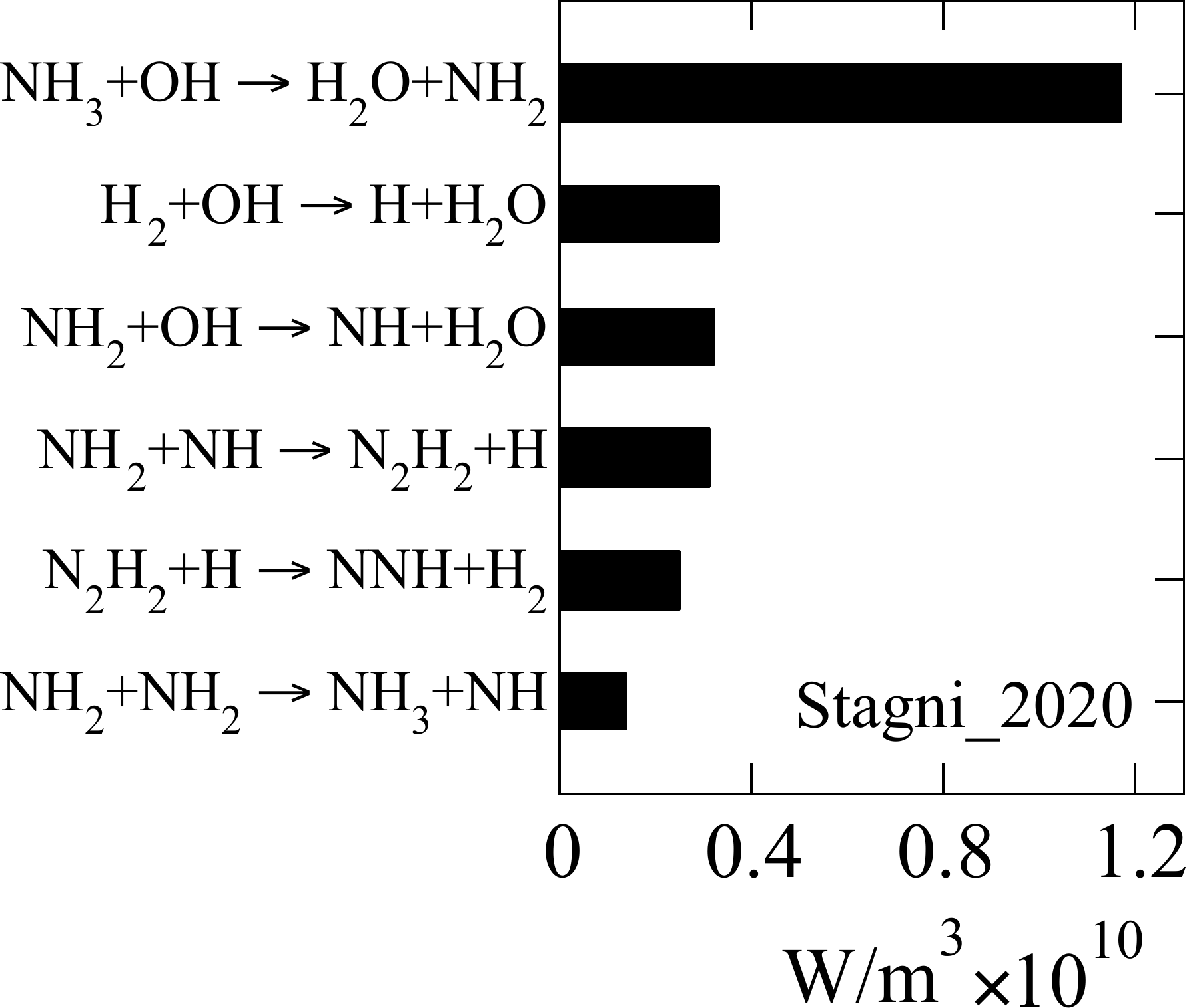} \hfill \hfill \hfill \\
 \vspace{14 pt}
 \includegraphics[scale=0.35]{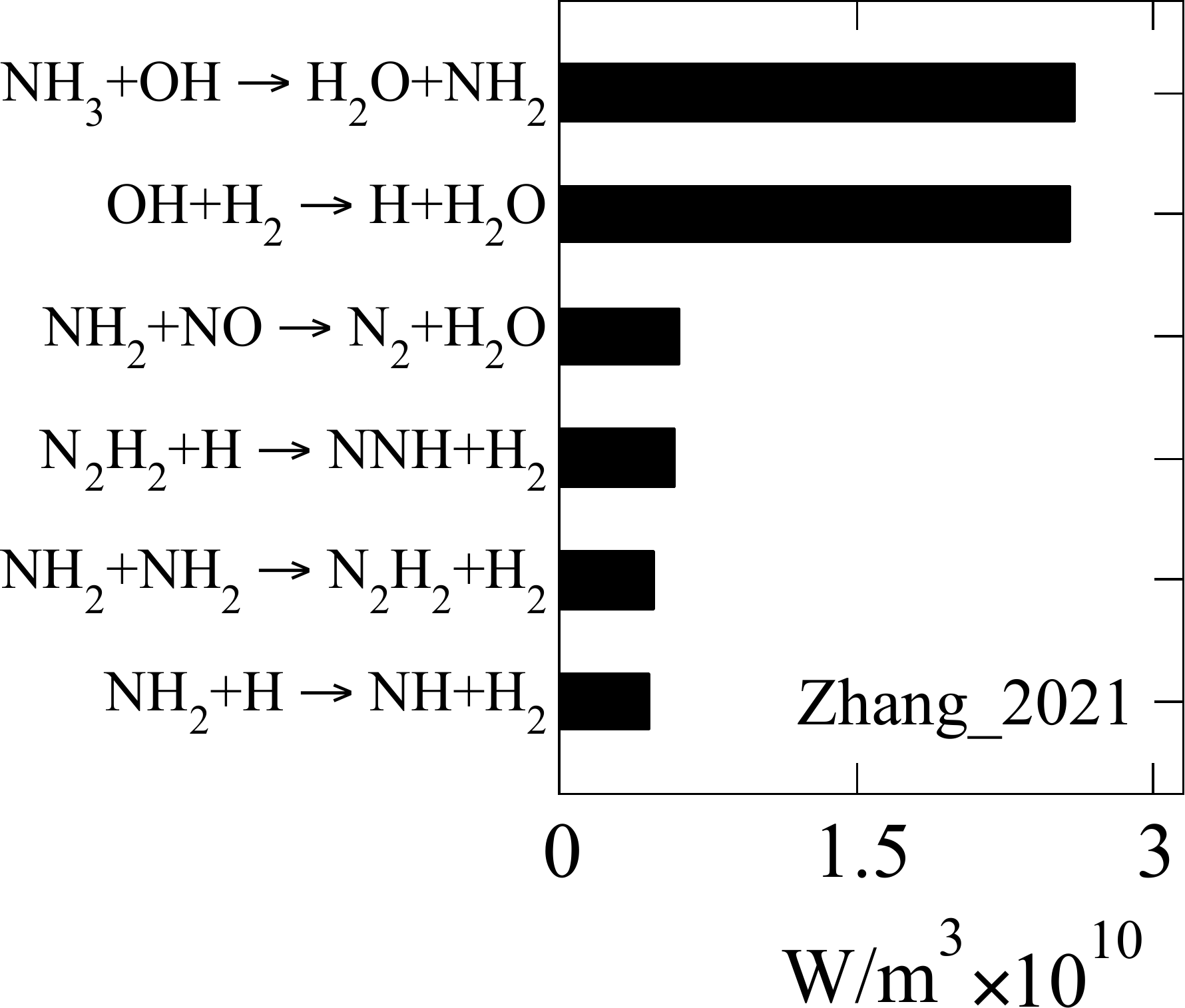} \\
\caption{The most exothermic reactions at P$_5$; T$_0$=1100 K, p$_0$=2 atm and $\phi$=1.0.}
\label{fig:exothermic}
\end{figure}

The role of the $OH$-producing reactions in promoting $\tau_{e}$ during the thermal runaway regime  is revealed when examining the reactions contributing the most to the rate of heat release.~Figure  \ref{fig:exothermic} displays the   reactions exhibiting the largest volumetric heat release rate at P$_5$ in all five cases considered; see Ref.~\citep{rabbani2022chemical} for the computation of the heat release rate of each of the two directions of a reaction.~As shown in Fig.~\ref{fig:Trunc_timescales}, at this  point the temperature undergoes a steep increase.~It is shown in Fig.~\ref{fig:exothermic} that the largest contributions to the heat release rate in all cases considered originate from the $OH$-consuming and $H_2O$-forming reactions $OH+H_2\rightarrow H+H_2O$ and $NH_3+OH\rightarrow NH_2+H_2O$.~The largest contribution is provided by the first reaction  in Glarborg$\_$2018 and Shrestha$\_$2018, by the second reaction in Stagni$\_$2020, while the two reactions provide similar contributions in Li$\_$2019 and Zhang$\_$2021.

As stated previously, reaction $OH+H_2\rightarrow H+H_2O$ contributes significantly to $\tau_{e}$, since it produces the radical for reaction $H+O_2\rightarrow O+OH$, which promotes $\tau_{e}$ the most.~However, this contribution is cancelled by its reverse direction.~In contrast, reaction $NH_3+OH\rightarrow NH_2+H_2O$ provides only negligible contribution to $\tau_{e}$.~These findings suggest that the explosive mode is not linked directly to the steep temperature rise that is recorded at the end of the thermal runaway; e.g, through an overall strongly exothermic reaction.~Instead, it is linked indirectly via the production of the $OH$-radicals by the reactions that support its explosive character, which are consumed by the most exothermic reactions in order to form the most stable species $H_2O$.~A similar relation of the explosive mode with the temperature evolution was also recorded in the super-adiabatic-temperature phenomenon in the post-ignition region of light alcohols \citep{rabbani2022chemical,manias2022effect}.

\begin{figure}[t]
\centering
\hfill \includegraphics[scale=0.26]{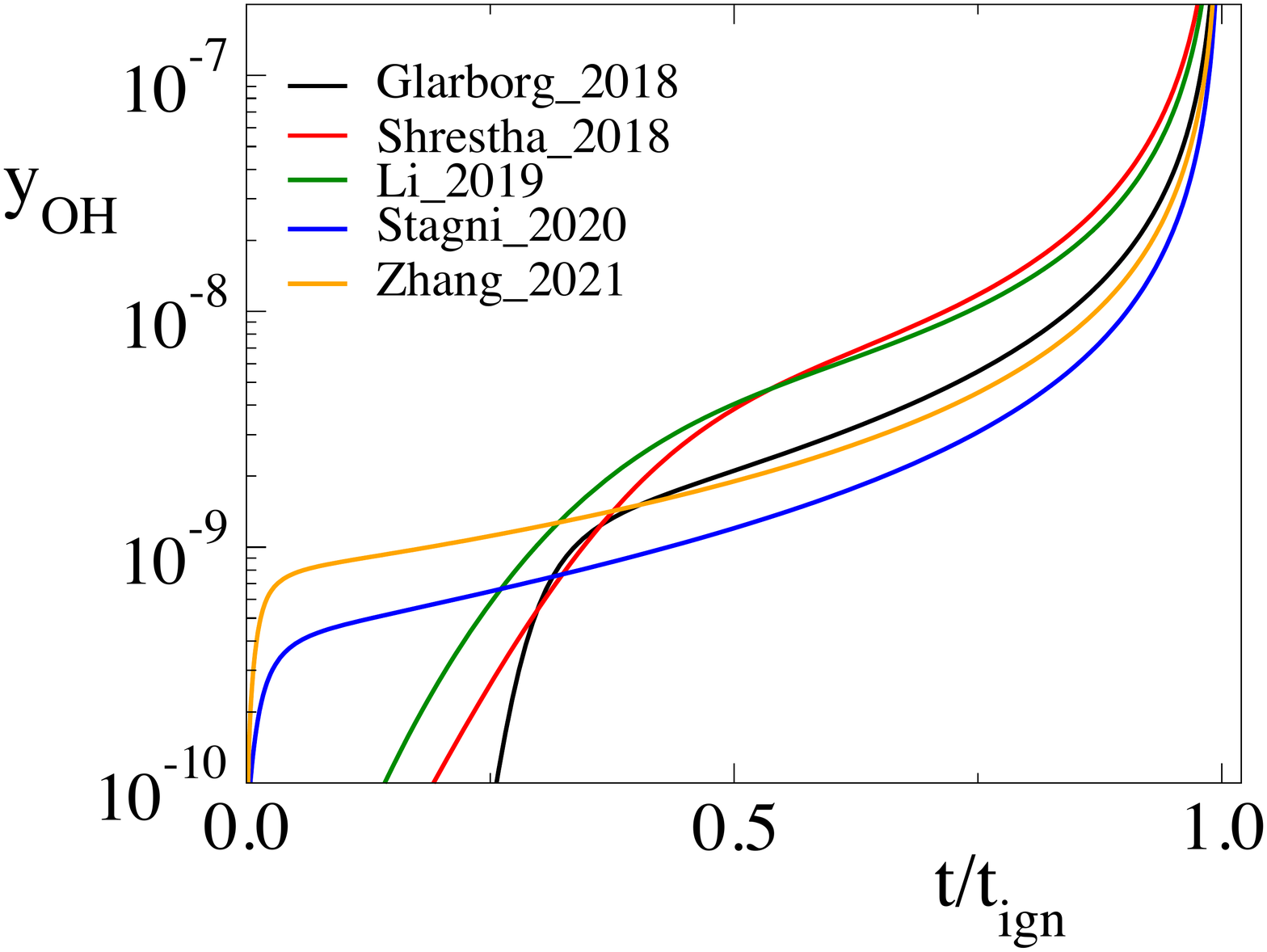}  \hfill\hfill  \includegraphics[scale=0.26]{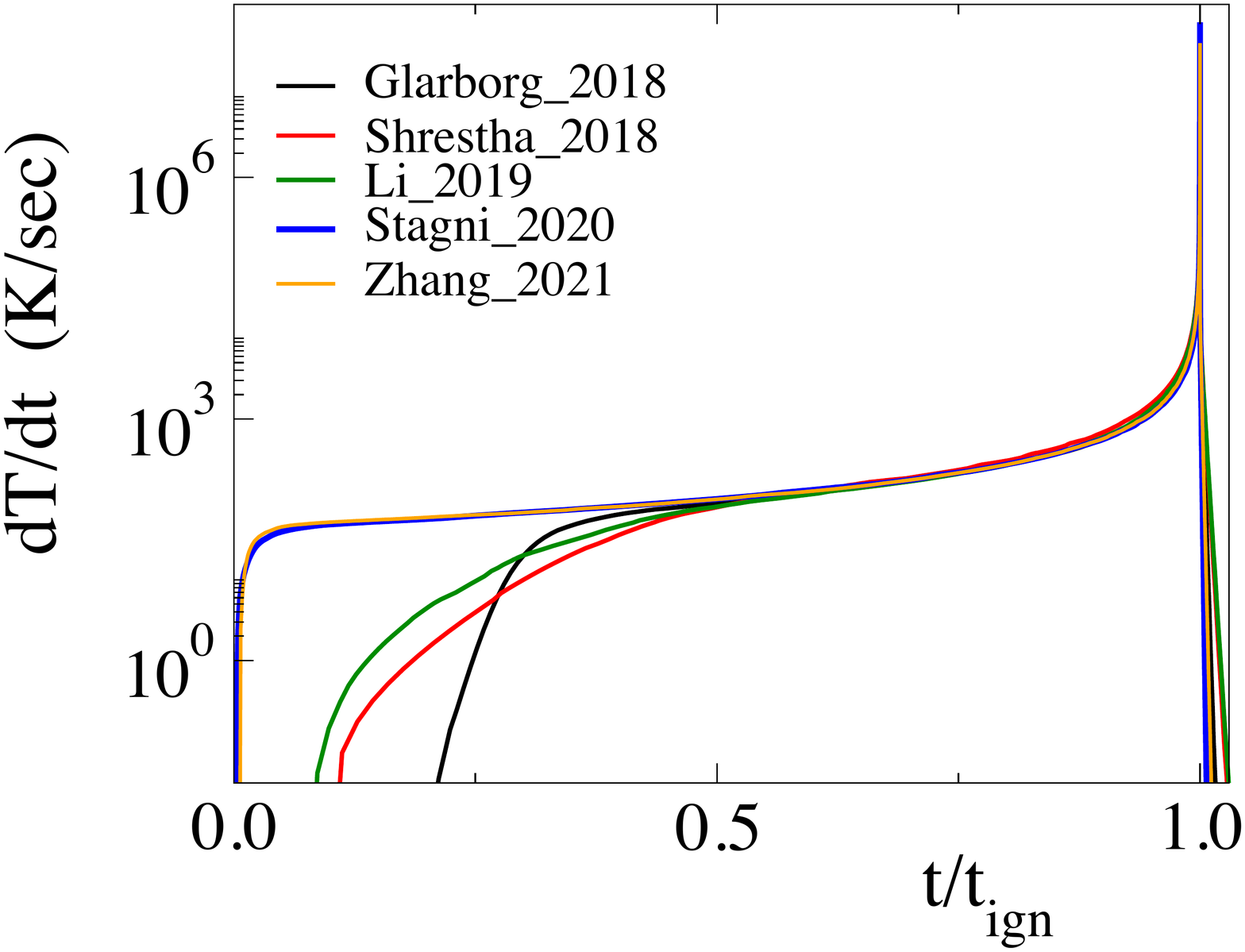}   \hfill \hfill 
\caption{Profiles of $OH$ mass fraction and of the rate of change of temperature vs time scaled with $t_{ign}$; T$_0$=1100 K, p$_0$=2 atm and $\phi$=1.0.}
\label{fig:OH_T}
\end{figure}

The influence of the generation of $OH$ on the rate of temperature rise is highlighted in Fig.~\ref{fig:OH_T}, where the profiles of the mass fraction of this species and of the rate of change of the temperature are displayed, against time scaled with $t_{ign}$.~The profiles in Fig.~\ref{fig:OH_T} are in agreement with the findings displayed in Fig.~\ref{fig:flow}, where it is shown that $OH$-producing reactions participating in the explosive dynamics show up the soonest in Stagni$\_$2020 and Zhang$\_$2021, in which cases the chemical runaway lasts for a very short period; see Fig.~\ref{fig:Trunc_timescales}.~In contrast, the $OH$-producing reactions show up the latest in Glarborg$\_$2018, which exhibits the longest chemical runaway.~The profiles displayed in the left panel of Fig.~\ref{fig:OH_T} validate these conclusions, since it is shown that a sizeable $OH$ pool develops earliest in Stagni$\_$2020 and Zhang$\_$2021 and last in Glarborg$\_$2018.~These findings are reflected in the right panel of Fig.~\ref{fig:OH_T}, where it is shown that the initial steep temperature rise is manifested first in Stagni$\_$2020 and Zhang$\_$2021 and last in Glarborg$\_$2018.

\begin{figure}[t]
\centering
\hfill \includegraphics[scale=0.28]{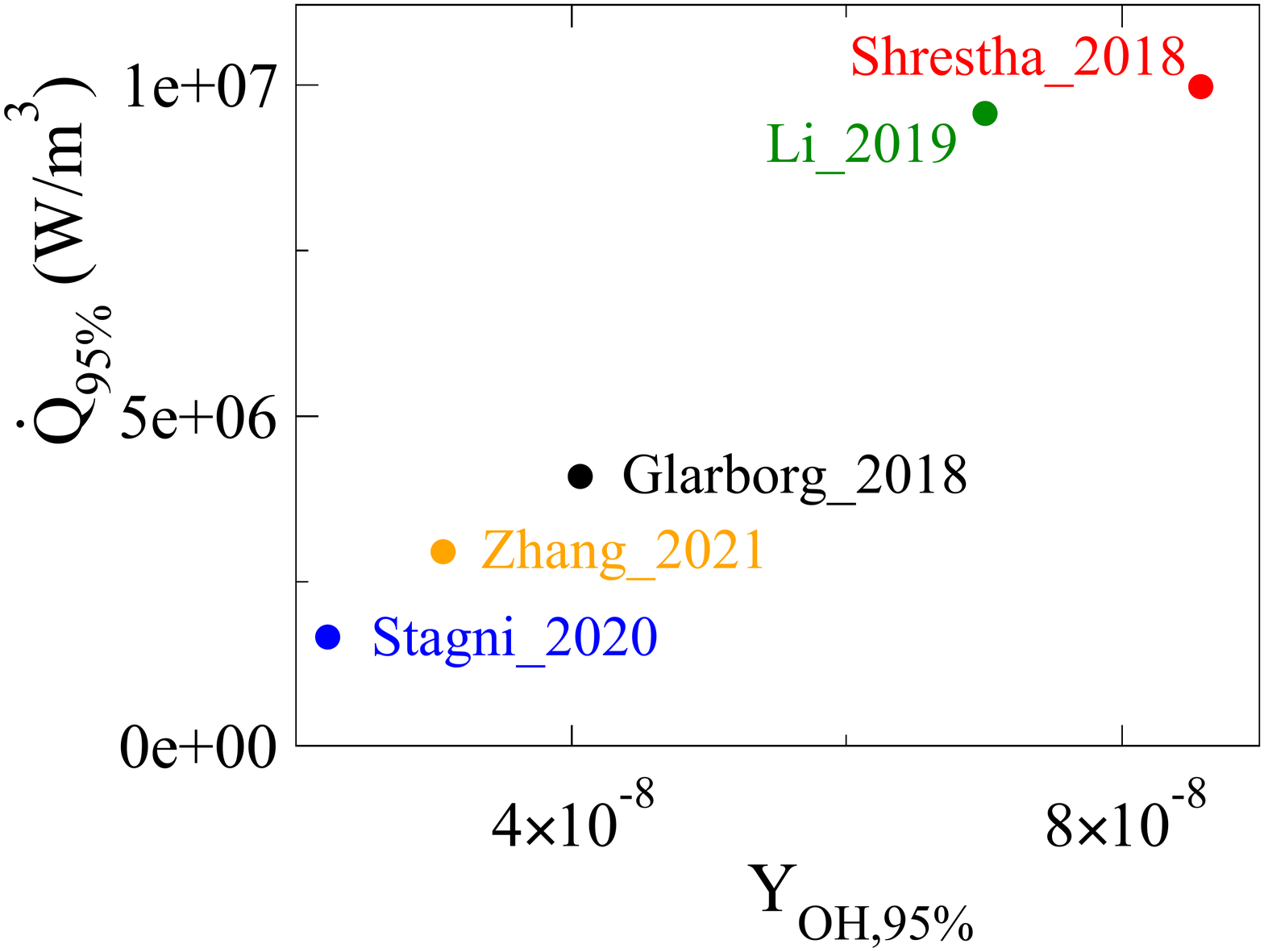} \hfill \includegraphics[scale=0.28]{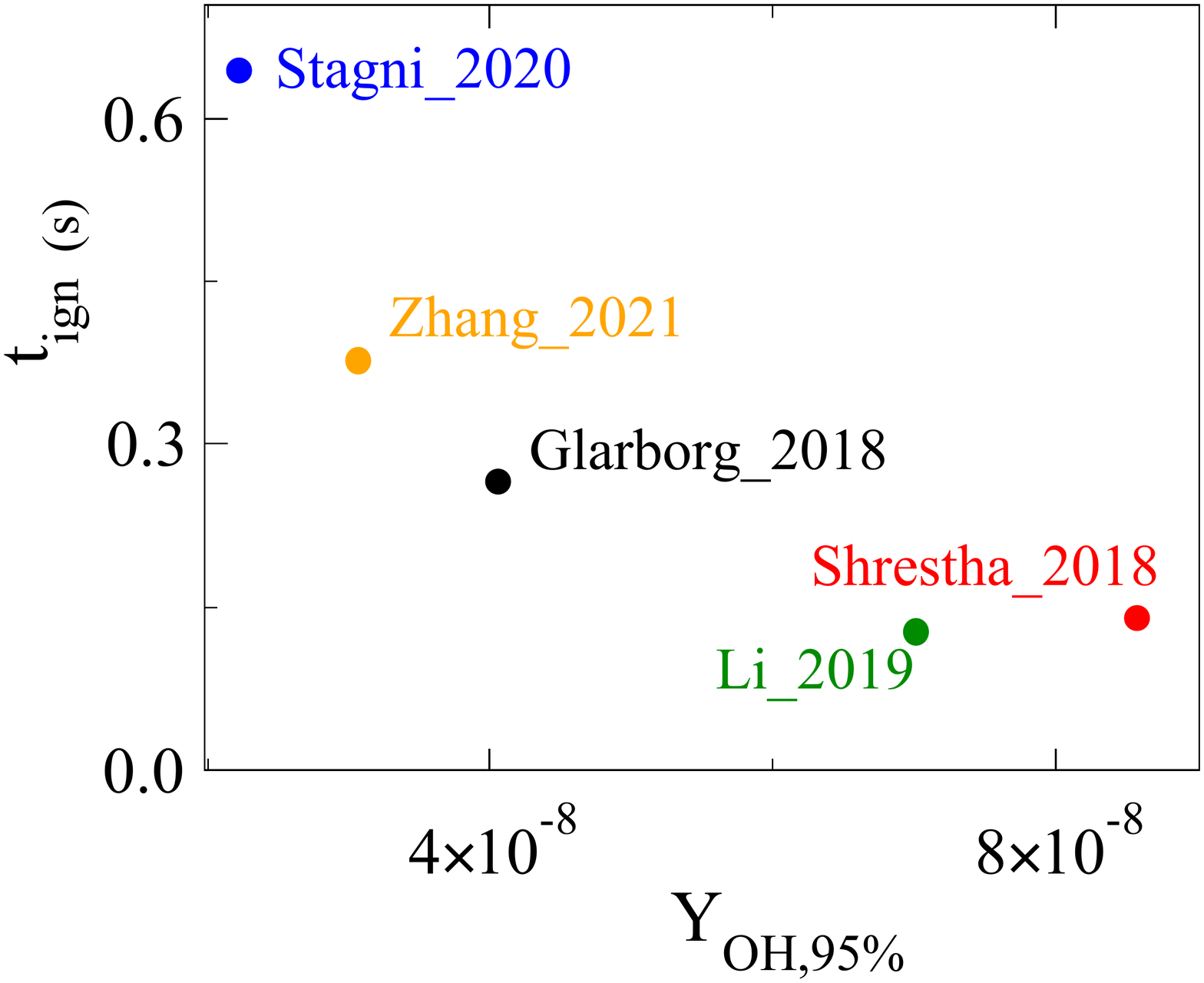} \hfill \hfill \hfill 
\caption{Values of the heat release rate (HRR) $\dot{Q}$ and of the ignition delay time (IDT) $t_{ign}$ vs.the values of the $OH$ mass fraction at 95\% of the IDT, $y_{OH,95\%}$; T$_0$=1100 K, p$_0$=2 atm and $\phi$=1.0.}
\label{fig:OH_IDT}
\end{figure}

Consideration of the  $t_{ign}$ values in Table \ref{tab:mechanisms}, in the light of the $OH$ profiles in the left panel of Fig.~\ref{fig:OH_T}, reveals that the mass fraction of $OH$ at the the last portion of the thermal runaway correlates strongly with the duration of IDT.~Specifically, Fig.~\ref{fig:OH_T} shows that near IDT the five mechanisms rank as follows in an ascending order of $y_{OH}$: the mechanism producing the smallest $y_{OH}$ is Stagni$\_$2020, followed by Zhang$\_$2021, then by Glarborg$\_$2018 and ultimately by Li$\_$2019 and Shrestha$\_$2018, the $y_{OH}$-profiles of which are very close to each other.~This is the way the five mechanisms rank in descending order of IDT, as shown in Table \ref{tab:mechanisms}.~This is another demonstration of the role of $OH$ in sustaining the thermal activity that leads to ignition, as shown further by the results in Fig.~\ref{fig:OH_IDT}.~There, values of heat release rate $\dot{Q}(t=0.95 t_{ign})$ and of ignition delay time $t_{ign}$  are displayed as a function of $y_{OH}(t=0.95 t_{ign})$.~The general trend is that the larger the mass fraction of $OH$ short before ignition, the higher the heat release and the shorter the ignition delay time.~Figure~\ref{fig:OH_IDT} shows that this conclusion clearly applies to Stagni$\_$2020, Zhang$\_$2021, Glarborg$\_$2018 and Li$\_$2019.~It is noted that, although  $y_{OH}(t=0.95 t_{ign})$ in Shrestha$\_$2018 is larger than in Li$\_$2019, the related values of $\dot{Q}(t=0.95 t_{ign})$ and $t_{ign}$ are very close.~This can be attributed to the difference in the Arrhenius constants of the most exothermic reaction close to ignition $OH+H_2 \rightarrow H+H_2O$ (see the panels of Fig.~\ref{fig:exothermic} related to these two mechanisms), resulting in a reaction rate constant in Shrestha$\_$2018 that is smaller  that in Li$\_$2019.

The discussion above leads to the following two major findings:
\begin{itemize}
    \item The early $OH$-formation in Stagni$\_$2020 and Zhang$\_$2021 activates the most exothermic reactions immediately after the start of the process.~This leads to the early completion of the chemical runaway regime and to the fastest transition to the thermal one.~In contrast, the late $OH$-formation in Glarborg$\_$2018 leads to the longest chemical runaway and the slowest transition to the thermal one.
    \item  The duration of IDT, $t_{ign}$, correlates strongly with the maximum value of $OH$ mass fraction that manifests itself in the last portion of the thermal runaway.
 \end{itemize}

\section{Conclusions} \label{sec:conclusions}

In Ref.~\citep{kawka2020comparison}, Kawka et.~al stated that ``\emph{the quantitative description of homogeneous ammonia combustion is still a challenge}".~This challenge was addressed here by analyzing five recently developed chemical kinetics mechanisms with the algorithmic tools of CSP.~The following conclusions were reached by analysing the explosive mode that was shown to drive the process:
\begin{itemize}
    \item The process is initially driven by nitrogen chemistry and transitions gradually to being driven by hydrogen chemistry towards the end of ignition delay time.~There is a general qualitative agreement, in terms of reactions and variables related to ignition, among the five mechanisms considered, as it relates to the period in which hydrogen chemistry dominates.~This agreement is demonstrated by the relatively small differences of $\tau_{e,min}$, displayed in Table \ref{tab:mechanisms}, which are recored at the end of IDT where hydrogen chemistry dominates.~In contrast, there are substantial differences in the period in which nitrogen chemistry dominates.~This disagreement is demonstrated by the very large differences of $t_{ign}$ displayed in the same Table.
    \item A major such difference relates to the influence of $N2$ chemistry  on the mechanism for ignition during the chemical runaway; such chemistry is active only in Shrestha$\_$2018, Li$\_$2019 and Zhang$\_$2021 and contributes to the decrease of $t_{ign}$.~This action of $N2$ chemistry is due to the generation of $H$ and to the fact that three $NH_2$ radicals are produced for each one consumed (see Fig.~\ref{fig:twopaths}).
    \item A serious issue relates to the duration of the chemical runaway; in Glarborg$\_$2018, Shrestha$\_$2018, Li$\_$2019 a sizeable chemical runaway develops, while in Stagni$\_$2020 and Zhang$\_$2021 is extremely short.~The cause of very short chemical runaways is the very fast generation of the radical pool, which slows down the process at the start of the thermal runaway, as it is evident in Figs.~\ref{fig:Trunc_timescales}, \ref{fig:Stagni_Timescales_t0}, \ref{fig:profiles1} and \ref{fig:OH_T},  through the activation of the inverse direction of the radical-producing reactions.~A typical example is  $NH_3+O_2 \rightarrow HO_2+NH_2$, which is the initiation reaction in all five mechanisms considered, the inverse of which exhibits a strong opposition to the explosive mode only in  Stagni$\_$2020 and Zhang$\_$2021.
    \item During the thermal runaway, in all five mechanisms considered the explosive mode is mainly generated by the $OH$-producing reactions $NH_2+NO\rightarrow NNH+OH$ and $NH_2+HO_2\rightarrow H_2NO+OH$ and by other reactions that support these two through the production of the required reactants $HO_2$ and $NO$.~These support reactions differ among the various mechanisms.
    \item The explosive mode is not linked directly to the steep temperature rise at the end of the thermal runaway; e.g, through an overall strongly exothermic reaction.~Instead, it is linked indirectly via the production of the $OH$-radicals by the reactions that support its explosive character, which are consumed by the most exothermic reactions in order to form the most stable species $H_2O$.
    \item Although temperature is the variable related the most to the explosive mode at the end of the thermal runaway, where  the steep temperature rise materialises, the $OH$-producing reactions that mainly support this mode are not the most exothermic.~Instead, in all five cases considered, the two most exothermic reactions in this period are the $OH$-consuming $OH+H_2 \rightarrow H+H_2O$ and $NH_3+OH \rightarrow NH_2+H_2O$.~The former reaction is the most exothermic in Glarborg$\_$2018, Shrestha$\_$2018 and Li$\_$2019 (where a sizeable chemical runaway develops) and the latter is the most exothermic one in Stagni$\_$2020 and Zhang$\_$2021 (where the chemical runaway is very short).~The less exothermic reactions differ among the mechanisms.
\end{itemize}

The purpose of this analysis was to identify thermochemical action that is either common or different among the five mechanisms considered and exerts significant influence on ignition delay.~A detailed examination was carried out for each mechanism and the output was synthesised  in order to reach conclusions on major features of the ignition process.~These conclusions point to specific directions for future research, in order to construct a reliable mechanism.~In particular, time-resolved measurements of $OH$ will clarify the issue of the duration of the chemical runaway.~Also, an accurate determination of the the Arrhenius parameters of the initiation reaction $NH_3+O_2\rightarrow NH_2+HO_2$ will be very important in this regard.~Finally, measurements of species like $N_2H_2$, $N_2H_3$ and $N_2H_4$ will determine the influence of $N2$ chemistry.

The present analysis considers only one set of values of T$_0$, p$_0$ and $\phi$.~It is reasonable to assume that the features  of the underlying chemistry (e.g., the role of $N2$ chemistry, the relative length of chemical runaway, etc) might vary with these parameters.~This dependance has to be explored in future research.


\section*{Acknowledgements}
The support from Khalifa University of Science and Technology, via projects CIRA-2019-033 and and RC2-2019-007, is gratefully acknowledged.

\clearpage
\newpage

\renewcommand{\thesection}{\@Appendix A}
\setcounter{equation}{0}\renewcommand{\theequation}{A.\arabic{equation}}
\setcounter{figure}{0}\renewcommand{\thefigure}{A.\arabic{figure}}
\setcounter{table}{0}\renewcommand{\thetable}{A.\arabic{table}}
\section{CSP Diagnostics for the Glarborg$\_$2018 mechanism}
\label{glarborg}

\begin{figure}[b]
\centering
\hfill\includegraphics[scale=0.355]{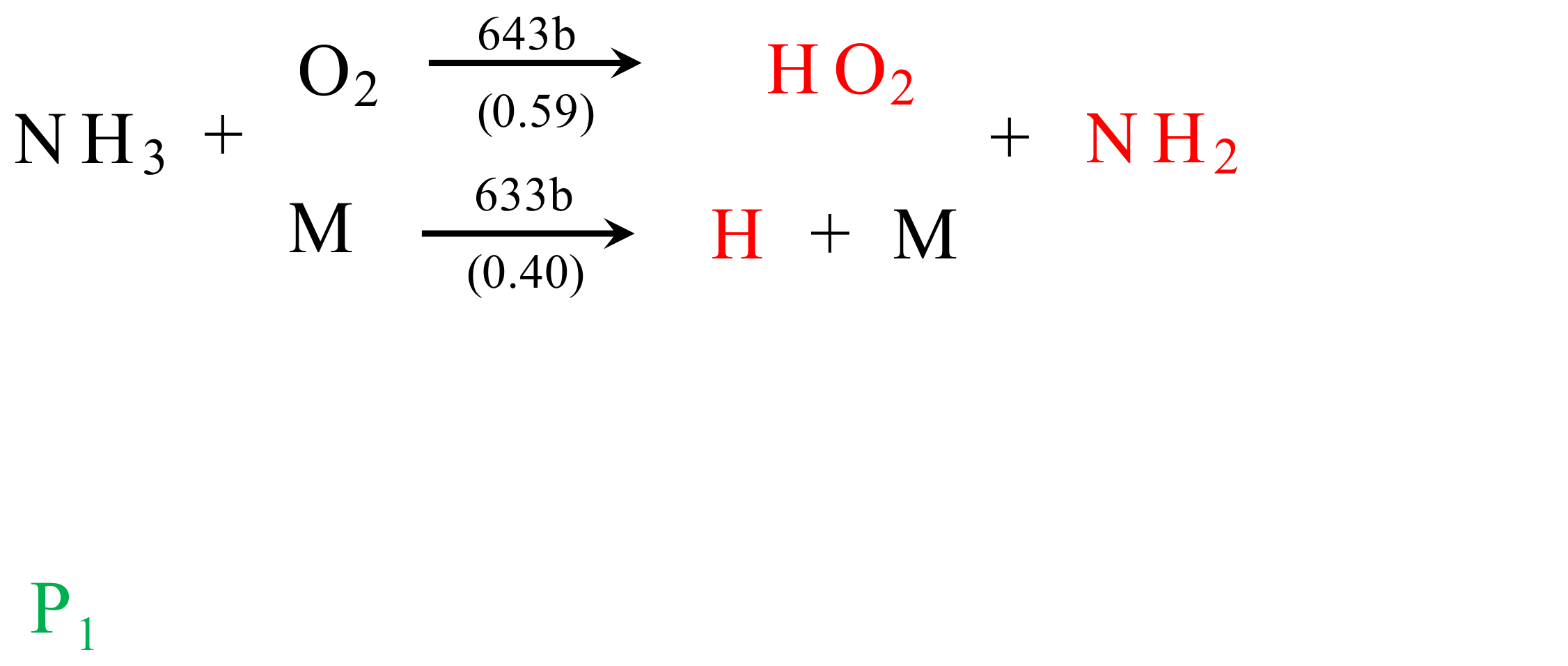}  \hfill \includegraphics[scale=0.355]{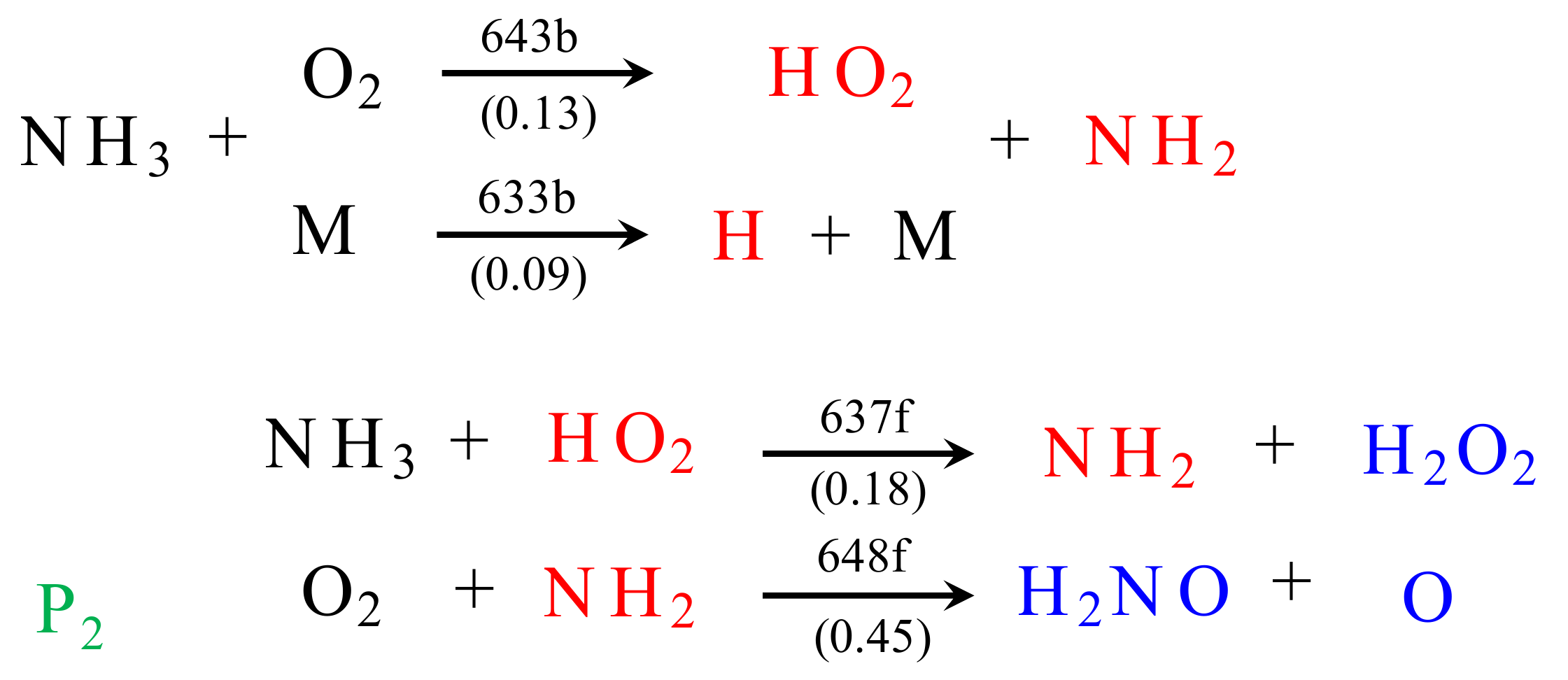} \hfill\hfill\hfill
\caption{Reactions exhibiting large APIs at points P$_1$ and P$_2$ in Glarborg$\_$2018; T$_0$=1100 K, p$_0$=2 atm and $\phi$=1.0.~Numbers in parentheses denote API values.~The different colors are meant to highlight several phases of the process as they are described in the text.}
\label{fig:API_Glarborg}
\end{figure}

The reactions related to the explosive dynamics during IDT in  Glarborg$\_$2018  are listed in Table \ref{tab:Glarborg_Reactions}.~The reactions exhibiting large APIs at P$_1$ and P$_2$ are shown in Fig.~\ref{fig:API_Glarborg}, while at points P$_3$ to P$_5$  they are the same with those exhibiting large TPIs.~All TPI and Po CSP diagnostics for the explosive mode, computed at points P$_1$ to P$_5$ shown in Fig.~\ref{fig:Trunc_timescales}, are displayed in Table \ref{tab:Glarborg_diagnostics}.

\begin{table}[h]
\caption{The reactions contributing the most to the explosive dynamics in Glarborg$\_$2018.~Bold/regular font represents exothermic/endothermic reactions.}
\begin{center}
\begin{tabular}{r@{~}c@{~}lr@{~}c@{~}l}
\hline
1f			&   :       &        $H$ + $ O_2$ $\rightarrow$ $O$ + $ OH$				&	648f			&   :       &        $NH_2$ + $ O_2$  $\rightarrow$  $H_2NO$ + $ O $			\\
3f			&   :       &        $O$ + $ H_2$ $\rightarrow$ $OH$ + $ H$				&	\textbf{654f}	&   :       &        $NH_2$ + $ NO$   $\rightarrow$  $N_2$ + $ H_2O$			\\
4\textbf{f}/b	&   :       &        $OH$ + $ H_2$ $\leftrightarrow$    $H$ + $ H_2O$		&	655f			&   :       &        $NH_2$ + $ NO$  $\rightarrow$    $NNH$ + $ OH$			\\
633b			&   :       &        $NH_2$ + $H (+M)$  $\leftarrow$ $NH_3(+M)$			&	699f/\textbf{b}	&   :       &        $H_2NO$ + $ M$ $\leftrightarrow$ $HNOH$  $+ M$			\\
634f/\textbf{b}	&   :       &        $NH_3$ + $ H$ $\leftrightarrow$    $NH_2$ + $ H_2$		&	705f			&   :       &        $H_2NO$ + $ O_2$ $\rightarrow$ $HNO$ + $ HO_2$			\\
637f			&   :       &        $NH_3$ + $ HO_2$  $\rightarrow$ $NH_2$ + $ H_2O_2$	&	\textbf{706f}	&   :       &        $H_2NO$ +  $NH_2$ $\rightarrow$  $HNO$ +  $NH_3$		\\
643b			&   :       &        $NH_2$ + $HO_2$  $\leftarrow$ $NH_3$ + $O_2$		&	\textbf{771f}	&   :       &        $HNOH$ +  $NH_2$ $\rightarrow$ $H_2NN$ + $ H_2O$		\\
\textbf{644f}	&   :       &        $NH_2$ + $ HO_2$  $\rightarrow$ $H_2NO$ + $ OH$	&	\textbf{772f}	&   :       &        $NH_2$ +  $NH_2 (+M)$ $\rightarrow$ $N_2H_4$ (+ $ M$)	\\
\hline
\end{tabular}
\label{tab:Glarborg_Reactions}
\end{center}
\end{table}

Figure \ref{fig:API_Glarborg} shows the way the oxidation process commences according to Glarborg$\_$2018. It is shown that at P$_1$ the reactions that exhibit large APIs  are 643b (0.59): $NH_2$ + $HO_2$  $\leftarrow$ $NH_3$ + $O_2$ and 633b (0.40): $NH_2$ + $H(+M)$  $\leftarrow$ $NH_3(+M)$, where the numbers in parentheses denote the API value.~At P$_2$ reactions 643b and 633b are still exhibiting large APIs, albeit smaller ones; i.e., (0.13) and (0.09) respectively.~In addition, the $HO_2$ produced by reaction 643b activates reaction 637f (0.18) $NH_3$ + $ HO_2$  $\rightarrow$ $NH_2$ + $ H_2O_2$, while the $NH_2$ produced by both 643b and 633b activates reaction 648f (0.45) $NH_2$ + $ O_2$  $\rightarrow$  $H_2NO$ + $ O $.~In summary, as shown in Fig.~\ref{fig:API_Glarborg}, the initiation of the oxidation process involves reactions producing initially $NH_2$  and $HO_2$ at P$_1$ and additionaly $H_2NO$, $H_2O_2$ and $O$ at P$_2$.~At P$_1$ and P$_2$ the Po indices displayed in Table \ref{tab:Glarborg_diagnostics} show that three of these species ($NH_2$, $HO_2$ and $H_2NO$) are the ones related the most to the explosive mode.~These species are reactants of the reactions 648f (0.70): $NH_2$ + $ O_2$  $\rightarrow$  $H_2NO$ + $ O $, 637f (0.23): $NH_3$ + $ HO_2$  $\rightarrow$ $NH_2$ + $ H_2O_2$ and 705f (0.07) $H_2NO$ + $ O_2$ $\rightarrow$ $HNO$ + $ HO_2$, which exhibit the largest TPIs, so they contribute the most to the explosive time scale $\tau_e$; i.e., they are the reaction that determine the time scale that characterizes the autoignition process there.

Figure \ref{fig:Trunc_timescales} shows that up to P$_3$, $\tau_e$ does not change much.~This is reflected in the CSP diagnostics at this point, displayed in Table \ref{tab:Glarborg_diagnostics}, which are not that much different from those in P$_1$ and P$_2$.~In particular, it is shown that, in addition to reactions 648f, 705f and 637f, a significant contribution to the explosive character of the mode is provided  by reaction 644f:  $NH_2$ + $ HO_2$  $\rightarrow$ $H_2NO$ + $ OH$ and a small opposition is provided by reaction 772f: $NH_2$ +  $NH_2(+M)$ $\rightarrow$ $N_2H_4$ (+ $ M$) that deprives $NH_2$ from the promoting reactions 648f and 644f.~

At P$_4$ there is a large number of reactions either favoring or opposing the explosive character of the mode, as shown in Table \ref{tab:Glarborg_diagnostics}.~Among those providing the largest contributions, are the two directions of reaction 699, which cancel each other.~Of the remaining reactions, those that promote the most the explosive character (positive TPIs) are  655f: $NH_2$ + $ NO$  $\rightarrow$    $NNH$ + $ OH$, 705f: $H_2NO$ + $ O_2$ $\rightarrow$ $HNO$ + $ HO_2$ and 1f: $H$ + $ O_2$ $\rightarrow$ $O$ + $ OH$.~The common feature of these reactions is that they produce radicals related to hydrogen chemistry.~The reactions that oppose the most the explosive character (negative TPIs) are reactions 654f, 706f and 771f, which (i) deprive the promoting reactions 655f and 705f  of the reactants of $NH_2$ and $H_2NO$ and (ii) lead to the formation of stable species, such that $N_2$, $NH_3$ and $H_2O$.~The variable associated the most to the explosive mode is temperature.~This is a reasonable finding, since P$_4$ lies in the thermal runaway, as shown in Fig.~\ref{fig:Trunc_timescales}.

\begin{table}[t]
\caption{The largest TPI and Po values for the explosive mode in Glarborg$\_$2018; T$_0$=1100 K, p$_0$=2 atm and $\phi$=1.0.~Numbers in parenthesis denote powers of ten.~Bold/regular font represents exothermic/endothermic reactions.}
\begin{center}
\begin{tabular}{rcccrcrcrc}
\hline
\multicolumn{2}{c}{\textbf{P$_1$}} & \multicolumn{2}{c}{\textbf{P$_2$}}                      & \multicolumn{2}{c}{\textbf{P$_3$}} & \multicolumn{2}{c}{\textbf{P$_4$}} & \multicolumn{2}{c}{\textbf{P$_5$}} \\ 
\hline
\multicolumn{2}{c}{t$_1$ =0.00(0)s}    & \multicolumn{2}{c}{t$_2$ =1.11(-2)s}                         & \multicolumn{2}{c}{t$_3$ =5.45(-2)s}    & \multicolumn{2}{c}{t$_4$ =1.84(-1)s}    & \multicolumn{2}{c}{t$_5$ =2.65(-1)s}    \\
\multicolumn{2}{c}{$\tau_{e,f}$=7.71(-3)s}    & \multicolumn{2}{c}{$\tau_{e,f}$=7.70(-3)s}                         & \multicolumn{2}{c}{$\tau_{e,f}$=6.46(-3)s}    & \multicolumn{2}{c}{$\tau_{e,f}$=7.32(-2)s}    & \multicolumn{2}{c}{$\tau_{e,f}$=1.32(-6)s}    \\ 
\hline
\multicolumn{10}{c}{\textbf{TPI}}                                                                                                                                                            \\ 
\hline
648f           & +0.60          & \multicolumn{1}{r}{648f} & \multicolumn{1}{l}{+0.60} & 648f               & +0.25      & 655f               & +0.14      & 1f                & +0.26       \\
637f           & +0.23          & \multicolumn{1}{r}{637f} & \multicolumn{1}{l}{+0.23} & \textbf{644f}      & +0.18      & \textbf{654f}      & -0.12      & \textbf{4f}       & +0.15       \\ 
705f           & +0.07          & \multicolumn{1}{r}{705f} & \multicolumn{1}{l}{+0.07} & 705f               & +0.16      & 705f               & +0.10      & 4b                & ~-0.14       \\
               &                &                          &                           & 637f               & +0.08      & 699f               & ~-0.06      & 3f                & +0.05       \\
               &                &                          &                           & \textbf{772f}      & ~-0.05      & \textbf{706f}      & ~-0.05      & 634f                   &~-0.04             \\
               &                &                          &                           &                    &            & \textbf{699b}      & +0.05      &                   &             \\
               &                &                          &                           &                    &            & 1f                 & +0.05      &                   &             \\
               &                &                          &                           &                    &            & \textbf{771f}      & ~-0.05      &                   &             \\ 
               &                &                          &                           &                    &            & 634f      & ~-0.04      &                   &             \\ 
               &                &                          &                           &                    &            & \textbf{644f}      & +0.04      &                   &             \\ 
                 \hline
\multicolumn{10}{c}{\textbf{Po}}                                                                                                                                                             \\ 
\hline
$NH_2$            & +0.65          & \multicolumn{1}{r}{$NH_2$}  & \multicolumn{1}{l}{+0.65} & $NH_2$                & +0.43      & $T$                  & +0.81      & $T$                 & +0.73       \\ 
$HO_2$            & +0.25          & \multicolumn{1}{r}{$HO_2$}  & \multicolumn{1}{l}{+0.25} & $HO_2$                & +0.26      & $H_2$                 & +0.16      & $NH_3$               & +0.49       \\ 
$H_2NO$           & +0.07          & \multicolumn{1}{r}{$H_2NO$} & \multicolumn{1}{l}{+0.07} & $H_2N$O               & +0.20      &                    &            & $H_2O$               & ~-0.40       \\ 
               &                &                          &                           &                    &            &                    &            & $H$                 & +0.14       \\ 
               \hline
\end{tabular}
\label{tab:Glarborg_diagnostics}
\end{center}
\end{table}

At the point P$_5$, where $\tau_e$ reaches minimum value, right before t$_{ign}$, Table \ref{tab:Glarborg_diagnostics} shows that the major contributions towards the explosive character of  $\tau_e$ originate from hydrogen chemistry.~In particular, the largest contribution is produced from reaction 1f: $H+O_2\rightarrow O+OH$.~The next contributions are produced from the two directions of reaction 4f: $OH+H_2\leftrightarrow H+H_2O$, which cancel each other.~The third largest contribution in promoting $\tau_{e}$ originates  from reaction 3f: $O+H_2\rightarrow H+OH$.~The major opposition to $\tau_e$ originates from reaction 634f:  $NH_3$ + $ H$ $\rightarrow$    $NH_2$ + $ H_2$, which consumes the reactant $H$ of the most promoting reaction 1f.~In summary, the peak of explosive dynamics associates with the production of mainly $OH$ and secondarily of $H$ and $O$ radicals, which are reactants of the $OH$-producing reactions 1f and 3f.

\clearpage
\newpage

\renewcommand{\thesection}{\@Appendix B}
\setcounter{equation}{0}\renewcommand{\theequation}{B.\arabic{equation}}
\setcounter{figure}{0}\renewcommand{\thefigure}{B.\arabic{figure}}
\setcounter{table}{0}\renewcommand{\thetable}{B.\arabic{table}}
\section{CSP Diagnostics for the Shrestha mechanism}
 \label{shrestha}

\begin{figure}[b]
\centering
\hfill\hfill\includegraphics[scale=0.355]{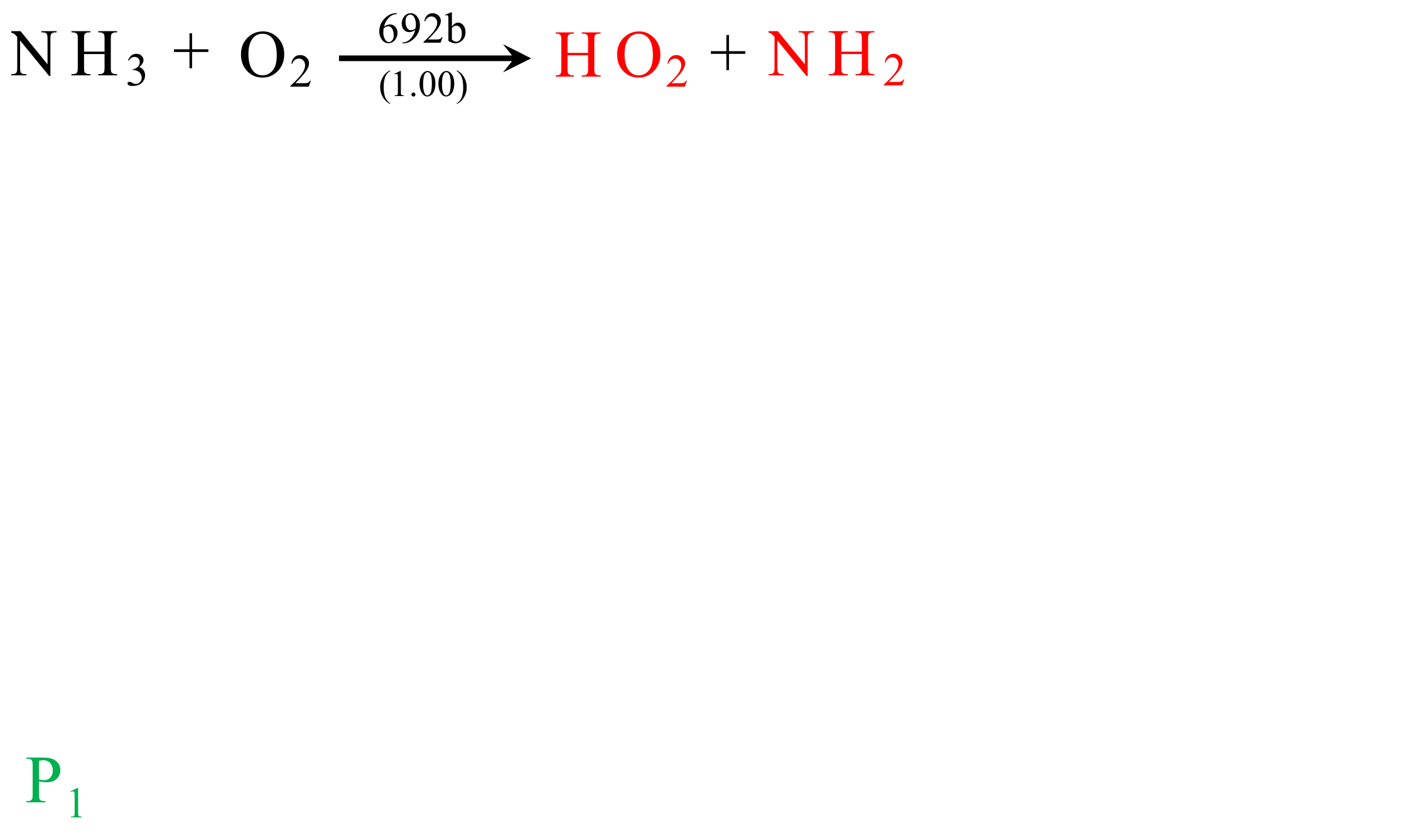}  \hfill \includegraphics[scale=0.355]{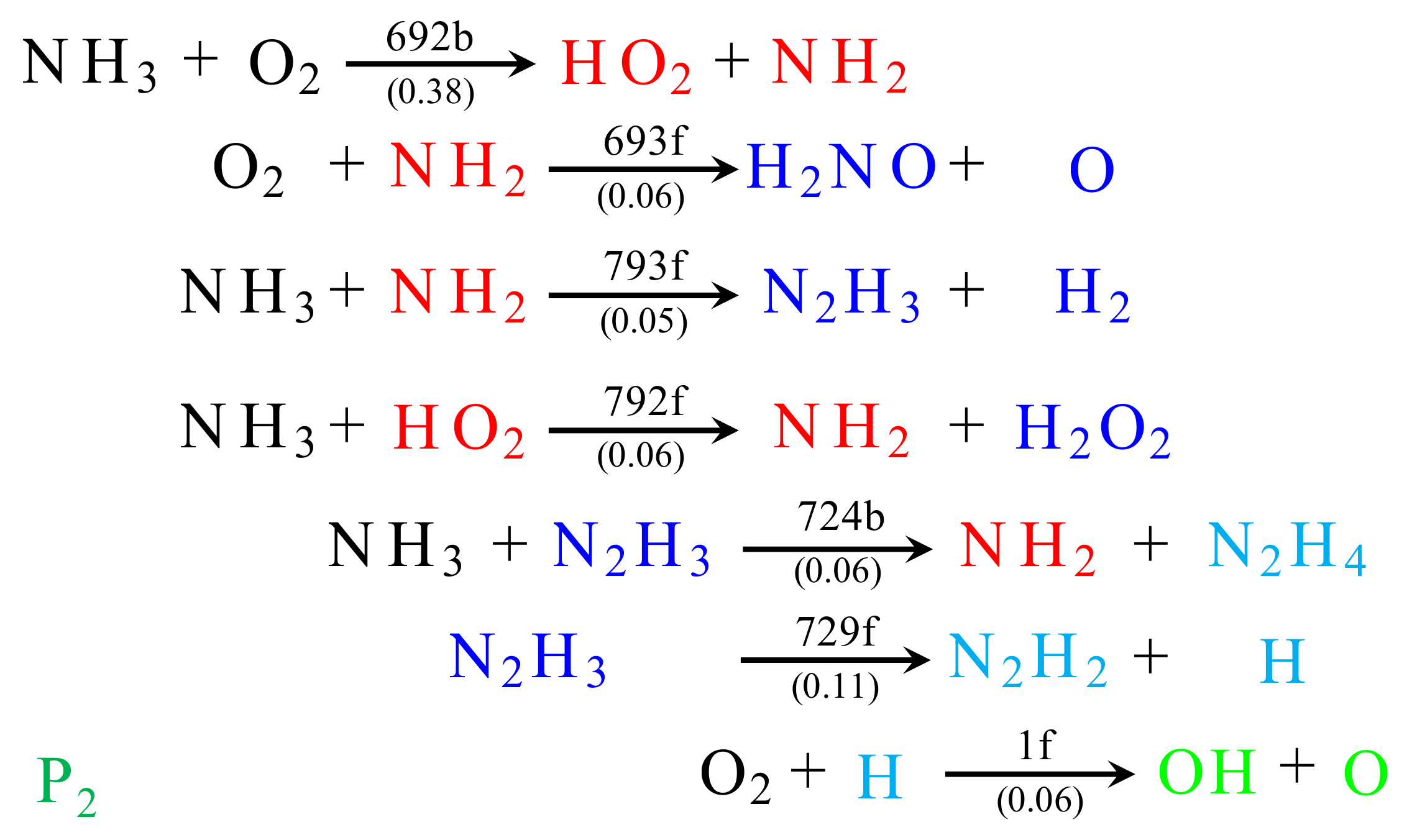} \hfill\hfill\hfill
\caption{Reactions exhibiting large APIs at points P$_1$ and P$_2$ in Shrestha$\_$2018; T$_0$=1100 K, p$_0$=2 atm and $\phi$=1.0.~Numbers in parentheses denote API values.~The different colors are meant to highlight several phases of the process as they are described in the text.}
\label{fig:API_Shrestha}
\end{figure}

The reactions contributing to the explosive dynamics during IDT in  Shrestha$\_$2018  are listed in Table \ref{tab:Shrestha_Reactions}.~Figure \ref{fig:API_Shrestha} displays the reactions exhibiting large APIs at P$_1$ and P$_2$.~At P$_3$ to P$_5$  the reactions exhibiting large APIs are similar to those exhibiting large TPIs.~Table \ref{tab:Shrestha_diagnostics} displays the TPI and Po indices for the explosive mode at P$_1$ to P$_5$, which are indicated in Fig.~\ref{fig:Trunc_timescales}.

\begin{table}[h]
\caption{The reactions contributing the most to the explosive time scale in Shrestha$\_$2018.~Bold/regular font represents exothermic/endothermic reactions.}
\begin{center}
\begin{tabular}{r@{~}c@{~}lr@{~}c@{~}l}
\hline
1f            		&: & $O_2$ + $H$ $\rightarrow$ $OH$ + $O$				&	\textbf{701f} 	&: & $NH_2$ + $NO_2$ $\rightarrow$ $N_2O$ + $H_2O$			\\
3f            		&: & $O$ + $H_2$ $\rightarrow$ $H$ + $OH$				&	\textbf{702f} 	&: & $NH_2$ + $NO_2$ $\rightarrow$ $H_2NO$ + $NO$			\\
4\textbf{f}/b 	&: & $OH$ + $H_2$ $\leftrightarrow$ $H$ + $H_2O$			&	704b          	&: & $NH_2$ + $HONO$ $\leftarrow$ $NH_3$ + $NO_2$			\\
\textbf{10f}	&: & $H$ + $O_2$ (+$M$) $\rightarrow$ $HO_2$ (+$M$)		&	718b          	&: & $NH_2$ + $NH_2$ (+$M$) $\leftarrow$ $N_2H_4$ (+$M$)	\\
17b           	& : &$O$ + $OH$ + $M$ $\leftarrow$ $HO_2$ + $M$		&	724b          	&: & $N_2H_4$ + $NH_2$ $\leftarrow$ $N_2H_3$ + $NH_3$		\\
\textbf{672f} 	&: & $NH$ + $OH$ $\rightarrow$ $HNO$ + $H$			&	729f          	&: & $N_2H_3$ $\rightarrow$ $N_2H_2$ + $H$				\\
\textbf{673f} 	&: & $NH$ + $OH$ $\rightarrow$ $NO$ + $H_2$			&	\textbf{742f} 	&: & $N_2H_3$ + $NH_2$ $\rightarrow$ $H_2NN$ + $NH_3$		\\
\textbf{691f} 	&: & $NH_2$ + $HO_2$ $\rightarrow$ $H_2NO$ + $OH$		&	789f          	&: & $NH_3$ + $H$ $\rightarrow$ $NH_2$ + $H_2$				\\
692b          	&: & $NH_2$ + $HO_2$ $\leftarrow$ $NH_3$ + $O_2$		&	792f			&: & $NH_3$ + $HO_2$ $\rightarrow$ $NH_2$ + $H_2O_2$		\\
693f          	&: & $NH_2$ + $O_2$ $\rightarrow$ $H_2NO$ + $O$		&	793f			&: & $NH_3$ + $NH_2$ $\rightarrow$ $N_2H_3$ + $H_2$		\\
\textbf{698f} 	&: & $NH_2$ + $NO$ $\rightarrow$ $N_2$ + $H_2O$		&	\textbf{852f} 	&: & $H_2NO$ + $NH_2$ $\rightarrow$ $HNO$ + $NH_3$		\\
699f          	&: & $NH_2$ + $NO$ $\rightarrow$ $NNH$ + $OH$			&	854f			&: & $H_2NO$ + $O_2$ $\rightarrow$ $HNO$ + $HO_2$			\\ 
\hline
\end{tabular}
\label{tab:Shrestha_Reactions}
\end{center}
\end{table}

Figure \ref{fig:API_Shrestha} shows that, according to Shrestha$\_$2018, the oxidation process at P$_1$ starts with reaction 692b (1.00): $NH_2$ + $HO_2$ $\leftarrow$ $NH_3$ + $O_2$, (the numbers in parentheses denote the API value).~At P$_2$ reaction 692b is still active, exhibiting the largest API.~The products of this reaction, $HO_2$ and $NH_2$, now activate reactions 693f (0.06): $NH_2$ + $O_2$ $\rightarrow$ $H_2NO$ + $O$, 793f (0.05) $NH_3$ + $NH_2$ $\rightarrow$ $N_2H_3$ + $H_2$ and 792f (0.06): $NH_3$ + $HO_2$ $\rightarrow$ $NH_2$ + $H_2O_2$.~The $N_2H_3$ generated activates reactions 724b (0.06): $N_2H_4$ + $NH_2$ $\leftarrow$ $N_2H_3$ + $NH_3$ and 729f (0.11): $N_2H_3$ $\rightarrow$ $N_2H_2$ + $H$.~In turn, the $H$ generated activates reaction 1f (0.06): $O_2$ + $H$ $\rightarrow$ $OH$ + $O$.~Therefore, the commencement of the oxidation process, according to Shrestha$\_$2018, initially involves a reaction producing $NH_2$ and $HO_2$ and then reactions producing $H_2NO$, $N_2H_3$, $N_2H_4$, $N_2H_2$, $OH$, $O$ and $H$.~Table \ref{tab:Shrestha_diagnostics} shows that  almost all of these products are identified by the Po as associated the most with the explosive mode at P$_1$ and P$_2$; e.g., $N_2H_3$ and $NH_2$ are the most pointed at P$_1$ and $H_2NO$, $N_2H_3$ and $NH_2$ at P$_2$.~The pointed species are reactants of the reactions with the largest TPIs, listed in Table \ref{tab:Shrestha_diagnostics}; e.g., $N_2H_3$,  $NH_2$ and $H_2NO$ are reactants of reactions 729f, 693f, 742f, 793f, 724b and 854f.

According to Fig.~\ref{fig:Trunc_timescales}, P$_3$ is positioned towards the end of the chemical runaway regime.~In comparison to P$_1$ and P$_2$, the CSP diagnostics on Table \ref{tab:Shrestha_diagnostics} show that  $N2$ chemistry is not influencing the explosive dynamics any longer, while the influence of $H_2NO$ and of $NO$ increases.~This is reflected in both the TPI and Po indices.~In particular, the reactions contributing the most to the explosive dynamics are 699f: $NH_2$ + $NO$ $\rightarrow$ $NNH$ + $OH$ and 854f: $H_2NO$ + $O_2$ $\rightarrow$ $HNO$ + $HO_2$, which involve as reactants $NO$ and $H_2NO$, that exhibit the largest Pos and produce $OH$ and $HO_2$.~The reactions providing the largest opposition are 698f: $NH_2$ + $NO$ $\rightarrow$ $N_2$ + $H_2O$ and 852f: $H_2NO$ + $NH_2$ $\rightarrow$ $HNO$ + $NH_3$, which consume $NO$ and $H_2NO$ towards mostly stable species.

This general pattern continues at P$_4$, which is located in the thermal runaway regime, where the $NH_2$ and $NO$ consuming reactions 699f and 698f keep providing the largest contribution and opposition to $\tau_e$, respectively.~A number of reactions that consume $H_2NO$ and $NH_2$ provide minor contributions.~The ones promoting $\tau_e$ generate $HO_2$ and $OH$ (854f, 691f, 1f), in tandem with the major contributor 699f that produces $OH$.~At this point, Po identifies the temperature as the variable related the most to the explosive mode.~This feature is in agreement with Fig.~\ref{fig:Trunc_timescales}, where it is shown that at about P$_4$ the temperature starts increasing sensibly.

\begin{table}[t]
\caption{The largest TPI and Po values for the explosive mode in Shrestha$\_$2018; T$_0$=1100 K, p$_0$=2 atm and $\phi$=1.0.~Numbers in parenthesis denote powers of ten.~Bold/regular font represents exothermic/endothermic reactions.}
\begin{center}
\begin{tabular}{rlrlrlrlrl}
\hline
\multicolumn{2}{c}{\textbf{P$_1$}} & \multicolumn{2}{c}{\textbf{P$_2$}} & \multicolumn{2}{c}{\textbf{P$_3$}} & \multicolumn{2}{c}{\textbf{P$_4$}} & \multicolumn{2}{c}{\textbf{P$_5$}} \\ \hline
\multicolumn{2}{c}{t$_1$ =0.00(0)s}    & \multicolumn{2}{c}{t$_2$ =5.24(-3)s}   & \multicolumn{2}{c}{t$_3$ =2.91(-2)s}    & \multicolumn{2}{c}{t$_4$ =9.80(-2)s}    & \multicolumn{2}{c}{t$_5$ =1.41(-1)s}    \\
\multicolumn{2}{c}{$\tau_{e,f}$=7.31(-3)s}    & \multicolumn{2}{c}{$\tau_{e,f}$=1.05(-2)s}    & \multicolumn{2}{c}{$\tau_{e,f}$=1.76(-2)s}    & \multicolumn{2}{c}{$\tau_{e,f}$=3.60(-2)s}    & \multicolumn{2}{c}{$\tau_{e,f}$=1.85(-6)s}    \\ \hline
\multicolumn{10}{c}{\textbf{TPI}}                                                                                                                                       \\ \hline
729f              	 & +0.20      	& 729f                   		& +0.12  & 699f               		& +0.18	& 699f          	& +0.21      	& 1f                	& +0.20       \\
1f                 	& +0.17      	& \textbf{742f}          		& ~-0.11  & 854f               		& +0.15	& \textbf{698f}	& ~-0.16      	& \textbf{4f}       & +0.13       \\
789f               	& ~-0.11      	& 854f                   		& +0.10  & \textbf{698f}      	& ~-0.11	& 854f        	& +0.05      	& 4b                	& ~-0.13       \\
724b               	& +0.09      	& 693f                   		& +0.08  & \textbf{852f}      	& ~-0.06	& \textbf{691f}	& +0.04           	& \textbf{673f}  	&  ~-0.04           \\
693f               	& +0.08      	& 724b                   		& +0.06  & \textbf{702f}      	& +0.06	&\textbf{702f}  	&  +0.04           	& 17b            	&  +0.03           \\
792f               	& +0.08      	& 1f                     		& +0.06  & \textbf{691f}      	& +0.05  	&\textbf{701f}  	& ~-0.04           	&  3f                 	&   +0.03          \\
718b               	& +0.07      	& 793f                   		& +0.06  & \textbf{701f}      	& ~-0.05 	&\textbf{852f}  	&   ~-0.03       	&\textbf{672f}   	&   ~-0.03          \\
793f               	& +0.07      	& 704b                   		& +0.05  &                    		&            	&    1f              	& +0.03          	&                   	&             \\
\textbf{10f}       	& ~-0.06      	& 789f                   		& ~-0.05  &                    		&            	&                    	&            		&                   	&             \\ \hline
\multicolumn{10}{c}{\textbf{Po}}                                                                                                                                        \\ \hline
$N_2H_3$ 	& +0.46      	& $H_2NO$                   	& +0.24  & $H_2NO$               	& +0.47  	& $T$         	& +0.88      	& $T$         	& +0.81       \\
$NH_2$     	& +0.22      	& $N_2H_3$                   	& +0.20  & $NO$                 	& +0.16   	& $H_2NO$  	& +0.05      	& $NH_3$  	& +0.43       \\
$HO_2$      	& +0.13      	& $NH_2$                    	& +0.18  & $NO_2$                	& +0.14   	& $NO$      	& +0.05      	& $H_2O$   	& ~-0.37       \\
$N_2H_4$     	& +0.12      	& $NO_2$                    	& +0.12  & $NH_2$                	& +0.07   	&                    	&            		& $N_2$        	& ~-0.17       \\
                   	&            		& $N_2H_2$                   	& +0.10  & $H_2O_2$               & +0.06    	&                    	&            		& $O_2$      	& +0.13       \\
                   	&            		& $N_2H_4$                   	& +0.09  &                    		&            	&                    	&            		& $H$    		& +0.07       \\
                   	&            		& \multicolumn{1}{l}{}   	&        &                    			&            	&                    	&            		& $OH$       	& +0.05       \\
                   	&            		& \multicolumn{1}{l}{}   	&        &                    			&            	&                    	&            		& $H_2$     	& +0.05       \\ \hline
\end{tabular}
\label{tab:Shrestha_diagnostics}
\end{center}
\end{table}
\vspace{0cm}

At P$_5$, Table \ref{tab:Shrestha_diagnostics} shows that hydrogen chemistry dominates the explosive dynamics, mainly through the $OH$-producing reaction 1f, while the next largest contributions originating from the two directions of reaction 4 cancel each other.~Several reactions provide minor contributions.~The ones supporting $\tau_e$ produce $OH$ (17b and 3f), while the ones opposing $\tau_e$ consume that species (673f and 672f).

\clearpage
\newpage

\renewcommand{\thesection}{\@Appendix C}
\setcounter{equation}{0}\renewcommand{\theequation}{C.\arabic{equation}}
\setcounter{figure}{0}\renewcommand{\thefigure}{C.\arabic{figure}}
\setcounter{table}{0}\renewcommand{\thetable}{C.\arabic{table}}
\section{CSP Diagnostics for the Li mechanism}
 \label{li}
 
\begin{figure}[b]
\centering
\hfill\hfill\includegraphics[scale=0.35]{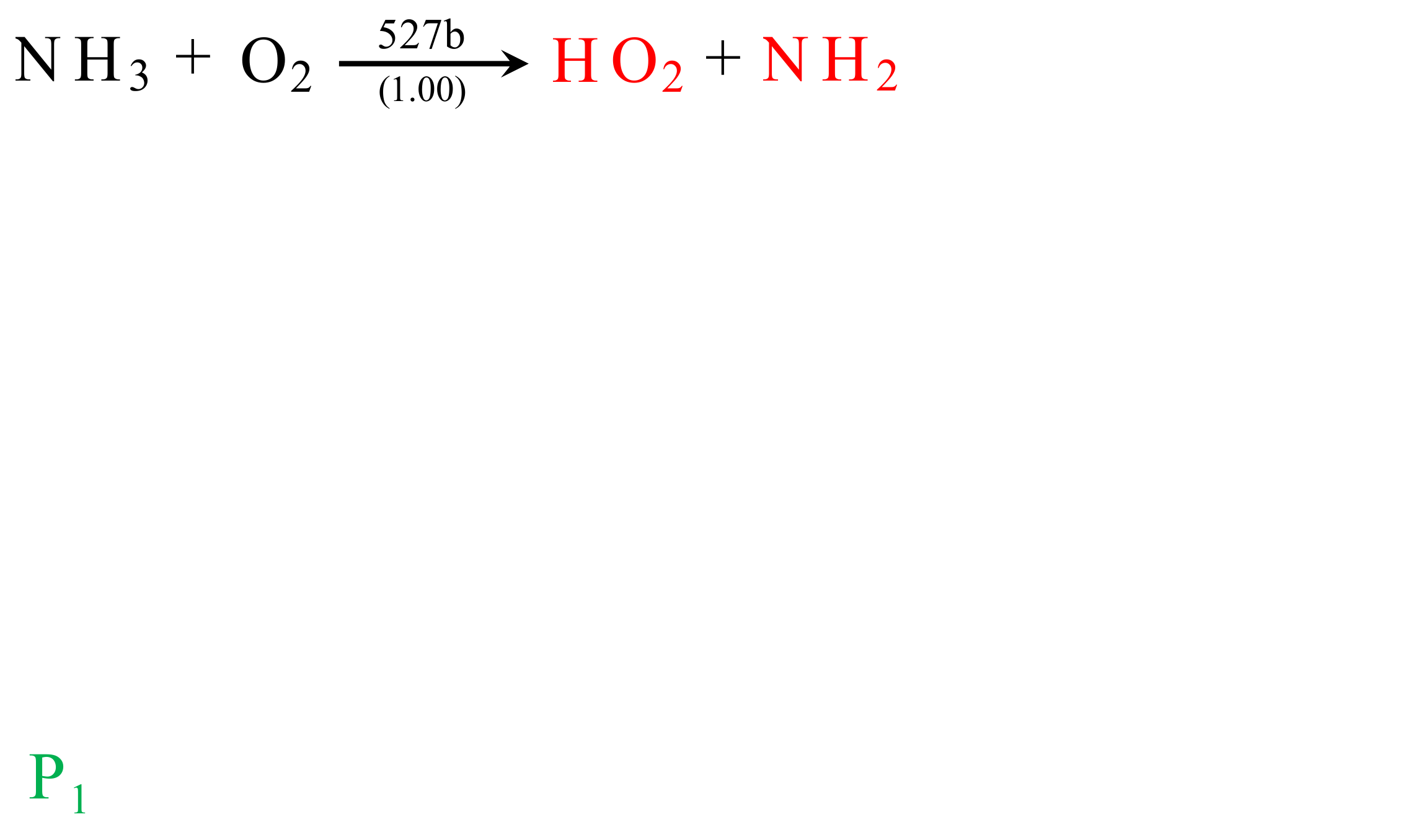}  \hfill \includegraphics[scale=0.35]{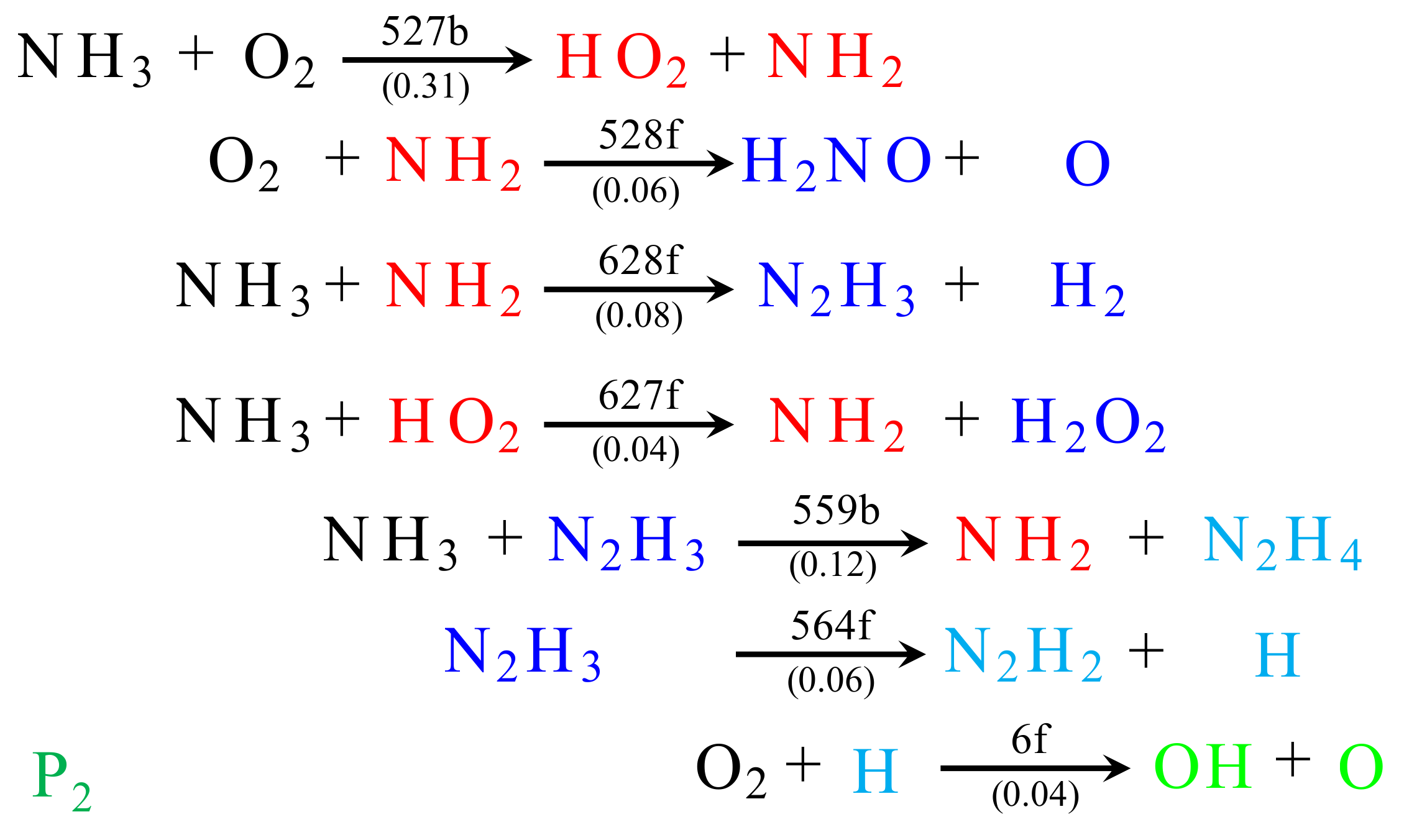} \hfill\hfill\hfill
\caption{Reactions exhibiting large APIs at points P$_1$ and P$_2$ in Li$\_$2019; T$_0$=1100 K, p$_0$=2 atm and $\phi$=1.0.~Numbers in parentheses denote API values.~The different colors are meant to highlight several phases of the process as they are described in the text.}
\label{fig:API_Li}
\end{figure}

 Table \ref{tab:Li_Reactions} displays the  reactions contributing to the explosive dynamics during IDT in Li$\_$2019.~The reactions exhibiting large APIs at P$_1$ and P$_2$ are displayed in Fig.~\ref{fig:API_Li}.~At P$_3$ to P$_5$  the reactions exhibiting large APIs are similar to those exhibiting large TPIs.~Table \ref{tab:Li_Diagnostics} displays the TPI and Po indices for the explosive mode at P$_1$ to P$_5$, shown in Fig.~\ref{fig:Trunc_timescales}.

\begin{table}[h]
\caption{The reactions contributing the most to the explosive time scale in Li$\_$2019.~Bold/regular font represents exothermic/endothermic reactions.}
\begin{center}
\begin{tabular}{r@{~}c@{~}lr@{~}c@{~}l}
\hline
3f			& : & $H_2$ + $O$  $       \rightarrow$ $H$ + $OH$			&	539b			& : & $NH_2$ + $HONO$ $\leftarrow$ $NH_3$ + $NO_2$			\\
4\textbf{f}/b	& : & $H_2$ + $OH$ $       \leftrightarrow$ $H$ + $H_2O$		&	553b			& : & $NH_2$ + $NH_2(+M)$ $\leftarrow$ $N_2H_4( + M)$		\\
6f			& : & $O_2$ + $H$ $        \rightarrow$ $O$ + $OH$			&	559b			& : & $N_2H_4$ + $NH_2$ $  \leftarrow$ $N_2H_3$ + $NH_3$		\\
\textbf{35f}	& : &  $N$ + $NO_2$      $\rightarrow$  $N_2O$ + $O$		&	564f			& : & $N_2H_3$ $           \rightarrow$ $N_2H_2$ + $H$			\\
\textbf{508f}	& : & $NH$ + $OH$  $       \rightarrow$ $NO$ + $H_2$		&	\textbf{577f}	& : & $N_2H_3$ + $NH_2$ $\rightarrow$ $H_2NN$ + $NH_3 $		\\
\textbf{526f}	& : & $NH_2$ + $HO_2$  $   \rightarrow$ $H_2NO$ + $OH$	&	\textbf{605f} 	& : & $HNOH$ + $NH_2$ $    \rightarrow$ $H_2NN$ + $H_2O$	\\
527b			& : & $NH_2$ + $HO_2$   $\leftarrow$ $NH_3$ + $O_2$		&	624f			& : & $NH_3$ + $H$ $       \rightarrow$ $NH_2$ + $H_2$			\\
528f			& : & $NH_2$ + $O_2$ $\rightarrow$ $H_2NO$ + $O$		&	627f			& : & $NH_3$ + $HO_2$      $\rightarrow$  $NH_2$ + $H_2O_2$	\\
\textbf{533f}	& : & $NH_2$ + $NO$  $     \rightarrow$ $N_2$ + $H_2O$	&	628f/\textbf{b}	& : & $NH_3$ + $NH_2$ $    \leftrightarrow$ $N_2H_3$ + $H_2$	\\
534f			& : & $NH_2$ + $NO$  $     \rightarrow$ $NNH$ + $OH$		&	\textbf{684f}	& : & $H_2NO$ + $NH_2$ $   \rightarrow$ $HNO$ + $NH_3$		\\
\textbf{536f}	& : & $NH_2$ + $NO_2$ $    \rightarrow$ $N_2O$ + $H_2O$	&	686f			& : & $H_2NO$ + $O_2$ $    \rightarrow$ $HNO$ + $HO_2$		\\ 
\textbf{537f}	& : & $NH_2$ + $NO_2$ $    \rightarrow$ $H_2NO$ + $NO$		\\
\hline
\end{tabular}
\label{tab:Li_Reactions}
\end{center}
\end{table}

A comparison of Fig.~\ref{fig:API_Li} and Table \ref{tab:Li_Diagnostics} with Fig.~\ref{fig:API_Shrestha} and Table \ref{tab:Shrestha_diagnostics}, respectively, reveals that the most important features related to the explosive dynamics of Li$\_$2019 are similar to those of Shrestha$\_$2018.

In particular, the oxidation process starts at P$_1$ with reaction 527b: $NH_2$ + $HO_2$ $\leftarrow$ $NH_3$ + $O_2$ and proceeds at P$_2$ with the addition of reactions 528f: $NH_2$ + $O_2$ $\rightarrow$ $H_2NO$ + $O$, 628f: $NH_3$ + $NH_2$ $\rightarrow$ $N_2H_3$ + $H_2$, 627f: $NH_3$ + $HO_2$ $\rightarrow$ $NH_2$ + $H_2O_2$, 559b: $N_2H_4$ + $NH_2$ $\leftarrow$ $N_2H_3$ + $NH_3$, 564f: $N_2H_3$ $\rightarrow$ $N_2H_2$ + $H$ and 6f: $O_2$ + $H$ $\rightarrow$ $OH$ + $O$.~Similarly to Shrestha$\_$2018, the products of these reactions at P$_1$ and P$_2$, notably among them $N_2H_3$, $H_2NO$ and $NH_2$, exhibit large Pos and are reactants of the reactions exhibiting the largest TPIs, as shown in Table \ref{tab:Li_Diagnostics}.~These reactions are similar to those exhibiting large TPIs at P$_1$ and P$_2$ in Shrestha$\_$2018, but re-ordered in terms of TPI.~For example, at P$_1$ the largest TPI in Li$\_$2019 is exhibited by reaction 559b: $N_2H_4$ + $NH_2$ $  \leftarrow$ $N_2H_3$ + $NH_3$, which has the fourth largest TPI in Shrestha$\_$2018, where it is denoted as reaction 724b.~At the same point, the largest TPI in Shrestha$\_$2018 is exhibited by reaction 729f: $N_2H_3$ $\rightarrow$ $N_2H_2$ + $H$, which has the fifth largest TPI in Li$\_$2019, where it is denoted as 564f.

From P$_3$ on, the similarities between Li$\_$2019 and Shrestha$\_$2018 keep increasing, in terms of both TPI and Po.~In particular, at P$_3$ and P$_4$, the major contributions to $\tau_e$ originate from reactions 534f: $NH_2$ + $NO$  $     \rightarrow$ $NNH$ + $OH$  and 686f: $H_2NO$ + $O_2$ $    \rightarrow$ $HNO$ + $HO_2$ (699f and 854f, respectively in Shrestha$\_$2018), while the major opposition originates from reaction 533f: $NH_2$ + $NO$  $     \rightarrow$ $N_2$ + $H_2O$ (698f in Shrestha$\_$2018).~Closely similar are the pointed variables; e.g. the most pointed $H_2NO$ at P$_3$ and $T$ at P$_4$.

Also, a stronger similarity between Li$\_$2019 and Shrestha$\_$2018 is manifested at P$_5$, where the explosive dynamics are dominated by hydrogen chemistry and $OH$ production and the most pointed variable is the temperature.

\begin{table}[h]
\caption{The largest TPI and Po values for the fast explosive mode in Li$\_$2019; T$_0$=1100 K, p$_0$=2 atm and $\phi$=1.0.~Numbers in parenthesis denote powers of ten.~Bold/regular font represents exothermic/endothermic reactions.}
\begin{center}
\begin{tabular}{rlrlrlrlrl}
\hline
\multicolumn{2}{c}{\textbf{P$_1$}} & \multicolumn{2}{c}{\textbf{P$_2$}} & \multicolumn{2}{c}{\textbf{P$_3$}} & \multicolumn{2}{c}{\textbf{P$_4$}} & \multicolumn{2}{c}{\textbf{P$_5$}} \\ \hline
\multicolumn{2}{c}{t$_1$ =0.00(0)s}    & \multicolumn{2}{c}{t$_2$ =5.15(-3)s}   & \multicolumn{2}{c}{t$_3$ =2.66(-2)s}    & \multicolumn{2}{c}{t$_4$ =9.00(-2)s}    & \multicolumn{2}{c}{t$_5$ =1.28(-1)s}    \\
\multicolumn{2}{c}{$\tau_{e,f}$=4.90(-3)s}    & \multicolumn{2}{c}{$\tau_{e,f}$=8.22(-3)s}    & \multicolumn{2}{c}{$\tau_{e,f}$=2.38(-2)s}    & \multicolumn{2}{c}{$\tau_{e,f}$=3.30(-2)s}    & \multicolumn{2}{c}{$\tau_{e,f}$=1.88(-6)s}    \\ \hline
\multicolumn{10}{c}{\textbf{TPI}}                                                                                                                                       \\ \hline
559b               	& +0.21      	& 628f                   		& +0.11  	& 534f               		& +0.20      		& 534f               	& +0.21	& 6f			& +0.24       \\
553b               	& +0.17      	& 539b                   		& +0.11  	& \textbf{533f}      	& ~-0.15      		& \textbf{533f}      	& ~-0.17	& \textbf{4f}	& +0.13       \\
628f               	& +0.15      	& 686f                   		& +0.09  	& 686f               		& +0.11      		& 686f               	& +0.05	& 4b			& ~-0.13       \\
6f                 	& +0.11      	& 559b                   		& +0.08  	& \textbf{684f}      	& ~-0.05      		& \textbf{536f}      	& ~-0.05	& \textbf{508f}	& ~-0.04            \\
564f               	& +0.10      	& 528f                   		& +0.08  	& \textbf{536f}      	& ~-0.05      		&  \textbf{537f}		& +0.04	& 3f			&   +0.04          \\
624f               	& ~-0.07      	& \textbf{577f}                   & ~-0.08  	& \textbf{526f}      	& +0.04            		&  \textbf{526f}         	&  +0.04	&  624f		& ~-0.03            \\
528f               	& +0.07      	& \textbf{628b}          		& ~-0.05  	& \textbf{605f}      	& ~-0.04            	&                    		&		&			&             \\
\textbf{35f}       	& ~-0.05      	&    					&        	& \textbf{537f}      	& +0.04   			&                    		&		&			&             \\ \hline
\multicolumn{10}{c}{\textbf{Po}}                                                                                                                                        \\ \hline
$N_2H_3$               & +0.41      & $NH_2$                    & +0.26  & $H_2NO$               & +0.42      & $T$                  & +0.86      & $T$                 & +0.80       \\
$NH_2$                & +0.28      & $H_2NO$                   & +0.25  & $NO$                 & +0.24      & $H_2NO$               & +0.05      & $NH_3$               & +0.46       \\
$N_2H_4$               & +0.22      & $NO_2$                    & +0.23  & $NO_2$                & +0.14      & $NO$                 & +0.05      & $H_2O$               & ~-0.39       \\
$HO_2$                & +0.06      & $N_2H_3$                   & +0.12  & $H_2$                 & ~-0.11      &                    &            & $N_2$                & ~-0.17       \\
                   &            & $N_2H_4$                   & +0.12  & $H_2O_2$               & +0.07      &                    &            & $O_2$                & +0.12       \\
                   &            & $H_2$                     & ~-0.09  & $NH_2$                & +0.07      &                    &            & $H$                 & +0.08       \\
                   &            & $N_2H_2$                   & +0.05  & $T$                  & +0.06      &                    &            & $OH$                & +0.05       \\ \hline
\end{tabular}
\label{tab:Li_Diagnostics}
\end{center}
\end{table}

\clearpage
\newpage

\renewcommand{\thesection}{\@Appendix D}
\setcounter{equation}{0}\renewcommand{\theequation}{D.\arabic{equation}}
\setcounter{figure}{0}\renewcommand{\thefigure}{D.\arabic{figure}}
\setcounter{table}{0}\renewcommand{\thetable}{D.\arabic{table}}
\section{CSP Diagnostics for the Stagni mechanism}
\label{stagni}

\begin{figure}[b]
\centering
\hfill\hfill\includegraphics[scale=0.36]{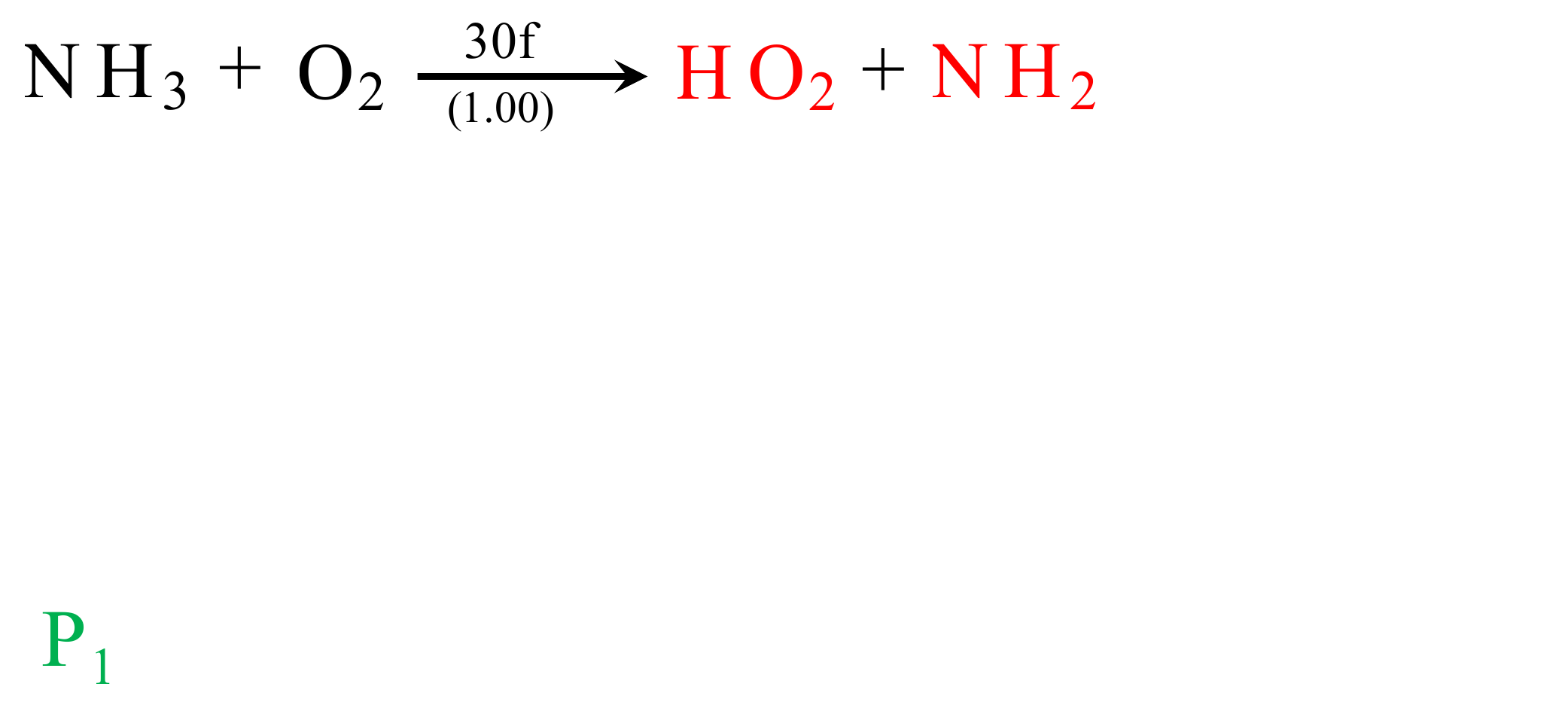}  \hfill \includegraphics[scale=0.36]{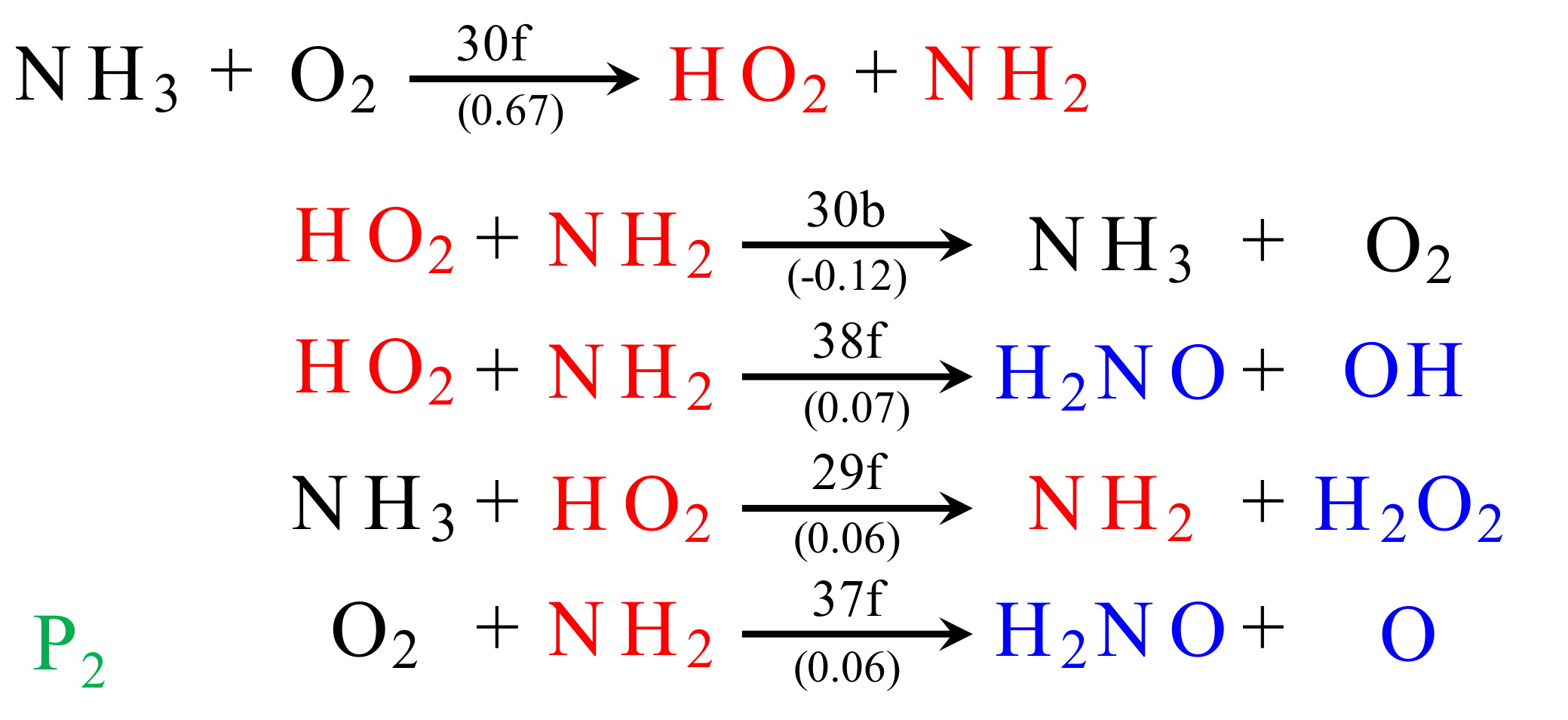} \hfill\hfill\hfill
\caption{Reactions exhibiting large APIs at points P$_1$ and P$_2$ in Stagni$\_$2020; T$_0$=1100 K, p$_0$=2 atm and $\phi$=1.0.~Numbers in parentheses denote API values.~The different colors are meant to highlight several phases of the process as they are described in the text.}
\label{fig:API_Stagni}
\end{figure}

 Table \ref{tab:Stagni_Reactions} lists the  reactions generating the explosive dynamics  in Stagni$\_$2020 during IDT. The reactions exhibiting large APIs at P$_1$ and P$_2$ are displayed in Fig.~\ref{fig:API_Stagni}, while at P$_3$ to P$_5$  the reactions exhibiting large APIs exhibit large TPIs as well.~Table \ref{tab:Stagni_diagnostics} displays the TPI and Po indices for the explosive mode at P$_1$ to P$_5$, that are indicated in Fig.~\ref{fig:Trunc_timescales}.

\begin{table}[h]
\caption{The reactions contributing the most to the explosive time scale in Stagni$\_$2020.~Bold/regular font represents exothermic/endothermic reactions.}
\begin{center}
\begin{tabular}{r@{~}c@{~}lr@{~}c@{~}l}
\hline
3\textbf{f}/b	&: & $H_2$ + $OH$ $    \leftrightarrow$ $H$ + $H_2O$			&	37f			&: & $NH_2$ + $O_2$ $  \rightarrow$ $H_2NO$ + $O$				\\
5f			&: & $O_2$ + $H$ $     \rightarrow$ $O$ + $OH$				&	\textbf{38f}	&: & $NH_2$ + $HO_2$ $ \rightarrow$ $OH$ + $H_2NO$				\\
9f			&: & $H_2O_2$ (+$M$) $     \rightarrow$ $OH$ + $OH$  (+$M$)	&	\textbf{74f}	&: & $NH_2$ + $NO_2$ $ \rightarrow$ $H_2NO$ + $NO$				\\
26f/\textbf{b}	&: & $NH_3$ + $H$ $     \leftrightarrow$ $H_2$ + $NH_2$			&	\textbf{76f}	&: & $NH_2$ + $NO$ $   \rightarrow$ $N_2$ + $H_2O$				\\
29f			&: & $NH_3$ + $HO_2$ $ \rightarrow$ $NH_2$ + $H_2O_2$		&	77f			&: & $NH_2$ + $NO$ $   \rightarrow$ $NNH$ + $OH$				\\
30f/\textbf{b}	&: & $NH_3$ + $O_2$ $  \leftrightarrow$ $NH_2$ + $HO_2$		&	\textbf{171f}	&: & $H_2NO$ + $NH_2$ $\rightarrow$ $HNO$ + $NH_3$			\\
\textbf{36f}	&: & $NH_2$ + $O_2$ $  \rightarrow$ $HNO$ + $OH$			&	172f			&: & $H_2NO$ + $O_2$ $ \rightarrow$ $HNO$ + $HO_2$				\\ 
\hline
\end{tabular}
\label{tab:Stagni_Reactions}
\end{center}
\end{table}

The explosive dynamics in Stagni$\_$2020 resembles to an extent that of Glarborg$\_$2018.~In particular, as shown in Fig.~\ref{fig:API_Stagni}, the oxidation process commences at P$_1$ with reaction 30f (1.00): $NH_3$ + $O_2$ $  \rightarrow$ $NH_2$ + $HO_2$, which is one of the two initiation reactions in  Glarborg$\_$2018; see Fig.~\ref{fig:API_Glarborg}.~At P$_2$, this reaction retains a dominant impact, assisted by reactions 38f (0.07): $NH_2$ + $HO_2$ $ \rightarrow$ $OH$ + $H_2NO$, 29f (0.06): $NH_3$ + $HO_2$ $ \rightarrow$ $NH_2$ + $H_2O_2$ and 37f (0.06): $NH_2$ + $O_2$ $  \rightarrow$ $H_2NO$ + $O$; the last two also encountered in Glarborg$\_$2018.~However, in this case there is a strong opposition to the impact of the explosive mode from 30b (-0.12): $NH_3$ + $O_2$ $  \leftarrow$ $NH_2$ + $HO_2$.~It is due to the action of this reaction that the chemical runaway is terminated very fast, as shown in Fig.~\ref{fig:Trunc_timescales}, since the action of the initiation reaction 30f is suppressed.~The products of reactions 30f, 38f, 29f and 37f, that promote the impact of the explosive mode (large and positive APIs) exhibit large Pos and are  reactants of the reactions exhibiting large TPIs, listed in Table \ref{tab:Stagni_diagnostics}.~For example, at P$_1$, $NH_2$ ($Po=0.71$) is the product of reaction 30f ($API=1.00$) and a reactant of reaction 37f: $NH_2$ + $O_2$ $  \rightarrow$ $H_2NO$ + $O$ ($TPI=0.64$).~Similarly, at P$_2$, $NH_2$ ($Po=0.38$) is a product of reactions 30f ($API=0.67$) and 29f ($API=0.06$) and a reactant of 30b: $NH_3$ + $O_2$ $  \leftarrow$ $NH_2$ + $HO_2$ ($TPI=-0.34$), 37f: $NH_2$ + $O_2$ $  \rightarrow$ $H_2NO$ + $O$ ($TPI=0.23$) and 38f: $NH_2$ + $HO_2$ $ \rightarrow$ $OH$ + $H_2NO$ ($TPI=0.1$6).

According to Table \ref{tab:Stagni_diagnostics}, at P$_1$ and P$_2$ the reactions promoting the explosive character of the mode (positive TPIs) the most  are 37f at P$_1$ and 37f and 38f at P$_2$.~What distinguishes this case from all previous ones, is the strong opposition to the explosive character of the mode (negative TPI) by reaction 30b: $NH_3$ + $O_2$ $  \leftarrow$ $NH_2$ + $HO_2$, which exhibits the largest TPI in magnitude right after the initiation of the process.~This is due to the fact that reactions 37f and 38f compete with reaction 30b for the same reactants and to the fact that the products of 30b are stable molecules, while those of 37f and 38f are not.~This action of reaction 30b is in contrast to the action of its forward direction, which is the one promoting the most the impact of the explosive  mode, as evidenced by the fact that it exhibits the largest API at both P$_1$ and P$_2$; see Fig.~\ref{fig:API_Stagni}.~As a result of this action of reaction 30b, $\tau_e$ decelerates abruptly, as shown in Fig.~\ref{fig:Trunc_timescales}.

\begin{table}[t!]
\caption{The largest TPI and Po values for the fast explosive mode in Stagni$\_$2020; T$_0$=1100 K, p$_0$=2 atm and $\phi$=1.0.~Numbers in parenthesis denote powers of ten.~Bold/regular font represents exothermic/endothermic reactions.}
\begin{center}
\begin{tabular}{rlrlrlrlrl}
\hline
\multicolumn{2}{c}{\textbf{P$_1$}} & \multicolumn{2}{c}{\textbf{P$_2$}} & \multicolumn{2}{c}{\textbf{P$_3$}} & \multicolumn{2}{c}{\textbf{P$_4$}} & \multicolumn{2}{c}{\textbf{P$_5$}} \\ \hline
\multicolumn{2}{c}{t$_1$ =0.00(0)s}    & \multicolumn{2}{c}{t$_2$ =3.40(-4)s}    & \multicolumn{2}{c}{t$_3$ =1.40(-1)s}    & \multicolumn{2}{c}{t$_4$ =4.53(-1)s}    & \multicolumn{2}{c}{t$_5$ =6.44(-1)s}    \\
\multicolumn{2}{c}{$\tau_{e,f}$=4.65(-3)s}    & \multicolumn{2}{c}{$\tau_{e,f}$=1.80(-2)s}    & \multicolumn{2}{c}{$\tau_{e,f}$=4.56(-1)s}    & \multicolumn{2}{c}{$\tau_{e,f}$=1.81(-1)s}    & \multicolumn{2}{c}{$\tau_{e,f}$=6.97(-6)s}    \\ \hline
\multicolumn{10}{c}{\textbf{TPI}}                                                                                                                                       \\ \hline
37f                	& +0.64      & \textbf{30b}          	 	& -0.34  	& \textbf{30b}	& -0.17     	& \textbf{30b}   	& -0.17     	&5f      	      	& +0.21       \\
172f               	& +0.14      & 37f                    		& +0.23  	& \textbf{38f}     & +0.13   	& \textbf{38f}   	& +0.12   	&\textbf{3f}      	& +0.11       \\
29f               	& +0.09      & \textbf{38f}          		& +0.16  	& 30f           	& +0.08   	& 30f                	& +0.09    	&3b     	 	& ~-0.10      \\
\textbf{36f}	& +0.05      & 172f                   		& +0.08  	& \textbf{76f}     & -0.08    	& \textbf{76f}   	& -0.08     	& 26f     	  	& ~-0.07   \\
9f                 	& +0.05      & \textbf{36f}		   	& +0.02    	& 172f               & +0.06   	& 77f                	& +0.06    	&\textbf{26b}  	& +0.05   \\
                   	&            	& \multicolumn{1}{l}{}   	&        	& \textbf{74f}     & +0.05   	& 172f               & +0.05    	&                   	&	             \\  \hline
\multicolumn{10}{c}{\textbf{Po}}                                                                                                                                        \\ \hline
$NH_2$		& +0.71	& $NH_2$                   	 & +0.38	& $T$	& +0.95	& $T$	& +0.98	& $T$		& +0.96       \\
$H_2NO$		& +0.13	& $H_2NO$                   	& +0.35	& $H_2$	& +0.06	& $H_2$	& +0.04	& $O_2$		& ~-0.59       \\
$HO_2$		& +0.10	& $NO_2$           	   	 & +0.11	&		&		&		&		& $NH_3$		& +0.38       \\
$H_2O_2$	& +0.05	& $N_2H_4$			& +0.07	&		&		&		&		& $H_2$		& +0.22       \\
			&		& \multicolumn{1}{l}{}		&        	&		&		&		&		& $H_2O$		& ~-0.16       \\
			&		& \multicolumn{1}{l}{}   	&        	&		&		&		&		& $H$		& +0.07       \\ 
			&		& \multicolumn{1}{l}{}   	&        	&		&		&		&		& $OH$		& +0.07       \\ \hline
\end{tabular}
\label{tab:Stagni_diagnostics}
\end{center}
\end{table}

The situation at P$_3$ and P$_4$ is an extension of that at P$_2$.~In particular, the opposing action of 30b is still present, counteracting 38f, which is the strongest proponent of explosive dynamics at  P$_3$ and P$_4$.~The influence of reaction 38f is still assisted by that of reactions 37f and 172f in producing $OH$ by recycling $HO_2$ and $H_2NO$.~This promoting action is further supported by reactions 30f (which produces the reactants of the most promoting reaction 38f), 74f (which produces the reactant of 172f)  and 77f (which produces OH).~The opposing action of 30b is supported by reaction 76f, which consumes $NH_2$ that is the reactant of the promoting reactions 38f, 74f and 77f.~According to Fig.~\ref{fig:Trunc_timescales}, P$_3$ and P$_4$ belong to the thermal runaway, therefore the most pointed variable is the temperature, as shown in Table \ref{tab:Stagni_diagnostics}.~In fact, at P$_4$ the combined action of the promoting reactions leads to $OH$ generation.~In particular, reactions 38f and 77f produce $OH$, while reactions 30f and 172f produce $NH_2$ and $HO_2$ that are reactants of 38f and 77f.~At P$_5$ hydrogen chemistry dominates and the $OH$ generation intensifies.~This species is generated by reaction 5f.~The $H$-consuming reactions 3b and 2f oppose the explosive character of the mode, since $H$ is reactant of the most promoting reaction 5f.~The most pointed variable is still temperature.

It is noted that in Stagni$\_$2020 there is no substantial influence of $N2$-chemistry.

\clearpage
\newpage

\renewcommand{\thesection}{\@Appendix E}
\setcounter{equation}{0}\renewcommand{\theequation}{E.\arabic{equation}}
\setcounter{figure}{0}\renewcommand{\thefigure}{E.\arabic{figure}}
\setcounter{table}{0}\renewcommand{\thetable}{E.\arabic{table}}
\section{CSP Diagnostics for Zhang mechanism}
\label{zhang}

\begin{figure}[b]
\centering
\hfill\hfill\includegraphics[scale=0.36]{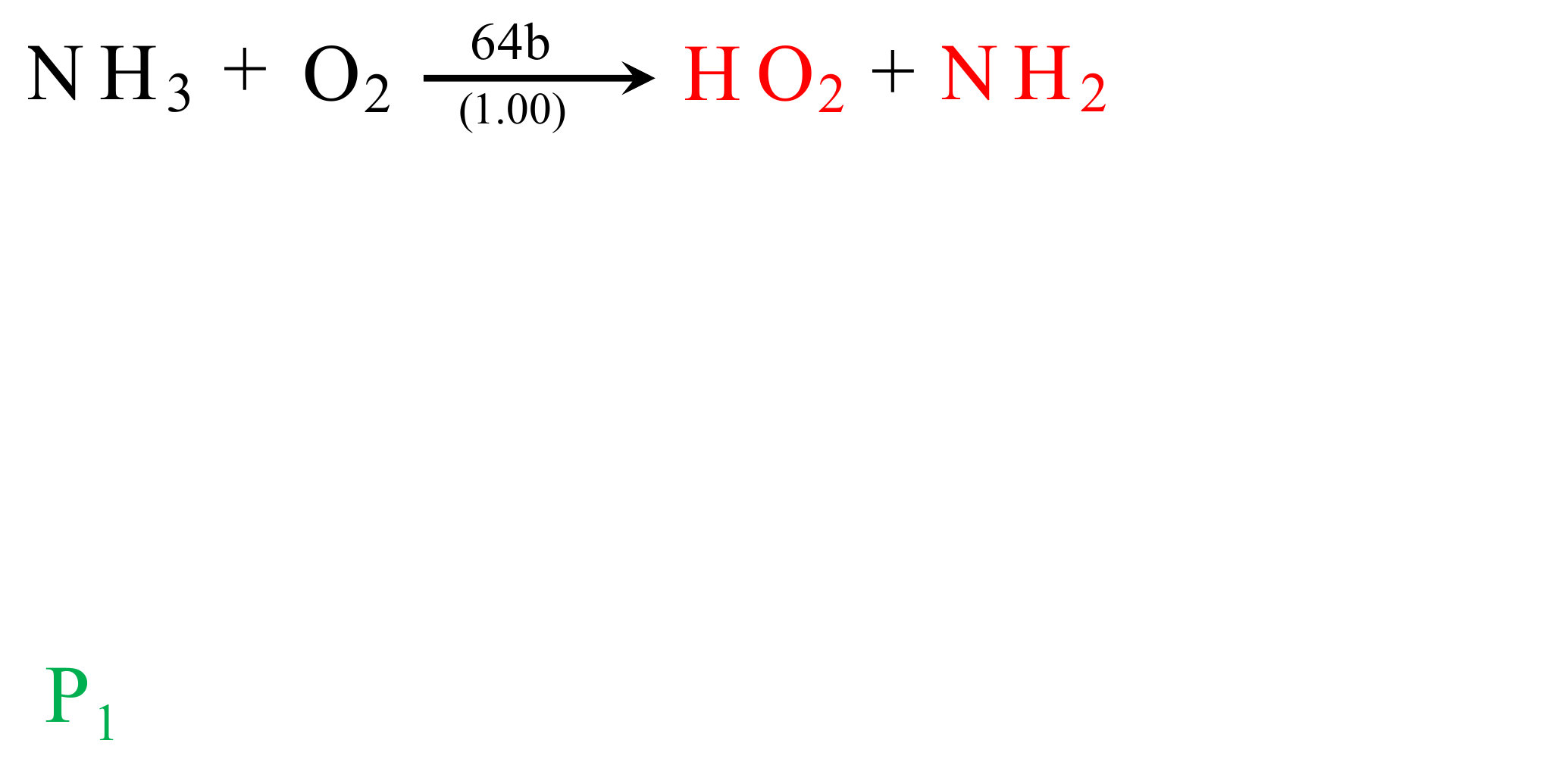}  \hfill \includegraphics[scale=0.36]{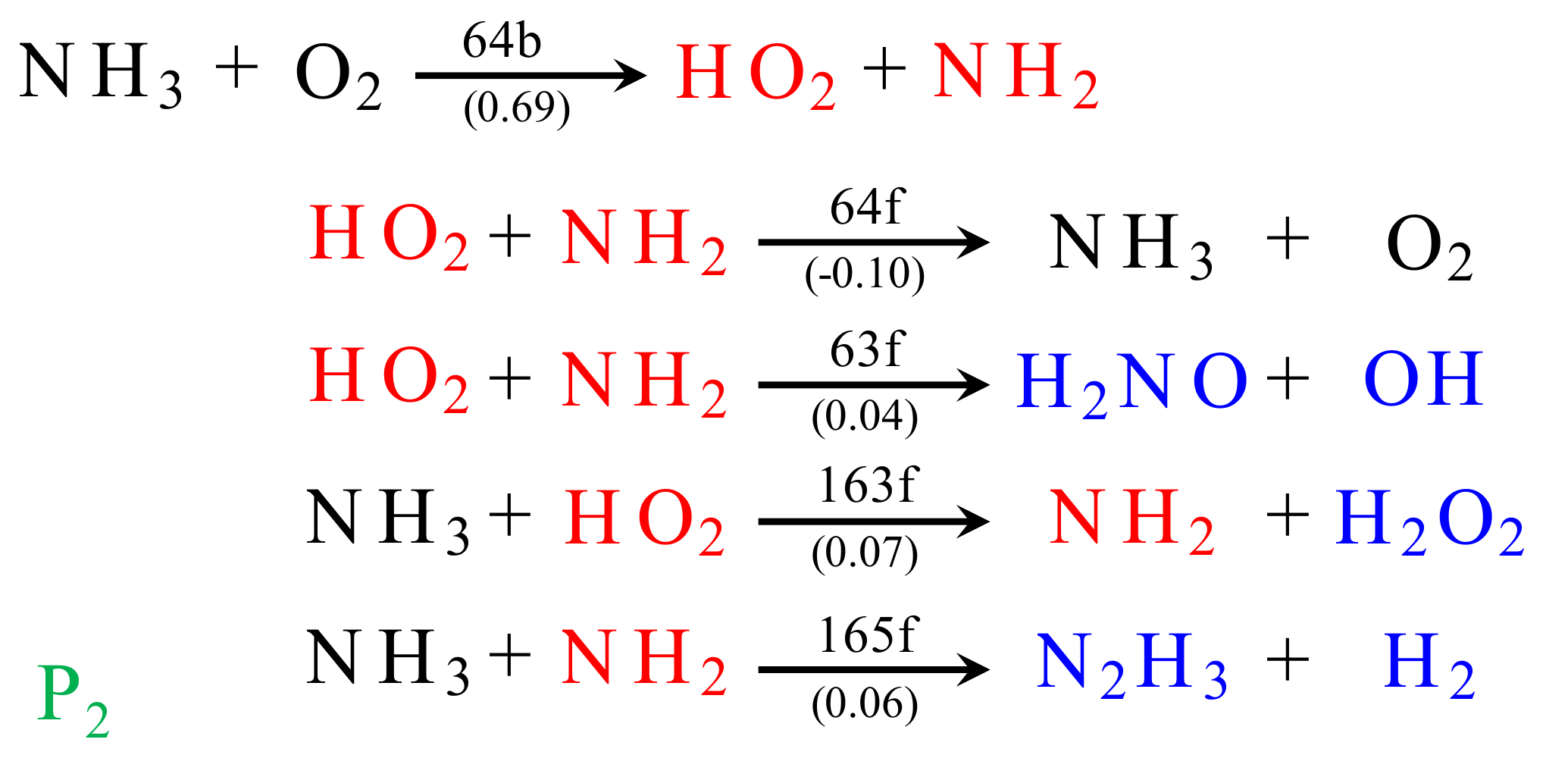} \hfill\hfill\hfill
\caption{Reactions exhibiting large APIs at points P$_1$ and P$_2$ in Zhang$\_$2021; T$_0$=1100 K, p$_0$=2 atm and $\phi$=1.0.~Numbers in parentheses denote API values.~The different colors are meant to highlight several phases of the process as they are described in the text.}
\label{fig:API_Zhang}
\end{figure}

The reactions contributing to the explosive dynamics during IDT in Zhang$\_$2021 are listed in Table \ref{tab:Zhang_Reactions}.~The reactions exhibiting large APIs at P$_1$ and P$_2$ are displayed in Fig.~\ref{fig:API_Zhang}, while those exhibiting large APIs at P$_3$ to P$_5$   are similar to those exhibiting large TPIs.~Table \ref{tab:Zhang_diagnostics} displays the TPI and Po indices for the explosive mode at P$_1$ to P$_5$, shown in Fig.~\ref{fig:Trunc_timescales}.

\begin{table}[h]
\caption{The reactions contributing the most to the explosive time scale in Zhang$\_$2021.~Bold/regular font represents exothermic/endothermic reactions.}
\begin{center}
\begin{tabular}{r@{~}c@{~}lr@{~}c@{~}l}
\hline
1f			&: & $H$ + $O_2$ $\rightarrow$ $O$ + $OH$           			&	\textbf{69f}	&: & $NH_2$ + $NO$ $\rightarrow$ $N_2$ + $H_2O$       		\\
3f			&: & $O$ + $H_2$ $\rightarrow$ $OH$ + $H$          			&	\textbf{89f}	&: & $N_2H_2$ + $HO_2$  $\rightarrow$ $NNH$ + $H_2O_2$		\\
4\textbf{f}/b	&: & $OH$ + $H_2$ $\leftrightarrow$ $H$ + $H_2O$        		&	103f			&: & $N_2H_3(+M)$ $\rightarrow$ $N_2H_2$ + $H(+M)$ 			\\
\textbf{58f}	&: & $NH_2$ + $H$ $\rightarrow$ $NH$ + $H_2$        		&	161f			&: & $NH_3$ + $H$ $\rightarrow$ $NH_2$ + $H_2$        			\\
\textbf{63f}	&: & $NH_2$ + $HO_2$ $\rightarrow$ $H_2NO$ + $OH$     	&	163f			&: & $NH_3$ + $HO_2$ $\rightarrow$ $NH_2$ + $H_2O_2$     	\\
64\textbf{f}/b	&: & $NH_2$ + $HO_2$ $\leftrightarrow$ $NH_3$ + $O_2$      	&	165f			&: & $NH_3$ + $NH_2$ $\rightarrow$ $N_2H_3$ + $H_2$     		\\
65f			&: & $NH_2$ + $O_2$ $\rightarrow$ $H_2NO$ + $O$       	&	221f			&: & $H_2NO$ + $O_2$ $\rightarrow$ $HNO$ + $HO_2$ 			\\   
68f			&: & $NH_2$ + $NO$ $\rightarrow$ $NNH$ + $OH$       			\\	\hline
\end{tabular}
\label{tab:Zhang_Reactions}
\end{center}
\end{table}

The explosive dynamics in Zhang$\_$2021 resembles closely  that of Stagni$\_$2020 and to an extent that of Glarborg$\_$2018.~As shown in Fig.~\ref{fig:API_Zhang}, the oxidation process commences at P$_1$ with reaction 64b (1.00): $NH_3$ + $O_2$ $  \rightarrow$ $NH_2$ + $HO_2$, which is the same as the initiation reaction in Stagni$\_$2020 and one of the two initiation reactions in  Glarborg$\_$2018; see Figs.~\ref{fig:API_Stagni} and \ref{fig:API_Glarborg}.~With one exception, the reactions contributing significantly to the impact of the explosive mode at P$_2$ (large APIs) are the same as in Stagni$\_$2020.~That is, the dominant impact is provided by reaction 64b (0.69), assisted by reaction 63f (0.04): $NH_2$ + $HO_2$ $\rightarrow$ $H_2NO$ + $OH$, 163f (0.07): $NH_3$ + $HO_2$ $\rightarrow$ $NH_2$ + $H_2O_2$ and 165f (0.06): $NH_3$ + $NH_2$ $\rightarrow$ $N_2H_3$ + $H_2$.~Reactions 64b, 63f and 163f are active at this point in Stagni$\_$2020, while reaction 165f is not.~It is noted that, contrary to Stagni$\_$2020, now there is substantial $N2$ chemistry via reaction 165f.~Similarly to Stagni$\_$2020, the major opposition to the impact of the explosive dynamics is provided by reaction 64f (-0.10): $NH_3$ + $O_2$ $  \leftarrow$ $NH_2$ + $HO_2$, which again leads to a fast termination of the chemical runaway, as shown in Fig.~\ref{fig:Trunc_timescales}, since it counteracts the influence of the most important initiation reaction 64b.

Similarly to all previous cases, the products of reactions 64b, 63f, 163f, 165f, which exhibit large and positive APIs, thus promoting the impact of the explosive mode, demonstrate large Pos and are  reactants of the reactions with large TPIs, listed in Table \ref{tab:Zhang_diagnostics}.~According to Table \ref{tab:Zhang_diagnostics}, $\tau_e$ at P$_1$ is supported by reactions 165f and 65f, which consume $NH_2$ that is produced there by the large-API reaction 64b.~Further support  to $\tau_e$ is provided by reaction 103f that consumes $N_2H_3$ produced by reaction 165f.~Reaction 103f also produces $H$, which feeds into reaction 1f that also supports $\tau_e$.~The main opposition to the explosive character of the mode at P$_1$ is provided by reaction 161f, which consumes $NH_3$ and $H$ that are reactants of the main supporting reactions.

At P$_2$ the reactions supporting $\tau_e$ the most are 165f: $NH_3$ + $NH_2$ $\rightarrow$ $N_2H_3$ + $H_2$   and 1f: $H$ + $O_2$ $\rightarrow$ $O$ + $OH$ that survive from P$_1$, in addition to 63f: $NH_2$ + $HO_2$ $\rightarrow$ $H_2NO$ + $OH$  and 221f: $H_2NO$ + $O_2$ $\rightarrow$ $HNO$ + $HO_2$.~Similarly to Stagni$\_$2020 and in contrast to all other cases considered, there is a strong opposition to the explosive character realised by reaction 64f: $NH_2$ + $HO_2$ $\leftrightarrow$ $NH_3$ + $O_2$, the reverse direction of which provides the largest contribution to the impact of the explosive mode (largest API).~Like Stagni$\_$2020, this action of reaction 64f decelerates the process.

\begin{table}[t!]

\caption{The largest TPI and Po values for the fast explosive mode in Zhang$\_$2021; T$_0$=1100 K, p$_0$=2 atm and $\phi$=1.0.~Numbers in parenthesis denote powers of ten.~Bold/regular font represents exothermic/endothermic reactions.}
\begin{center}
\begin{tabular}{rlrlrlrlrl}
\hline
\multicolumn{2}{c}{\textbf{P$_1$}} & \multicolumn{2}{c}{\textbf{P$_2$}} & \multicolumn{2}{c}{\textbf{P$_3$}} & \multicolumn{2}{c}{\textbf{P$_4$}} & \multicolumn{2}{c}{\textbf{P$_5$}} \\ \hline
\multicolumn{2}{c}{t$_1$ =0.00(0)s}    & \multicolumn{2}{c}{t$_1$ =3.24(-4)s}    & \multicolumn{2}{c}{t$_3$ =7.64(-2)s}    & \multicolumn{2}{c}{t$_4$ =2.65(-1)s}    & \multicolumn{2}{c}{t$_5$ =3.78(-1)s}   \\
\multicolumn{2}{c}{$\tau_{e,f}$=2.52(-3)s}    & \multicolumn{2}{c}{$\tau_{e,f}$=9.45(-3)s}    & \multicolumn{2}{c}{$\tau_{e,f}$=2.81(-1)s}    & \multicolumn{2}{c}{$\tau_{e,f}$=1.03(-1)s}    & \multicolumn{2}{c}{$\tau_{e,f}$=2.23(-6)s}    \\ \hline
\multicolumn{10}{c}{\textbf{TPI}}                                                                                                                                       \\ \hline
161f    	&  ~-0.19  & \textbf{64f}	& ~-0.24	& \textbf{63f}	& +0.19  	& \textbf{64f}           & ~-0.18  & 1f                     & +0.23  \\
165f        	& +0.18  	& 165f               & +0.16 	& \textbf{64f}  	& ~-0.18  	& \textbf{63f}           & +0.17  & \textbf{4f}            & +0.14  \\
103f     	& +0.18  	& \textbf{63f} 	& +0.09 	& 64b           	& +0.10  	& 64b                    & +0.10  & 4b                     & ~-0.13  \\
1f    		& +0.17  	& 161f    		& ~-0.08	& \textbf{69f}   	& ~-0.09  	& \textbf{69f}           & ~-0.09  & \textbf{58f}                        & ~-0.04        \\
65f   		& +0.11  	& 1f        		& +0.07	& 221f     		& +0.06  	& 221f                   & +0.06  & 3f                       & +0.04        \\
163f        	& +0.06  	& 221f    		& +0.07 	& 68f     		& +0.04	& 68f                        &+0.05        &161f                        &~-0.04        \\
		&		&\textbf{89f}	&+0.04	& \textbf{219f}   & ~-0.04	&                          &         &                         &         \\ \hline
\multicolumn{10}{c}{\textbf{Po}}                                                                                                                                                 \\ \hline
$NH_2$                    & +0.49  & $N_2H_2$               & +0.38      & $T$                      & +1.01  & $T$                      & +0.99  & $T$                      & +0.82  \\
$N_2H_3$                   & +0.29  & $H_2NO$               & +0.33      &                      &   &                  &   & $NH_3$                   & +0.53  \\
$HO_2$                    & +0.10  & $NH_2$                & +0.20      &                     &   &                        &        & $H_2O $                   & ~-0.46  \\
$H_2O_2$                   & +0.05  & $H_2$                 & ~-0.19      &                        &        &                        &        & $N_2$                     & ~-0.18  \\
\multicolumn{1}{l}{}   &        & $N_2H_3$              & +0.17      & \multicolumn{1}{l}{}   &        & \multicolumn{1}{l}{}   &        & \multicolumn{1}{l}{}   &        \\ \hline
\end{tabular}                                                                                                               
\label{tab:Zhang_diagnostics}
\end{center}
\end{table}

A comparison of the CSP diagnostics in  Stagni$\_$2020 and Zhang$\_$2021, displayed in Tables \ref{tab:Stagni_diagnostics} and \ref{tab:Zhang_diagnostics}, reveals that columns P$_2$ to P$_4$ display similar results, with the exception of the most promoting reaction in P$_2$, which is 37f: $NH_2$ + $O_2$ $  \rightarrow$ $H_2NO$ + $O$ in  Stagni$\_$2020 and 165f: $NH_3$ + $NH_2$ $\rightarrow$ $N_2H_3$ + $H_2$ in Zhang$\_$2021.~In particular, at P$_3$ and P$_4$ the opposing influence of reaction 64f persists, assisted by reaction 69f, both consuming $NH_2$.~The major proponent of $\tau_e$ are the $NH_2$ and $HO_2$-consuming reaction 63f, the $NH_3$ and $O_2$ consuming reaction 64b, the $H_2NO$-consuming reaction 221f and the $NH_2$-consuming reaction 68f.~As in Stagni$\_$2020, the combined action of the promoting reactions 63f, 64b, 221f and 68f leads to $OH$-generation.~This is achieved directly by reactions 63f and 68f and indirectly by 64b and 221, which produce $NH_2$ and $HO_2$ that are reactants of 63f and 68f.~P$_3$ and P$_4$ belong to the thermal runaway, as shown in Fig.~\ref{fig:Trunc_timescales}.~Therefore, the variable associated the most to the explosive mode is the temperature, as shown in Table \ref{tab:Zhang_diagnostics}.

At P$_5$, where $\tau_e$ records minimum value, the reactions supporting $\tau_e$ the most relate to hydrogen chemistry, similarly to all other cases considered.~In particular, the largest contribution originates from reaction $H$ + $O_2$ $\rightarrow$ $O$ + $OH$, assisted by reaction 3f: $O$ + $H_2$ $\rightarrow$ $OH$ + $H$ , while the contributions of the two directions of reaction 4: $OH$ + $H_2$ $\leftrightarrow$ $H$ + $H_2O$  effectively cancel.~The major opposition originate from reactions 58f and 161f, which relate to nitrogen chemistry and consume $H$, which is reactant the major proponent to $\tau_e$, reaction 1f.~The temperature remains the most pointed variable.
\clearpage
\newpage



\bibliography{Ammonia_Ref.bib}

\bibliographystyle{unsrtnat}






\end{document}